\documentclass[conference]{IEEEtran}

\ifCLASSINFOpdf
\else
\fi

\usepackage{cite}
\usepackage{amsmath,amssymb,amsfonts}
\usepackage{algorithmic}
\usepackage{graphicx}
\usepackage{textcomp}
\usepackage{bmpsize}
\usepackage[svgnames]{xcolor}
\usepackage{colortbl}
\usepackage{lipsum}
\usepackage{comment}
\usepackage{cancel}
\usepackage{subcaption}
\usepackage{xspace}
\usepackage{hhline}
\usepackage[utf8]{inputenc}

\usepackage{tikz}
\usepackage{paralist}
\usepackage{tabularx}
\usepackage{booktabs}
\usepackage{comment}
\usepackage{multirow}
\usepackage{subcaption}
\usepackage{url}
\usepackage[most]{tcolorbox}
\usepackage{enumitem}

\usepackage{fontawesome5}
\usepackage{pifont}

\newcommand{\sysname}[0]{\textsf{VR ProfiLens}}

\newcommand{\parheading}[1]{\vspace{2pt}\noindent{}\textbf{{#1}}}

\newcommand{\eg}{e.g.,}
\newcommand{\ie}{i.e.,}

\newcommand{\etal}{et al.}
\newcommand{\wrt}{w.r.t.}
\newcommand{\xmark}{\textcolor{red}{\ding{55}}}

\newcommand{\steam}{SteamVR}
\newcommand{\users}{20}
\newcommand{\apps}{10}
\newcommand{\attributes}{29}
\newcommand{\vrdevice}{Quest Pro}
\newcommand{\alvr}{ALVR}
\newcommand{\bodydata}{body motion}
\newcommand{\eyedata}{eye gaze}
\newcommand{\handdata}{hand joints}
\newcommand{\facedata}{facial expression}

\newcommand{\Eyedata}{Eye Gaze}
\newcommand{\Handdata}{Hand Joints}
\newcommand{\Facedata}{Facial Expression}
\newcommand{\taxonomy}{VR User Profiling Taxonomy}
\usepackage[colorlinks=true,urlcolor=black]{hyperref}
\def\BibTeX{{\rm B\kern-.05em{\sc i\kern-.025em b}\kern-.08em
    T\kern-.1667em\lower.7ex\hbox{E}\kern-.125emX}}

\begin{document}
\title{\sysname: User Profiling Risks in Consumer Virtual Reality Apps}

\author{\IEEEauthorblockN{Ismat Jarin\IEEEauthorrefmark{2}\IEEEauthorrefmark{4},
Olivia Figueira\IEEEauthorrefmark{2},
Yu Duan\IEEEauthorrefmark{2}, 
Tu Le\IEEEauthorrefmark{3} and
Athina Markopoulou\IEEEauthorrefmark{2}}
\IEEEauthorblockA{\IEEEauthorrefmark{2}University of California, Irvine}
\IEEEauthorblockA{\IEEEauthorrefmark{3}The University of Alabama}
\IEEEauthorblockA{\IEEEauthorrefmark{4}Email: ijarin@uci.edu}
}

\IEEEoverridecommandlockouts
\makeatletter\def\@IEEEpubidpullup{6.5\baselineskip}\makeatother
\IEEEpubid{\parbox{\columnwidth}{
		Symposium on Usable Security and Privacy (USEC) 2026 \\
		27 February 2026, San Diego, CA, USA \\
		ISBN 978-1-970672-07-7 \\
		https://dx.doi.org/10.14722/usec.2026.23003 \\
		www.ndss-symposium.org, https://www.usablesecurity.net/USEC/
}
\hspace{\columnsep}\makebox[\columnwidth]{}}

\maketitle

\IEEEpeerreviewmaketitle

\begin{abstract}
Virtual reality (VR) platforms and apps collect user sensor data, including motion, facial, eye, and hand data, in abstracted form. These data may expose users to unique privacy risks without their knowledge or meaningful awareness, yet the extent of these risks remains understudied. To address this gap, we propose \sysname{}, a framework to study user profiling based on VR sensor data and the resulting privacy risks across consumer VR apps. To systematically study this problem, we first develop a taxonomy rooted in CCPA definition of personal information and expanded it by sensor, app, and threat contexts to identify user attributes at risk. 
Then, we conduct a user study in which we collect VR sensor data from \textit{four} sensor groups from real users interacting with \apps{} popular consumer VR apps, followed by a survey.
We design and apply an analysis pipeline to demonstrate the feasibility of inferring user attributes using these data. 
Our results demonstrate the feasibility of user attribute inference, including sensitive personal information, have a moderately high to high risk (with up to $\sim90\%$ F1 score) of being inferred from the abstracted sensor data. Through feature analysis, we further identify correlations among app groups and sensor groups in inferring user attributes. 
Our findings highlight risks to users, including privacy loss, tracking, targeted advertising, and safety threats. Finally, we discuss both design implications and regulatory recommendations to enhance transparency and better protect users' privacy in VR.

\end{abstract}

\section{Introduction}\label{sec: intro}
Virtual reality (VR) provides immersive, interactive experiences for users across diverse apps, such as gaming, education and remote work~\cite{vrforeducation, vrforwork, vrforentertainment}. As part of the broader Metaverse, which connects virtual, augmented, and mixed reality--also known as extended reality (XR)~\cite{ieeesp_RoesnerK24_SPMetaVerse}, the VR market continues to grow rapidly~\cite{grandview2024vrgrowth}, driven by major platforms such as the Meta~\cite{metastore}, SteamVR~\cite{steamvr}. Though VR offers substantial benefits~\cite{vrforeducation,vrforwork,vrforentertainment}, its extensive data collection and immersive nature introduce unique privacy and security challenges for users.

\parheading{User Privacy and Security in VR.} 
VR apps collect diverse sets of sensor data that may contain users' biometrics or behavioral fingerprints. 
Compared to mobile and web platforms, users have far less control over their privacy decisions in VR~\cite{USECRoesnerXRPermission2025,wapo_vr_privacy_2022, eff_vr_warrant_2020}. Independent reviews of major VR platforms reveal weak security controls and vague privacy policies, limiting users’ ability to make informed decisions regarding sharing their sensitive data~\cite{popets25AbhinayaUserTrust,zhan2024vpvet}.
Meta, for example, discloses in their privacy policy that platforms and app developers can collect and use ``abstracted'' data derived from raw inputs~\cite{metapolicy}, which may seem less privacy-sensitive. However, prior works show that sensor data\footnote{``Abstracted" sensor data refers to processed
telemetry derived from raw sensor inputs (\eg{} image, video), available to platforms and app developers~\cite{garrido2023sok,metapolicy}. We use abstracted sensor data and sensor data interchangeably.} can uniquely identify users~\cite{miller2020personal,nair2023unique,jarin2025behavr}. As the Metaverse incorporates advertising and marketing campaigns~\cite{meta_vr_ad_campaigns}, sensor data can be repurposed for behavioral targeting. Additionally, AI agents capable of impersonating users or influencing user judgment and behavior~\cite{EmbodiedAgents2025} may further expand the attack surface for manipulation and identity misuse. These privacy concerns have escalated into lawsuits and enforcement actions targeting biometric data practices in immersive ecosystems, prompting stronger regulatory scrutiny and compliance expectations~\cite{reuters2024meta, ftc2025biometric, texas2024cubisettlement, faegredrinker2025ftc}.

\parheading{Problem Statement.} Due to pervasive data collection and users’ limited control over data sharing in VR, ``abstracted" sensor data can be exploited for user profiling and other non-functional purposes. We define {\em user profiling} as inference of private user attributes \cite{garrido2023sok}, and we use the two terms interchangeably. Understanding the extent to which users can be profiled from ``abstracted'' sensor data across consumer VR apps\footnote{Consumer VR apps are designed for naive purposes such as social interaction (\eg{} $a_1$) or gaming (\eg{} $a_3$), not for intentional attacks.}, and the risks such profiling entails remains a critical yet understudied privacy problem.%
We aim to answer the following question, which is later expanded into six research questions in Section~\ref{sec:evaluation}: \textit{to what extent can VR users be profiled using \textbf{only} ``abstracted'' sensor data across consumer VR apps, and how does this impact users' VR privacy and safety?}

\parheading{Research Gaps.} 
Recent studies show that ``abstracted" sensor data enables unique identification~\cite{jarin2025behavr,nair2023unique,miller2020personal} and user tracking across apps~\cite{jarin2025behavr,acrossapp2025}. However, user profiling using sensor data remains understudied, and existing studies exhibit several limitations. First, while prior work has proposed a VR user attribute taxonomy~\cite{garrido2023sok}, their focus was VR literature only, which limits the taxonomy's scope. As a result, this taxonomy does not explain the privacy nor regulatory significance of user attributes, nor does it capture attributes relevant to broader threat scenarios such as targeted advertising, safety and harm, or app contexts. Rather than focusing on available VR data as a basis (\ie{} bottom-up approach), we construct our taxonomy %
from privacy law and expand it with threat scenarios and app specific attributes. We subsequently analyze which attributes are currently applicable in the VR context (\ie{} top-down approach), considering that other attributes included in the taxonomy may become relevant as VR ecosystem evolve. %
Second, prior works focused on profiling using a single sensor (\ie{} body motion~\cite{nair2022metadata, Nair1kPersonal2023}), leaving the risks associated with other sensors and multi-sensor combinations underexplored. Existing studies rely on custom adversarial app~\cite{TricomiNPCG23CanNotHide,nair2022metadata} or a single consumer app (\eg{} Beat Saber~\cite{beatsaber}~\cite{Nair1kPersonal2023}), limiting diversity of user activities. Moreover, prior work underexplores the privacy and safety implications of inferred attributes and offers limited user-centric, sensor- or app-specific mitigation and regulatory guidance.

\begin{figure*}[t!]
	\centering
	\includegraphics[width=.96\linewidth]{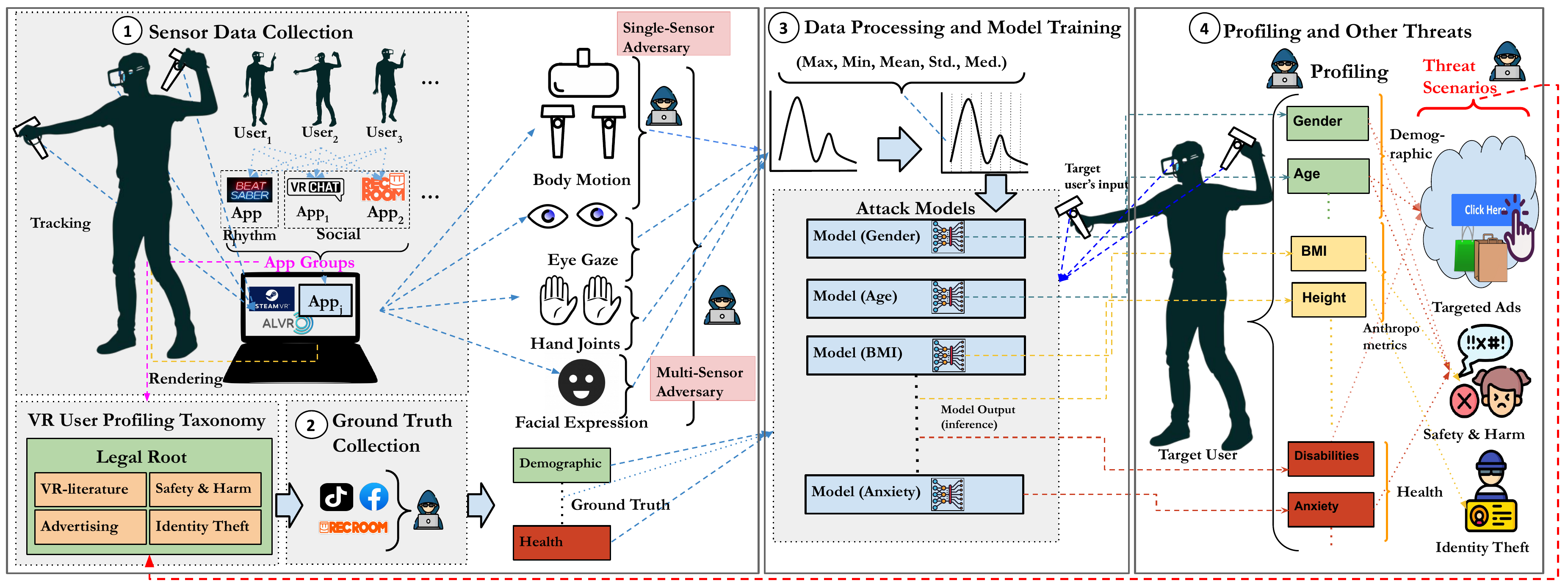}
	\caption{\textbf{ Overview of \sysname{}.} 
        (1) \textbf{Sensor Data Collection using BehaVR\cite{jarin2025behavr} setup:} %
        Each user interacts with \apps{} consumer apps using \vrdevice{} while \textit{four} sensor groups are recorded; Single-Sensor Adversaries have access to one sensor group, and Multi-Sensor Adversaries have access to multiple sensor groups; we grouped apps based on similarities of activities and emotional states.
        (2) \textbf{Ground Truth Collection:} Ground truth collection from other platforms or apps which is guided by \taxonomy{}. Our taxonomy is rooted in law and expanded by threat scenarios and app groups as indicated by arrows. 
        (3) \textbf{Data Processing and Model Training}: Sensor Data processing, feature engineering, and inference attack model training using sensor data and ground truth.
        (4) \textbf{Profiling and Other Threats:} An adversary can take only VR sensor data as model input to infer user attributes. Next, they initiate further attacks aligned with threat scenarios in Section~\ref{subsubsec:ThreatModel_ThreatScenarios}.
        }
	\label{fig:vrprofilens-workflow}
\end{figure*}

\parheading{Approach.} 
To address prior research gaps, we develop \sysname{}, a framework
for systematically investigating user profiling risks in consumer VR apps, as depicted in Figure~\ref{fig:vrprofilens-workflow}. 
Overall, we make the following key contributions:

\textit{(1) VR User Profiling Taxonomy (Section~\ref{subsec:method_profilingTaxonomy}).} We introduce a novel \taxonomy{} that is rooted in privacy law, namely the California Consumer Privacy Act's (CCPA) definition of personal information~\cite{CCPA2018}, enabling systematic reasoning about privacy and regulatory relevance of user attributes. The taxonomy is further expanded across diverse threat scenarios, including targeted advertising, identity theft, and safety and harm, as well as prior literature related to VR profiling and app contexts, enabling us to identify and analyze user attributes across different threat scenarios and app groups.  
Our taxonomy enables us to analyze the relationships among user attributes, sensor groups, app groups, and threat scenarios, and we utilize superscripts (see Section~\ref{subsec:method_profilingTaxonomy} and Table~\ref{tab:full_taxonomy}) to indicate each attribute’s associated threats and legal origin for interpretability and traceability. Since we utilized a top-down approach in developing our taxonomy, it serves as a global taxonomy that can be further expanded and generalized following our methodology, for example, as new privacy laws, VR threats, and apps are introduced.

\textit{(2) Methodology for Investigating VR User Profiling (Sections~\ref{sec:methodology} and \ref{sec: experimental_setup}).}
We design a methodology to investigate user profiling in VR, including a user study to collect users' data, %
practical threat model, an analysis pipeline that evaluates profiling from sensor data under multiple threat scenarios within multiple apps, examining both individual sensor and their combinations---an unexplored approach. Our methodology can be generalized to other platforms that collect similar sensor data and/or other apps aligned with our app groups.  

\textit{(3) Empirical Evaluation of VR User Profiling (Section~\ref{sec:evaluation}).}
We apply our methodology to quantify the feasibility of inferring user attributes from abstracted sensor data across \apps{} consumer VR apps and to assess profiling risk under different settings. Further, our findings highlight how sensor and app groups influence user profiling risk.

\textit{(4) Design Implications and Mitigation Insights (Section~\ref{Discussion}).} %
We discuss potential design implications for enhancing user's privacy in VR, including user-centered mitigation strategies tailored to both sensor and app groups, as well as regulatory and compliance recommendations.

\parheading{Paper Outline.} The rest of the paper is organized as follows: Section~\ref{sec:background_relatedwork} discusses related work, Section~\ref{sec:methodology} outlines our methodology, Section~\ref{sec: experimental_setup} presents data collection and analysis pipeline, Section~\ref{sec:evaluation} details the outcomes of our experimental evaluation on user profiling, Section~\ref{Discussion} discuss the implications of our findings, and Section~\ref{sec:conclusion} concludes our study.

\section{Related Work}\label{sec:background_relatedwork}

\subsection{Privacy and Security Threats in VR}\label{subsec:Related_Security_Safety}

\parheading{User Profiling.} %
\label{subsec:Related_profiling} While VR sensor data has been widely studied for unique identification~\cite{miller2020personal,miller2022combining,nair2023unique,jarin2025behavr}, their influence for user profiling remains underexplored.%
Prior work demonstrates attribute inference from body motion and eye tracking, but is limited to age and gender\cite{TricomiNPCG23CanNotHide}. Other works extracted 25 attributes by creating an adversarial VR game~\cite{nair2022metadata} or 40 user attributes using single-consumer app settings~\cite{Nair1kPersonal2023}. 

\parheading{Other Privacy and Safety Threats.} Prior studies have identified security and safety threats in VR, including attribute-driven risks. One potential threat is identity theft~\cite{IdentityTheftLinL22,IdentityTheft_purchage11,IdentityTheftMcAmis24,qin2025identity}, raising concerns about whether avatars accurately represent their real-world users. Identity verification in sensitive settings (\eg{} virtual courtrooms or age-restricted spaces) may rely on sensory attributes, can be exploited by attackers~\cite{IdentityTheftMcAmis24}.
Studies have highlighted VR safety concerns, including virtual shock~\cite{MhaidliRFHS24VRADs}, harassment~\cite{CSCW23_VRHarrasment,VRHarassmentUSENIX24_Abh, FreemanZMA22Safety, SchulenbergFLB23Safety, CHIFreemanFMGL024Safety}, cyberbullying, and discrimination~\cite{FaceReaderCCS23},with heightened impact on youth (\eg{} under 18), who are more vulnerable to harmful consequences from such experiences~\cite{hinduja2024metaverseSafety}. Attackers may gather user attributes to steer users toward unnecessary purchases~\cite{CHI21TargetedAds}, such as through targeted ads~\cite{meta_vr_ad_campaigns}, and distressing shockvertisements~\cite{CHI21TargetedAds, MhaidliRFHS24VRADs}.

 \subsection{VR Taxonomies} \label{subsec:Related_Taxonomy} 
 Prior work proposed VR taxonomies but remained domain-specific. Garrido \etal{} \cite{garrido2023sok} derived a taxonomy solely from VR literature, while legal-domain taxonomies grounded in the CCPA~\cite{CCPA2018} focused narrowly on children’s privacy~\cite{DiffAudit24}.

\section{Methodology}\label{sec:methodology}
This section outlines our %
methodology for investigating user profiling, %
including %
VR sensor data and device (\ref{subsec:VR_Device&Sensors}) we studied, %
our selected VR apps and app groups (\ref{subsec:VR_apps}), our threat model and threat scenarios (\ref{subsec:Threat_Model}), and the development of our \taxonomy{} (\ref{subsec:method_profilingTaxonomy}).

\subsection{VR Devices and Sensors} \label{subsec:VR_Device&Sensors}
VR platforms vary widely in software and hardware configurations. In this study, we focus on \steam{}, the leading VR gaming platform with over 7,000 applications~\cite{steamvrnumberofapps} and millions of users~\cite{steamvr_user_count}. We use the Meta Quest Pro for its comprehensive sensor suite, including body motion and eye gaze (also supported by older devices such as Quest 2), as well as hand joints and facial expression, which are increasingly supported by newer AR/VR devices~\cite{applevisionproapps,quest3}.
We explore the following \textit{four} VR sensor groups: (1) \bodydata{} (BM)~\cite{bodytracking, openxrcoordinate}, (2) \eyedata{} (EG)~\cite{eyetracking, eyegazepositionrotationkhronos}, (3) \handdata{} (HJ)~\cite{handtracking, openxrhandtracking}, and (4) \facedata{} (FE)~\cite{facetracking, openxrfacetracking}. These sensor groups are available to developers through the OpenXR APIs~\cite{openxrstandard}, which offer a common interface across different VR devices. 
We adopt the data structure definitions from the OpenXR standard~\cite{openxrstandard}. Details regarding the sensor data structure are described in Appendix~\ref{app:sensor_data_structure}.

\subsection{VR Apps} \label{subsec:VR_apps}
\subsubsection{App Selection}
VR apps span multiple %
platforms, including SteamVR~\cite{steamvr}, Apple Vision Pro~\cite{applevisionproapps}, Meta’s Oculus VR~\cite{metastore}, and HTC Viveport~\cite{viveportapps}. We select \apps{} apps from the top 100 apps on the SteamVR store~\cite{apprankings, steamvrstore}. %
To ensure coverage, we include one to two of the most popular apps from each defined app group (see Section~\ref{subsec:method_app_group}). We refer apps as $a_1,...,a_{10}$ (see Appendix~\ref{app:Apps}).

\subsubsection{App Groups} \label{subsec:method_app_group}
Seven app groups, as detailed in Appendix~\ref{app:App groups} and Table~\ref{tab:Apps-Grouping}, are defined based on similarities in user activities, predominantly representing BM, HJ, and user emotional (valence-arousal~\cite{suhaimi2018modeling_valence}) states induced by apps, namely FE.\footnote{While our prior work~\cite{jarin2025behavr} touched app-grouping, this study defines and develops a broader, more structured, and expandable categorization.} The app groups are: Social (\ie{} social activities, positive emotional state), Flight Simulation (\ie{} flying aircraft, mostly negative emotion), Interactive Navigation (IN) (\ie{} frequent user-object interactions, neutral emotional states), Knuckle-Walking (KW) (\ie{} gorilla movement, positive emotional states), Rhythm (\ie{} fast dance-like movements, mixed emotional states), and Shooting \& Archery (\ie{} shooting targets, mostly negative states). We group apps: (1) %
to generalize insights to other apps within the same group; (2) to support deeper investigations, such as identifying data collection patterns in VR (see Section~\ref{subsec:method_dataCollectionPractices}) and building taxonomies (see Section~\ref{subsec:taxonomy_VR_UserProfiling}); and (3) to provide design insights for privacy-preserving, usable defenses. %

\subsubsection{Sensor Data Collection Practices} \label{subsec:method_dataCollectionPractices} 
We examined sensor data collection practices of 20 popular apps in Oculus Quest\cite{metastore} and SteamVR\cite{steamvr}. We found that data collection practices and disclosures vary across platforms: Oculus provides more transparency and permission controls than SteamVR. Few app's data collection practices align with their functionality, while others collect more sensor data or lack privacy disclosures.
Future VR apps may collect all sensor data to support richer, multiplayer interactions (see Appendix \ref{app:data_collection_practices}).

\subsection{Threat Model} \label{subsec:Threat_Model}

\subsubsection{Adversary Capabilities}\label{subsubsec:ThreatModel_Adversaries}
The primary goal of \sysname{} adversaries is to infer private user attributes from sensor data. We consider app developers, companies, or third parties with equivalent permissions (\eg{} Unity~\cite{Unity}), corresponding to app/client or server adversaries in prior work~\cite{garrido2023sok,jarin2025behavr}. Such adversaries may collect sensor data paired with available attributes (\eg{} gender) as ground truth, train ML models, and then profile a \textit{new set of users} using only sensor data at inference time (Section~\ref{subsec:model}). Based on adversarial knowledge and sensor access, we define two types of adversary:

\parheading{Single-Sensor Adversary.}
Our single-sensor adversary has access to only one sensor group as %
certain sensor groups may be unavailable due to data loss, limited availability for the third party, or selective sensor data sharing by the users. 

\parheading{Multi-Sensor Adversary.} 
This adversary has access to multiple sensor groups, may use either a single sensor or a combination of sensors (\eg{} BM, FE together) for attack. This adversary has been underexplored due to limited access to multi-sensor data. Prior work~\cite{TricomiNPCG23CanNotHide} combined eye and body data but focused on a narrow set of attributes (age and gender). 

Both adversaries may leverage one or more user attributes to facilitate additional attacks described next.

\subsubsection{Threat Scenarios} 
\label{subsubsec:ThreatModel_ThreatScenarios} 

Our threat model examines potential threat scenarios driven by inferred attributes. It is motivated by recent work on security, privacy, and safety challenges in Metaverse~\cite{ieeesp_RoesnerK24_SPMetaVerse}, and subsequent studies on privacy risks~\cite{NairROS24_TruthInMotion}, targeted advertising~\cite{MhaidliRFHS24VRADs}, identity theft~\cite{IdentityTheftMcAmis24}, and safety and harm~\cite{OHaganGMBJM24Safety}. Our goal is not to present an exhaustive threat model, but to establish a flexible framework that can be expanded to incorporate new threats. While prior work studies user profiling~\cite{Nair1kPersonal2023,nair2022metadata}, it offers limited analysis of adversarial misuse and threat scenarios, while other studies on attribute-specific threats~\cite{popets25AbhinayaUserTrust, CSCW23_VRHarrasment, IdentityTheftLinL22} do not demonstrate how such attributes can be inferred from implicit identifiers (\ie{} sensor data). We therefore propose a practical threat model assuming adversaries exploit attributes inferred from sensor data to enable attacks aligned with threat scenarios, described next.

\parheading{Honest-but-Curious Adversary.} 
An honest-but-curious adversary~\cite{paverd2014modelling} records a few minutes of sensor traces %
and could infer one or more private attributes (\eg{} gender, health conditions) while the user remains anonymous and only shares sensor data. Revealing such information not only compromises individual privacy rights~\cite{CCPA2018_law,GDPR2016}, also undermines trust in VR apps and platforms~\cite{popets25AbhinayaUserTrust}.
The attributes inferable by this adversary in VR are detailed in Sections~\ref{subsec:taxonomy_VR_literature} and \ref{subsec:taxonomy_VR_UserProfiling} and are shown in our \taxonomy{} (Table~\ref{tab:full_taxonomy}) marked by the superscript `5' or `6', or both (\eg{} race$^{\text{4,5,6}}$).

\parheading{Targeted Advertising.} 
As discussed in Section \ref{subsec:Related_Security_Safety}, targeted advertising is a growing concern in VR.
According to Meta's VR advertising documentation \cite{meta_vr_ad_campaigns}, advertisers can utilize various user attributes to target advertisements, including location, age, gender, device identifiers, and interactions with Meta services. Alternatively, if advertisers can profile VR users with implicit identifiers (\ie{} sensor data), allowing them to bypass advertising services and cut costs by targeting users directly. Targeted advertising in immersive environments may %
exploit user's vulnerabilities, leading to manipulative or harmful purchase~\cite{CHI21TargetedAds}.
Such attributes are discussed in Sections~\ref{subsec:taxonomy_advertising}, \ref{subsec:taxonomy_VR_literature}, and \ref{subsec:taxonomy_VR_UserProfiling}, and are marked with the superscript `1' or `2' (or both) in Table~\ref{tab:full_taxonomy} (\eg{} chronic illness$^{1,5,6}$).

\parheading{Identity Theft.} 
With user's information, adversaries can initiate identity theft attack by impersonating a user’s identity through bots to access secure or confidential areas. Identity theft may extend across digital platforms (\eg{} Facebook~\cite{IdentityTheftVelayudhanB19}) and physical world, where attackers could misuse digital identities to compromise user's privacy and security. The attributes related to identity theft are discussed in Section~\ref{subsec:taxonomy_IdentityTheft} and marked by the superscript `3' in Table~\ref{tab:full_taxonomy} (\eg{} IPD$^{\text{3,5,6}}$).

\parheading{Safety and Harm.} \label{subsec:Related_manipulation}
VR users may experience harassment and safety threats, such as hate speech, violence, virtual crashing, and sexual harassment, based on attributes such as gender, race, and physical characteristics, as discussed in Section~\ref{subsec:Related_Security_Safety}. An adversary may infer those attributes from sensor data, even when users do not disclose them through their account or avatar choices (\eg{} selecting an avatar of a different gender).
VR can further deliver more immersive and targeted ads %
based on relevant attributes: for example, if a user’s fear of heights is inferred, the adversary could deliver an immersive experience involving a virtual fall from a building~\cite{MhaidliRFHS24VRADs}
These attributes %
are included in our taxonomy (see Section~\ref{subsec:taxonomy_safetyHarm}) and marked by the superscript `4' in Table~\ref{tab:full_taxonomy} (\eg{} stress$^{\text{4,6}}$).

\subsection{\taxonomy{}} \label{subsec:method_profilingTaxonomy}
 
In this section, we discuss our \taxonomy{} (Table~\ref{tab:full_taxonomy}) and how it was developed.
As discussed in Section~\ref{sec:background_relatedwork}, prior work has identified various attributes that can be deduced from VR sensor data and user behaviors, but their taxonomies were limited to the VR context as they utilized a bottom-up approach in their development, namely starting from VR sensor data as a basis and analyzing related attributes. In order to develop a comprehensive and generalizable taxonomy that can enable the analysis of user profiling risk in VR in various contexts, we utilized a top-down approach instead. %
We present a new \taxonomy{} that is rooted in the CCPA definition of personal information~\cite{CCPA2018} (Section~\ref{subsec:taxonomy_legal}), which we further expand by incorporating attributes from advertising domains (Section~\ref{subsec:taxonomy_advertising}), identity theft domains (Section~\ref{subsec:taxonomy_IdentityTheft}), VR safety and harm literature (Section~\ref{subsec:taxonomy_safetyHarm}), and VR profiling literature (Section~\ref{subsec:taxonomy_VR_literature}). 
We root our taxonomy in the CCPA definition of personal information, which specifies legally protected categories, and group all newly introduced attributes within these existing legal categories. The VR taxonomy~\cite{garrido2023sok} from prior work forms a subset of our taxonomy, as it is derived solely from VR literature. In contrast, by incorporating attributes from diverse sources beyond VR literature, our taxonomy is substantially more in-depth and encompasses various threat scenarios, which enables us to identify sensitive attributes that are shared among different threats. Our taxonomy serves as a global taxonomy as it includes a broad scope of attributes, and it can be further expanded following our methodology and generalized as new threats and app groups are introduced.

To develop our taxonomy, two researchers jointly decided on an approach and then independently worked to create the taxonomy.
Once completed, discrepancies were discussed and resolved by consensus on the final taxonomy, presented in Table~\ref{tab:full_taxonomy}. Next, we discuss the taxonomy development process, including incorporating attributes across sources (Sections~\ref{subsec:taxonomy_legal}-\ref{subsec:taxonomy_VR_literature}), and explain superscripts used in Table~\ref{tab:full_taxonomy}.

\subsubsection{Legal Domain Attributes}\label{subsec:taxonomy_legal} We began with the CCPA definition of personal information, which is %
defined as ``information that identifies, relates to, describes, is reasonably capable of being associated with, or could reasonably be linked, directly or indirectly, with a particular consumer or household.''\footnote{CAL. CIV. Code § 1798.140(v)(1)} The definition includes 12 categories of personal information, such as identifiers, biometric information, geolocation data, and inferences. 
Following the category names and organization from the CCPA-based ontology in %
~\cite{DiffAudit24}, %
we reorganized our table in a similar manner so that the categories were more specific to the attributes they contain, since some of the categories from the CCPA %
include sub-categories. We maintained references to other laws and sub-definitions included in the CCPA text, such as the Family Educational Rights and Privacy Act (FERPA)\footnote{20 U.S.C. Sec. 1232g; 34 C.F.R. Part 99}, personal information described in subdivision (e) of Section 1798.80, and sensitive personal information in the CCPA, which are identified with the superscripts `P', `F', and `S', respectively. We discuss more details %
in Appendix~\ref{app:taxonomy_details}.

\subsubsection{Advertising Domain Attributes}\label{subsec:taxonomy_advertising} Next, we incorporated attributes from advertising domains, namely the Interactive Advertising Bureau (IAB) Tech Lab Audience Taxonomy ~\cite{iab_data_transparency} and the  Meta VR advertising documentation ~\cite{meta_vr_ad_campaigns}. The IAB taxonomy attributes are grouped into three categories regarding user audiences that can be used for targeted advertising: demographics, purchase intent, and interests. For each demographics attributes, we added it to our taxonomy or marked the attributes that were already present in our taxonomy with a superscript `1', as shown in Table~\ref{tab:full_taxonomy}. Purchase intent and interest categories already existed in our taxonomy, and these categories from the IAB taxonomy jointly contain over 1,400 attributes, which we leave out of the taxonomy due to space, except for a few attributes that are relevant to our study (\eg{} caffeine or alcohol consumption). %
The Meta VR advertising documentation includes attributes that can be used for targeted advertising on Meta platforms, such as user location, age, gender and service interactions, which we incorporate into the corresponding taxonomy categories and mark with the superscript “2”.

\subsubsection{Identity Theft Attributes}\label{subsec:taxonomy_IdentityTheft}Next, we studied attributes that could be misused for identity theft as described in Section \ref{subsubsec:ThreatModel_ThreatScenarios}. We utilized the California penal code regarding identity theft, which defines various attributes that may uniquely identify an individual and be misused for identity theft~\cite{CA_penal_code_identity_theft, DOJ_victims_identity_theft}. Attributes include name, address, date of birth, unique biometric data, unique telecommunication data, and generally any ``equivalent form of identification.''\footnote{CAL. PEN. Code § 530.55(b)} We incorporated these attributes and marked them with the superscript `3'.

\subsubsection{Safety \& Harm Literature Attributes}\label{subsec:taxonomy_safetyHarm}
Next, we incorporated attributes that can be misused for harassment and endanger user's safety, as described in Section \ref{subsubsec:ThreatModel_ThreatScenarios}. We extracted attributes from prior VR literature on harassment, abuse, stalking, AI-driven harms, and shock advertising (\ie{} ``shockvertisements'') that may incite fear, distress, or manipulation through targeted content or malware~\cite{VRHarassmentUSENIX24_Abh, CSCW23_VRHarrasment, FreemanZMA22Safety, hinduja2024metaverseSafety,CHIFreemanFMGL024Safety, OHaganGMBJM24Safety, SchulenbergFLB23Safety, MhaidliRFHS24VRADs}. Identified attributes include demographic characteristics, avatar features revealing user identity, anxiety, and interests, among others, %
and we marked them with the superscript `4'.

\subsubsection{VR Literature Attributes}\label{subsec:taxonomy_VR_literature}Next, we analyzed attributes derived from prior VR literature~\cite{garrido2023sok,Nair1kPersonal2023}. The taxonomy in~\cite{garrido2023sok} categorizes VR %
attributes based on a review of 75 privacy attack and defense studies, while~\cite{Nair1kPersonal2023,NairSurvey23} identifies app-specific features in a single commercial app, \textit{BeatSaber}~\cite{beatsaber}. 
We marked all related 
attributes %
with the superscript `5'.

\subsubsection{VR User Profiling Attributes}\label{subsec:taxonomy_VR_UserProfiling}
In this step, we identify and expand %
attributes %
that can %
be captured from VR ecosystem (using multiple sensor data, app groups and threats). First, attributes derived from safety, harm, and VR literature (see Sections~\ref{subsec:taxonomy_safetyHarm} and \ref{subsec:taxonomy_VR_literature}) are automatically included %
as they directly within the VR context. 
Next, we added more attributes related to each of all app groups (see Section \ref{subsec:method_app_group}), following a method used by ~\cite{Nair1kPersonal2023,NairSurvey23} for BeatSaber~\cite{beatsaber}. We expanded app-group specific attributes inspired by prior research~\cite{Nair1kPersonal2023,NairSurvey23}. For example, in the social group, we added attributes such as social media usage and activity preference as they are relevant to our social apps' activities.
We marked attributes associated with this step using the superscript `6'. 
\subsubsection{Final Attribute Selection for \sysname{}}\label{subsec:final_attribute_list}

Finally, we select a subset of attributes from our taxonomy to demonstrate user profiling risk. We consider both explicit and implicit attributes that are directly or indirectly mapped to our four sensor groups (HMD controllers, sensors, IMU, and observations) and app groups. Some attributes are (\eg{} marital status, income) have correlation with other attributes, such as age or gender, which are inferable from sensor or behavioral data. Next, we excluded certain attributes based on our experimental setup. We omitted homogeneous user attributes, such as geolocation (\ie{} participants were in the same location) as well as device and account related attributes (\ie{} participants used the same device and account).
We mark the final attributes in \textbf{bold text} in Table~\ref{tab:full_taxonomy}, resulting in 48 attributes from 5 categories, namely Demographics, Health, Anthropometrics (included in the Biometrics category), 
Personal History, and User Interests and Behaviors. More details are in Appendix~\ref{app:Attributes_and_statistics} Table ~\ref{tab:Attributes_and_statistics}.

\section{Experimental Setup}\label{sec: experimental_setup} This section outlines our experimental setup and analysis pipeline, including sensor data collection from \apps{} consumer VR apps, user study protocol (\ref{subsec:profiling_dataset}), data processing (\ref{subsec:data_processing}), model training for attribute inference (\ref{subsec:model}), and feature organization and analysis (\ref{subsec:feature_analysis}) to support our evaluation.

\subsection{\sysname{} Dataset} \label{subsec:profiling_dataset}
To study VR user profiling, we require users' sensor data and ground truth attributes. Thus, we conduct an IRB-approved user study with \users{} participants, including VR sensor data collection followed by a survey. Participants were compensated at a \$10/hour rate, and their data were stored using unique random IDs. We utilize the sensor dataset collected in our prior work~\cite{jarin2025behavr} and augment it with more sensor data from the same participants across expanded app settings. Next, we will discuss our sensor data collection and survey procedures.

\newcommand\cprot{\text{P}}
\newcommand\ferpa{\text{F}}
\newcommand\sens{\text{S}}

\newcommand\mwid{0.9cm} %
\newcommand\lwid{0.9cm} %
\newcommand\nwid{1.75cm} %
\newcommand\rwid{12cm} %

\begin{table*}[t!]
\centering
\caption{\textbf{\taxonomy{}.}
\small{{The taxonomy is developed from Legal, Advertising, Identity Theft, VR Safety/Harm, and VR User Profiling Literature domains and includes our Proposed VR User Profiling Attributes. The taxonomy is rooted in the CCPA definition of personal information~\cite{CCPA2018}. Superscript values identify attributes that are included from other domains or legal texts (see Section~\ref{subsec:method_profilingTaxonomy}). Attributes in \textbf{bold text} are studied in this work. If all attributes within a category have the same superscripts, they are marked on the category.}}}
\scriptsize{
   \begin{tabular}{|p{\mwid}|p{\lwid}|p{\nwid}|p{\rwid}|}\hline
    \bf{Level 1} & \bf{Level 2} & \bf{Level 3} & \bf{Level 4} \\\hline
    \multirow{9}{\mwid}{Identifiers} &  \multirow{6}{\lwid}{Personal Identifiers}   &  Name & Name$^{\text{2,3,4,5,6,\ferpa}}$, signature \\\hhline{~~--}
                                 &                                                  &  Linked Personal Identifiers$^{\text{3}}$  & Social security number$^{\text{\ferpa,\sens}}$, driver's license number$^{\text{\sens}}$, passport number$^{\text{\sens}}$, state identification card number$^{\text{\sens}}$, taxpayer identification number, US citizenship and immigration services-assigned number, birth or death certificate information (\eg{} place of birth$^{\text{\ferpa}}$) \\\hhline{~~--}
                                 
                                 &                                                  &  Contact Information &  Telephone numbers$^{\text{2,3}}$, postal/home address$^{\text{2,3,4,\ferpa}}$, email address$^{\text{2,5,6}}$ \\\hhline{~~--}
                                 
                                 &                                                  &  Reasonably Linkable Personal Identifiers & IP Address$^{\text{5,6}}$, unique personal identifier$^{\text{3,4,5,6}}$, Online identifier$^{\text{3,4,5,6}}$, aliases$^{\text{3,4,5,6}}$, unique pseudonym$^{\text{3,4,5,6}}$\\\hhline{~~--}
                                 
                                 &                                                  &  User/Customer Numbers  & Account name$^{\text{2,4,5,6}}$, customer number$^{\text{3}}$, insurance policy number$^{\text{3}}$ (\eg{} health insurance), bank account number$^{\text{3,\sens}}$ (\eg{} demand deposit, savings, checking account), credit card number$^{\text{3,\sens}}$, debit card number$^{\text{\sens}}$, student/school identification number$^{\text{3,\ferpa}}$, employee identification number$^{\text{3}}$, professional or occupation number$^{\text{3}}$ \\\hhline{~~--}
                                 
                                 &                                                  &  Login Information$^{\text{3,\sens}}$ & Account log-in, security or access code, password, or credentials \\\hhline{~---}
                                 
                                 & \multirow{3}{\lwid}{Device Identifiers}          &  Device Hardware Identifiers & Device identifier$^{\text{2,3,5,6}}$ (\eg{} IMEI, MAC address), serial number$^{\text{5,6}}$ \\\hhline{~~--}
                                 
                                 &                                                  &  Device Software Identifiers & Cookies$^{\text{5,6}}$, beacons, pixel tags$^{\text{2,6}}$, mobile advertising identifiers$^{\text{2}}$, or similar technology$^{\text{3}}$ \\\hhline{~~--}
                                 
                                 &                                                  &  Device Informa\-tion \& Specific\-ations$^{\text{5,6}}$ & Refresh rate, tracking rate, field of view, resolution, CPU/GPU power, CPU brand, logical cores, CPU speed, graphics card, system version, form factor, operating system, system memory, drive space, base stations\\\hline

    \multirow{12}{\mwid}{Personal Informa\-tion} & \multirow{6}{\lwid}{Personal Characteristics \& History}   
                                        &   Demographic Information$^{\text{\cprot}}$  & \textbf{Race$^{\text{4,5,6,\sens}}$}, color, \textbf{religion$^{\text{4,5,6,\sens}}$}, \textbf{sex/gender$^{\text{1,2,4,5,6}}$}, sexual orientation/preference$^{\text{4,5,6}}$, \textbf{marital status$^{\text{1,2,5,6}}$}, military or veteran status, \textbf{ethnicity/national origin$^{\text{4,5,6,\sens}}$}, language$^{\text{1,2,5,6}}$, ancestry, \textbf{age/date of birth$^{\text{1,2,3,4,5,6,\ferpa}}$} \\\hhline{~~--}
    
                                            &                                                               &   Health Information &  Medical conditions$^{\text{4,\cprot}}$ (\eg{} \textbf{illness/chronic illness$^{\text{1,5,6}}$, color blindness$^{\text{1,5,6}}$ close/distance vision and lenses$^{\text{1,5,6}}$}, acuity$^{\text{5,6}}$, \textbf{motion sickness$^{\text{4,6}}$}, substance/drug use$^{\text{4,5,6}}$, \textbf{sleepiness$^{\text{5,6}}$}, dental care$^{\text{1}}$), disability$^{\text{4,\cprot}}$ (\eg{} \textbf{mental disability$^{\text{4,5,6}}$}, \textbf{physical disability$^{\text{4,5,6}}$}, HIV/AIDS, cancer), genetic characteristics \& information$^{\text{\cprot,\sens}}$, medical history (\eg{} medical leave$^{\text{\cprot}}$, family care leave$^{\text{\cprot}}$, pregnancy disability leave$^{\text{\cprot}}$, retaliation for reporting patient abuse$^{\text{\cprot}}$, vaccines$^{\text{1}}$), physical health$^{\text{4,5,6}}$ (\eg{} physical fitness$^{\text{1,4,5,6}}$, women’s health$^{\text{1}}$, weight loss$^{\text{1}}$), mental health$^{\text{4,5,6}}$ (\eg{} \textbf{anxiety$^{\text{4,6}}$, stress$^{\text{4,6}}$, height phobia$^{\text{4,6}}$})\\\hhline{~~--}

                                            &                                                               &   Biometric Information$^{\text{\ferpa,\sens}}$ &  \textbf{Anthropometric information} (\eg{} \textbf{hand shape/length$^{\text{3,5,6}}$, face length$^{\text{3,5,6}}$, height$^{\text{3,5,6}}$, limb length$^{\text{3,5,6}}$ (arms, feet, etc.), interpupillary distance (IPD)$^{\text{3,5,6}}$, body measurements and relationships$^{\text{3,5,6}}$ (\eg{} body ratios$^{\text{3,5}}$, wingspan$^{\text{3,5,6}}$, BMI$^{\text{3,6}}$), weight$^{\text{3,4,6}}$, reaction time$^{\text{3,5,6}}$, physical characteristics or description$^{\text{3,4,6}}$}), physiological characteristics$^{\text{3,4,6}}$ (\eg{} heart rate, neural data$^{\text{\sens}}$), biological characteristics$^{\text{3}}$, behavioral characteristics$^{\text{3}}$, DNA$^{\text{3}}$, imagery of the iris$^{\text{3,4,5,6}}$ (\eg{} eye color), retina$^{\text{3,4,5,6}}$, fingerprint$^{\text{3,6}}$, face (\eg{} facial features$^{\text{3,5,6}}$, facial movement$^{\text{3,5,6}}$, eye movement$^{\text{3,4,5,6}}$, faceprint$^{\text{3,5,6}}$), hand$^{\text{3,6}}$, palm$^{\text{3,6}}$, and vein patterns$^{\text{3,6}}$, keystroke patterns/rhythms$^{\text{3,6}}$, biometric rhythms$^{\text{3,4,5,6}}$ (\eg{} gestures$^{\text{3,4,5,6}}$, biometric movement$^{\text{3,5,6}}$), gait patterns/rhythms$^{\text{3,5,6}}$, sleep, health, or exercise data that contain identifying information$^{\text{3,6}}$, voice recordings$^{\text{3,4,5,6}}$ (\eg{} voiceprint$^{\text{3}}$, bone and air-borne vibrations$^{\text{5}}$) 
                                            \\\hhline{~~--}

                                          & &  Personal History 
                                          & Education information$^{\text{1,5,6}}$ (\eg{} \textbf{education (highest level), academic interests}), employment information \& History (\eg{} employment role$^{\text{1,6}}$, employment sector/industry$^{\text{1}}$, employment status$^{\text{1,5}}$, working preference$^{\text{1,6}}$ (\eg{} remote working), place of employment$^{\text{3}}$), financial information (\eg{} \textbf{income$^{\text{1,5,6}}$}, wealth$^{\text{5,6}}$, personal level affluence or band$^{\text{1}}$, household income$^{\text{1}}$, monthly housing payment$^{\text{1}}$, median home value$^{\text{1}}$), health insurance information, household data$^{\text{1,6}}$ (\eg{} number of adults, children, individuals in household), citizenship or immigration status$^{\text{\sens}}$, union membership$^{\text{\sens}}$, family members’ names$^{\text{3,\ferpa}}$ (\eg{} mother’s maiden name)\\\hhline{~---}
                                            
                                            &  \multirow{2}{\lwid}{Geoloca\-tion}                             &   Precise Geolocation$^{\text{\sens}}$  &  Precise geolocation$^{\text{2,4,5,6,\sens}}$ (\eg{} GPS location, postal/home address$^{\text{2,3,4,\ferpa}}$, coordinates (latitude, longitude))\\\hhline{~~--}
                                            
                                            &                                                               &   Coarse Geolocation   & Coarse geolocation$^{\text{1,2,4,5,6}}$ (\eg{} home location$^{\text{1}}$, country extension$^{\text{1}}$, region/state extension$^{\text{1}}$, city Extension$^{\text{1}}$, metro/DMA extension$^{\text{1}}$, zip or postal code extension$^{\text{1}}$)\\\hhline{~---}
                                            
                                            &  \multirow{2}{\lwid}{User Commun\-ications}                     &   Communications  & Contents of mail, email, and text messages and conversations$^{\text{2,5,6,\sens}}$ (\eg{} text content and semantic meaning, audio/speech content transcription)\\\hhline{~~--}
                                            
                                            &                                                               &   Internet Activity  &  Internet or other electronic network activity information$^{\text{2,3,5,6}}$, network bandwidth$^{\text{2,3,5,6}}$, network latency$^{\text{2,3,5,6}}$, browsing history, search history, unique electronic data including information identification number assigned to the person, address or routing code$^{\text{3}}$\\\hhline{~---}
                                            
                                            &  \multirow{4}{\lwid}{User Interests \& Behaviors}             &   Purchasing Habits \& Histories & Commercial information, records of personal property$^{\text{1}}$ (\eg{} length of residence, single or multi generation household, household ownership, property type, urbanization), products or services purchased, obtained, or considered$^{\text{1}}$, other purchasing or consuming histories or tendencies$^{\text{1}}$ \\\hhline{~~--}
                                            
                                            &                                                               &   Sensor Data & Audio (\eg{} user’s voice$^{\text{4,5,6}}$, bystanders’ voices$^{\text{6}}$), electronic and thermal (\eg{} physiological signals$^{\text{5,6}}$ (brain activity, electrothermal activity, skin galvanic response), IMU data$^{\text{5,6}}$ (device angular velocity, orientation, proper acceleration)), visual$^{\text{4,5,6}}$ (\eg{} surrounding real-world space, physical objects, bystanders, room dimensions, play area), olfactory, or similar information \\\hhline{~~--}
                                            
                                            &                                                               &   App or Service Usage \& Interaction    &  Information regarding a consumer’s interaction with an internet website application, or advertisement$^{\text{2,4,5,6}}$ (\eg{} user-app interactions$^{\text{2,4,5,6}}$, app name$^{\text{5,6}}$, app preferences$^{\text{6}}$, VR location$^{\text{4,6}}$, usage time$^{\text{5,6}}$, session info$^{\text{5,6}}$, digital presence$^{\text{5,6}}$ (\eg{} avatars$^{\text{4,5,6}}$ and digital assets$^{\text{5,6}}$ (\eg{} objects, currency, real estate)), user-to-user interaction$^{\text{4,5,6}}$, density of friendship$^{\text{6}}$, \textbf{social interaction$^{\text{6}}$}, user-object interaction$^{\text{6}}$, field of view in VR$^{\text{5,6}}$, mobile device used$^{\text{2,6}}$)\\\hhline{~~--}
                                            
                                            &                                                               &   Inferences & Interests$^{\text{1,2,4,5,6}}$ (\eg{} hobbies), preferences/predispositions$^{\text{1,2,4,5,6}}$ (\eg{} \textbf{political orientation$^{\text{1,4,5,6}}$}, background/experience (\eg{} \textbf{musical instrument$^{\text{1,5,6}}$, dance$^{\text{1,5,6}}$, VR games $^{\text{5,6}}$, athletics/sports$^{\text{1,4,5,6}}$), violence tolerance$^{\text{1,4,6}}$, working preferences (\eg{} remote working)$^{\text{6}}$, organizational preferences$^{\text{6}}$, activity preference (\eg{} indoor/outdoor)$^{\text{6}}$}, travel preferences$^{\text{2,6}}$, clothing preferences$^{\text{5}}$ (\eg{} lower/upper body clothing, footwear), characteristics \& attitudes$^{\text{5,6}}$ (\eg{} \textbf{introvert/extrovert$^{\text{6}}$, openness$^{\text{6}}$, conscientiousness$^{\text{6}}$, agreeableness$^{\text{6}}$}), psychological trends (\eg{} \textbf{emotions$^{\text{4,5,6}}$, emotional stability$^{\text{6}}$}, cognitive processes$^{\text{5,6}}$, \textbf{attention/concentration$^{\text{5,6}}$}, fear of death$^{\text{4,6}}$), behavior (\eg{} \textbf{handedness$^{\text{5,6}}$}, physical preparation$^{\text{5,6}}$, drug consumption$^{\text{4,5,6}}$, \textbf{alcohol consumption$^{\text{1,5,6}}$, caffeine consumption$^{\text{1,5,6}}$, social media usage$^{\text{4,6}}$}), aptitudes \& abilities$^{\text{5}}$ (\eg{} \textbf{reasoning \& problem solving abilities$^{\text{6}}$, shooting experience$^{\text{1,4,6}}$}, video game aptitude $^{\text{4,6}}$, intelligence$^{\text{6}}$)\\\hhline{----}

  \end{tabular}
}
\scriptsize{
   \begin{tabular}{lll}
   {Attribute Sources Legend:} && \\
     \textbf{1}: IAB Advertising Audience Taxonomy & \textbf{4}: VR Safety \& Harm Threat Scenario & \textbf{\cprot}: Protected Classifications in CA \\
     
     \textbf{2}: Meta VR Advertising Campaign & \textbf{5}: VR Literature & \textbf{\ferpa}: FERPA Definition of PII \\
     
     \textbf{3}: Identity Theft & \textbf{6}: Proposed for VR User Profiling & \textbf{\sens}: Sensitive Personal Info. in CCPA \\

    \end{tabular}
}
  \label{tab:full_taxonomy}
\end{table*}

\subsubsection{Sensor Data Collection} \label{sensor_data_collection}

We collect the following four groups of sensor data, using the data collection setup from our prior work~\cite{jarin2025behavr}: body motion (BM), eye gaze (EG), hand joint (HJ), and facial expressions (FE), recorded as time series. 
Each participant provided 5-6 minutes of data per sensor group per app, totaling to 3-4 hours per user for 10 apps, which required three months to collect for all participants. See Appendix~\ref{app:dataset} for details.

\subsubsection{Survey Protocol} \label{subsec:survey}
Participants also completed a survey\footnote{Our survey questionnaire is available at \url{https://osf.io/wnue5/?view_only=004315c37a9d471fb9cfe2dbee62018e}.}, which collected user attributes derived from our taxonomy (see Section~\ref{subsec:final_attribute_list}), as listed next, serving as the ground truth for model training. %
The survey included the following: 

\begin{itemize}[leftmargin=*,topsep=0pt]
{
    \item {\em Demographics:} Participants were asked demographic questions based on the US Census~\cite{2020USCensus} (\eg{} gender, age).
     
     \item {\em Personal History:} Participants were asked questions relevant to their personal history (\eg{} income). %
    \item{\em Anthropometrics:} Participants either voluntarily visited the lab, where the lead researcher collected anthropometric attributes (\eg{} height, IPD), or they were self-reported. %
    \item{\em Health:} 
    Participants answered questions regarding their mental and physical health status, disabilities and other VR-relevant health factors, such as fear of heights for flight apps.
    
    \item{\em User Interests and Behaviors:} 
   Participants were asked about their general behaviors, interests, and attributes relevant to each app group, such as social media usage (related to social apps) and organizational preferences (related to IN apps).

    }
\end{itemize}

\subsubsection{Dataset Summary and Size}\label{subsec:final_dataset} 
Our final dataset links each user’s sensor data with their ground truth survey responses. Thus, it includes \users{} participants, each with four sensor groups across \apps{} apps (from seven app groups) and 48 ground-truth attributes (from five categories), as identified in Section \ref{subsec:final_attribute_list}. Overall, it contains 200 data records (\users{} for 10 apps) per attribute, thus 200 $\times$ 48 records for all participants across all attributes. More details regarding participants' statistics and attributes are discussed in Table~\ref{tab:Attributes_and_statistics} and Appendix~\ref{app:Attributes_and_statistics}.

\subsection{Data Processing and Feature Engineering} \label{subsec:data_processing}
We process users' raw sensor streams (\ie{} sensor data collected over time, discussed in Section~\ref{sensor_data_collection}) into 1-second blocks and summarize each reading using five statistics, yielding feature vectors for BM, EG, HJ, and FE that follow the OpenXR standard~\cite{openxrstandard}. This results in 165 BM, 46 EG, 1,820 HJ, and 320 FE features per block for downstream model training. Additional details are provided in Appendix~\ref{app:data_processing}.

\subsection{Inference Attack Models} \label{subsec:model}

\subsubsection{Ground Truth Selection} \label{subsec:experiment_attribute_final}
We initially considered all 48 attributes identified in Section~\ref{subsec:final_attribute_list} and collected corresponding user responses through our survey. We then removed attributes with non-discriminative responses (\eg{} no reports of physical disabilities), consistent with prior work~\cite{Nair1kPersonal2023}. Next, we excluded attributes with limited responses (\eg{} most users did not disclose their IPD) and those that were highly correlated or produced duplicate responses. Finally, we selected \attributes{} attributes as the ground truth for model training and inference, as listed in Table~\ref{tab:Attributes_and_statistics} in Appendix~\ref{app:Attributes_and_statistics} (marked with \checkmark). %

\subsubsection{Model Selection and Training} \label{subsec:model_training}
We formulate inference tasks as either categorical classification (\eg{} gender) or continuous regression (\eg{} height), depending on the attribute. We choose two types of ML models: classification for categorical (\eg{} gender) and regression for continuous (\eg{} height) attributes.
We explore %
Random Forest (RF)~\cite{RF}, Gradient Boosting (XGB)~\cite{XGB}, Light GBM (LGB)~\cite{LGB}, and Support Vector Machine (SVM)~\cite{SVM} for classification.
After our initial analysis, we choose RF and LGB based on performance, which also align with our objective to conduct feature analysis. For the regression, %
we explore Linear Regression (LR)~\cite{LR} and Random Forest Regression (RFR)~\cite{RFR}. Based on our empirical analysis on model performance, we select RFR. We evaluated acceptable error margins ($\pm 5$ cm, $\pm 5$ lb, $\pm$ 2.5 cm , $\pm 0.5$ cm for height, weight, hand, and face length respectively).

Our attack classifier design follows our threat model and participant response distribution; details are provided in Appendix~\ref{app:classifier}.
We train multiple attack models, %
one per attribute for each sensor group and combinations of multiple sensor groups. This includes single-sensor models, which have one model per attribute (\eg{} gender) per sensor group (\eg{} BM) for the single-sensor adversary, and multi-sensor models, which have one model per attribute per sensor combination (\eg{} BM and FE) for the multi-sensor adversary (see Section~\ref{subsubsec:ThreatModel_Adversaries}); \eg{} BM gender inference model is trained on the BM sensor with gender serving as the ground truth.

\subsubsection{Inference} \label{subsec:inference_task}
Finally, we evaluate the feasibility of user profiling using 5-fold cross-validation, ensuring that users in the test fold are never present in the corresponding training folds. We report average performance to minimize bias from particular user subsets. While the participant count is limited, each user contributes multiple samples across apps, enabling per-sample evaluation under a strict user-disjoint setting. We report our final results per-sample basis (1s per chunk) rather than per user to reduce bias.

\subsection{Feature Organization and Analysis} \label{subsec:feature_analysis}

\subsubsection{Feature Importance and Ranking}
We evaluate feature importance for each attribute-inference model using information gain~\cite{2013feature}. Then, we rank features as either high importance (HI), medium high importance (MH), medium importance (MI), or low importance (LI) by extracting three elbow points~\cite{Elbow_ShiWWWLL21} per attribute per app group, %
based on sorted importance scores. These elbow points mark where feature scores sharply decline, serving as thresholds for ranking. %

\subsubsection{Feature Interpretation}
Features from sensor groups are important for describing user actions.
Prior works focused on a single attribute (\eg{} identification \cite{jarin2025behavr, nair2023unique, miller2020personal}) or did not analyze features (\eg{}~\cite{Nair1kPersonal2023,nair2022metadata}), thus used ad hoc feature sets. %
Our study contains multiple dimensions (\eg{} attribute vs. app vs. sensor group), requiring a more systematic and automated method. 
We provide an automated pipeline that enhances the interpretability of sensor-derived features. Details can be found in Appendix \ref{app:feature_analysis}, Tables~\ref{tab:feature_interpretation_BM} (BM), \ref{tab:feature_interpretation_FE} (FE), and \ref{tab:feature_interpretation_hj} (HJ).

\section{Evaluation }\label{sec:evaluation}
This section details our experimental evaluation on user profiling,
including our evaluation metrics (\ref{subsec:evaluation_metrics}), research questions that guide our evaluation and discussion (\ref{subsec:rqs}), and attribute inference results for our adversarial settings (\eg{} sensor groups, app groups, adversaries) (\ref{subsec:eval_profiling_quant}-\ref{subsec:eval_Privileged_Adversary}).

\subsection{Evaluation Metrics}\label{subsec:evaluation_metrics} We evaluate our results using two evaluation metrics:

\parheading{Attack Performance.}
We adopt F1 score metric to measure the performance of the inference attack models.

\parheading{Risk Assessment.} Leveraging industry and NIST practices~\cite{encord, arize, NIST}, we map F1 scores into four risk levels to interpret profiling risk: High/Very High Risk (F1=80-100\%, purple), Moderately High Risk (F1=70-80\%, orange), Moderate Risk (F1=50-70\%, blue), and Low Risk ($F1<50\%$, gray), as detailed in Appendix \ref{app:Risk Assessments}. Colors indicate associated risk levels in our result tables (\eg{} Table~\ref{tab:f1score-bm}). Additionally, we assume each attribute’s risk level maps into its corresponding threat scenarios; \eg{} if gender$^{\text{1,2,4,5,6}}$ inference via BM yields an F1 of 85\%, placing gender$^{\text{1,2,4,5,6}}$ at high-risk level and, users sharing BM are likely at high risk for associated threats (\eg{} targeted advertising, safety/harm; see Section~\ref{subsubsec:ThreatModel_ThreatScenarios}) as indicated by the attribute’s superscripts in Table~\ref{tab:full_taxonomy}.

\begin{table*}[!ht]
\centering
\renewcommand{\arraystretch}{0.78}
\footnotesize
\caption{\textbf{User Profiling Using BM Sensor Data Across 7 App Groups.} The color code, designed to be color-blind friendly~\cite{colorBlindFriendly}, represents four risk levels based on F1: High/Very High Risk (F1=80-100\%,
purple), Moderate-High Risk (F1=70-80\%, orange), Moderate Risk (F1=50-70\%, light-blue), and Low Risk ($F1<50\%$, gray) as per~\ref{subsec:evaluation_metrics}. Attribute superscripts indicate associated threats according to our threat scenarios (see Section~\ref{subsubsec:ThreatModel_ThreatScenarios}) and taxonomy (see Section~\ref{subsec:taxonomy_VR_UserProfiling}).} 
\label{tab:f1score-bm}
\resizebox{\textwidth}{!}{%
\begin{tabular}{|l|l|lllllll|}
\hline
\multicolumn{1}{|c|}{\multirow{2}{*}{\textbf{Attribute Groups}}} &
  \multicolumn{1}{c|}{\multirow{2}{*}{\textbf{Attributes}}} &
  \multicolumn{7}{c|}{\textbf{App Groups}} \\ \cline{3-9}
\multicolumn{1}{|c|}{} &
  \multicolumn{1}{c|}{} &
  \multicolumn{1}{c|}{Social} &
  \multicolumn{1}{c|}{Flight} &
  \multicolumn{1}{c|}{Shooting} &
  \multicolumn{1}{c|}{Rhythm} &
  \multicolumn{1}{c|}{IN} &
  \multicolumn{1}{c|}{KW} &
  \multicolumn{1}{c|}{Archery} \\ \hline
\multirow{5}{*}{Demographics} &
  Gender$^{\text{1,2,4,5,6}} $ &
  \multicolumn{1}{l|}{\cellcolor{purple!50}85} &
  \multicolumn{1}{l|}{\cellcolor{teal!20}65} &
  \multicolumn{1}{l|}{\cellcolor{purple!50}90} &
  \multicolumn{1}{l|}{\cellcolor{purple!50}94} &
  \multicolumn{1}{l|}{\cellcolor{orange!50}75} &
  \multicolumn{1}{l|}{\cellcolor{orange!50}78} &
  \cellcolor{purple!50}92 \\ \cline{2-9}
 &
 Age$^{\text{1,2,3,4,5,6,\ferpa}}$ &
  \multicolumn{1}{l|}{\cellcolor{teal!20}68} &
  \multicolumn{1}{l|}{\cellcolor{orange!50}70} &
  \multicolumn{1}{l|}{\cellcolor{orange!50}70} &
  \multicolumn{1}{l|}{\cellcolor{teal!20}68} &
  \multicolumn{1}{l|}{\cellcolor{teal!20}67} &
  \multicolumn{1}{l|}{\cellcolor{teal!20}63} &
  \cellcolor{orange!50}70 \\ \cline{2-9}
 &
  Ethnicity$^{\text{4,5,6,\sens}}$  &
  \multicolumn{1}{l|}{\cellcolor{orange!50}70} &
  \multicolumn{1}{l|}{\cellcolor{orange!50}71} &
  \multicolumn{1}{l|}{\cellcolor{teal!20}58} &
  \multicolumn{1}{l|}{\cellcolor{teal!20}51} &
  \multicolumn{1}{l|}{\cellcolor{orange!50}70} &
  \multicolumn{1}{l|}{\cellcolor{teal!20}62} &
  \cellcolor{orange!50}73 \\ \cline{2-9}
 &
  Marital status$^{\text{1,2,5,6}}$ &
  \multicolumn{1}{l|}{\cellcolor{teal!20}58} &
  \multicolumn{1}{l|}{\cellcolor{teal!20}52} &
  \multicolumn{1}{l|}{\cellcolor{teal!20}54} &
  \multicolumn{1}{l|}{\cellcolor{teal!20}55} &
  \multicolumn{1}{l|}{\cellcolor{teal!20}63} &
  \multicolumn{1}{l|}{\cellcolor{teal!20}55} &
  \cellcolor{gray!45}42 \\ \cline{2-9}
  \hline
  
\multirow{6}{*}{Anthropometrics} &
 Height$^{\text{3,5,6}}$ &
  \multicolumn{1}{l|}{\cellcolor{purple!50}80} &
  \multicolumn{1}{l|}{\cellcolor{gray!45}44} &
  \multicolumn{1}{l|}{\cellcolor{purple!50}90} &
  \multicolumn{1}{l|}{\cellcolor{purple!50}100} &
  \multicolumn{1}{l|}{\cellcolor{teal!20}52} &
  \multicolumn{1}{l|}{\cellcolor{teal!20}69} &
  \cellcolor{purple!50}90 \\ \cline{2-9}
 &
  Reaction Time$^{\text{3,5,6}}$ &
  \multicolumn{1}{l|}{\cellcolor{purple!50}85} &
  \multicolumn{1}{l|}{\cellcolor{purple!50}90} &
  \multicolumn{1}{l|}{\cellcolor{purple!50}90} &
  \multicolumn{1}{l|}{\cellcolor{purple!50}95} &
  \multicolumn{1}{l|}{\cellcolor{purple!50}97} &
  \multicolumn{1}{l|}{\cellcolor{purple!50}85} &
  \cellcolor{purple!50}90 \\ \cline{2-9}
 &
  Face Length$^{\text{3,5,6}}$ &
  \multicolumn{1}{l|}{\cellcolor{teal!20}58} &
  \multicolumn{1}{l|}{\cellcolor{gray!45}40} &
  \multicolumn{1}{l|}{\cellcolor{gray!45}40} &
  \multicolumn{1}{l|}{\cellcolor{teal!20}56} &
  \multicolumn{1}{l|}{\cellcolor{teal!20}52} &
  \multicolumn{1}{l|}{\cellcolor{gray!45}49} &
  \cellcolor{teal!20}61 \\ \cline{2-9} &
    Arm Length$^{\text{3,5,6}}$ &
  \multicolumn{1}{l|}{\cellcolor{orange!50}75} &
  \multicolumn{1}{l|}{\cellcolor{gray!45}45} &
  \multicolumn{1}{l|}{\cellcolor{teal!20}66} &
  \multicolumn{1}{l|}{\cellcolor{orange!50}71} &
  \multicolumn{1}{l|}{\cellcolor{orange!50}70} &
  \multicolumn{1}{l|}{\cellcolor{teal!20}50} &
  \cellcolor{gray!45}49 \\ \cline{2-9}
 &
  Weight$^{\text{3,4,6}}$ &
  \multicolumn{1}{l|}{\cellcolor{orange!50}75} &
  \multicolumn{1}{l|}{\cellcolor{gray!45}46} &
  \multicolumn{1}{l|}{\cellcolor{purple!50}80} &
  \multicolumn{1}{l|}{\cellcolor{orange!50}72} &
  \multicolumn{1}{l|}{\cellcolor{orange!50}73} &
  \multicolumn{1}{l|}{\cellcolor{teal!20}69} &
  \cellcolor{orange!50}75 \\ \cline{2-9}
 &
  BMI$^{\text{3,6}}$ &
  \multicolumn{1}{l|}{\cellcolor{orange!50}70} &
  \multicolumn{1}{l|}{\cellcolor{teal!20}60} &
  \multicolumn{1}{l|}{\cellcolor{orange!50}73} &
  \multicolumn{1}{l|}{\cellcolor{teal!20}65} &
  \multicolumn{1}{l|}{\cellcolor{purple!50}81} &
  \multicolumn{1}{l|}{\cellcolor{orange!50}73} &
  \cellcolor{teal!20}61 \\ \hline
\multirow{6}{*}{Health} &

  Close / Distance vision and lenses$^{1,5,6}$ &
  \multicolumn{1}{l|}{\cellcolor{teal!20}58} &
  \multicolumn{1}{l|}{\cellcolor{teal!20}61} &
  \multicolumn{1}{l|}{\cellcolor{teal!20}53} &
  \multicolumn{1}{l|}{\cellcolor{teal!20}60} &
  \multicolumn{1}{l|}{\cellcolor{teal!20}60} &
  \multicolumn{1}{l|}{\cellcolor{teal!20}51} &
  \cellcolor{teal!20}65
   \\ \cline{2-9} 
 &
  Physical fitness$^{\text{1,4,5,6}}$ &
  \multicolumn{1}{l|}{\cellcolor{purple!50}80} &
  \multicolumn{1}{l|}{\cellcolor{teal!20}65} &
  \multicolumn{1}{l|}{\cellcolor{orange!50}71} &
  \multicolumn{1}{l|}{\cellcolor{purple!50}86} &
  \multicolumn{1}{l|}{\cellcolor{orange!50}70} &
  \multicolumn{1}{l|}{\cellcolor{purple!50}80} &
  \cellcolor{purple!50}86 \\ \cline{2-9}
 &
  Anxiety$^{4,6}$ &
  \multicolumn{1}{l|}{\cellcolor{purple!50}85} &
  \multicolumn{1}{l|}{\cellcolor{purple!50}87} &
  \multicolumn{1}{l|}{\cellcolor{teal!20}65} &
  \multicolumn{1}{l|}{\cellcolor{orange!50}71} &
  \multicolumn{1}{l|}{\cellcolor{orange!50}72} &
  \multicolumn{1}{l|}{\cellcolor{teal!20}67} &
  \cellcolor{orange!50}70 \\ \cline{2-9}
 &
  Stress$^{4,6}$ &
  \multicolumn{1}{l|}{\cellcolor{purple!50}88} &
  \multicolumn{1}{l|}{\cellcolor{purple!50}90} &
  \multicolumn{1}{l|}{\cellcolor{purple!50}89} &
  \multicolumn{1}{l|}{\cellcolor{purple!50}81} &
  \multicolumn{1}{l|}{\cellcolor{purple!50}83} &
  \multicolumn{1}{l|}{\cellcolor{purple!50}86} &
  \cellcolor{purple!50}87 \\ \cline{2-9}
 &
  Height phobia$^{4,6}$  &
  \multicolumn{1}{l|}{\cellcolor{orange!50}70} &
  \multicolumn{1}{l|}{\cellcolor{purple!50}81} &
  \multicolumn{1}{l|}{\cellcolor{teal!20}62} &
  \multicolumn{1}{l|}{\cellcolor{teal!20}59} &
  \multicolumn{1}{l|}{\cellcolor{purple!50}81} &
  \multicolumn{1}{l|}{\cellcolor{orange!50}74} &
  \cellcolor{orange!50}72 \\ \cline{2-9}
 &
  Motion sickness$^{4,6}$ &
  \multicolumn{1}{l|}{\cellcolor{gray!45}45} &
  \multicolumn{1}{l|}{\cellcolor{teal!20}60} &
  \multicolumn{1}{l|}{\cellcolor{gray!45}45} &
  \multicolumn{1}{l|}{\cellcolor{gray!45}48} &
  \multicolumn{1}{l|}{\cellcolor{teal!20}60} &
  \multicolumn{1}{l|}{\cellcolor{orange!50}70} &
  \cellcolor{teal!20}55 \\ \hline
\multirow{12}{*}{User Interests \& Behaviors} &   
Problem Solving Abilities$^{6}$ &  
  \multicolumn{1}{l|}{\cellcolor{orange!50}74} &
  \multicolumn{1}{l|}{\cellcolor{teal!20}62} &
  \multicolumn{1}{l|}{\cellcolor{teal!20}55} &
  \multicolumn{1}{l|}{\cellcolor{gray!45}42} &
  \multicolumn{1}{l|}{\cellcolor{gray!45}43} &
  \multicolumn{1}{l|}{\cellcolor{orange!50}78} &
  \cellcolor{orange!50}79   \\ 
\cline{2-9}  &    
Alcohol consumption$^{1,5,6}$ &  
  \multicolumn{1}{l|}{\cellcolor{teal!20}64} &
  \multicolumn{1}{l|}{\cellcolor{orange!50}71} &
  \multicolumn{1}{l|}{\cellcolor{teal!20}59} &
  \multicolumn{1}{l|}{\cellcolor{teal!20}59} &
  \multicolumn{1}{l|}{\cellcolor{teal!20}59} &
  \multicolumn{1}{l|}{\cellcolor{teal!20}60} &
  \cellcolor{teal!20}62    \\ \cline{2-9} &    
VR Experience$^{5,6}$ &   
  \multicolumn{1}{l|}{\cellcolor{orange!50}77} &
  \multicolumn{1}{l|}{\cellcolor{orange!50}73} &
  \multicolumn{1}{l|}{\cellcolor{teal!20}65} &
  \multicolumn{1}{l|}{\cellcolor{orange!50}77} &
  \multicolumn{1}{l|}{\cellcolor{teal!20}66} &
  \multicolumn{1}{l|}{\cellcolor{orange!50}74} &
  \cellcolor{orange!50}78    \\ \cline{2-9}   &    
Activity preference$^{6}$ &   
  \multicolumn{1}{l|}{\cellcolor{teal!20}59} &
  \multicolumn{1}{l|}{\cellcolor{orange!50}70} &
  \multicolumn{1}{l|}{\cellcolor{teal!20}53} &
  \multicolumn{1}{l|}{\cellcolor{teal!20}60} &
  \multicolumn{1}{l|}{\cellcolor{teal!20}69} &
  \multicolumn{1}{l|}{\cellcolor{teal!20}53} &
  \cellcolor{teal!20}65  \\ \cline{2-9}        &    
Shooting Experiences$^{1,4,6}$ &  
  \multicolumn{1}{l|}{\cellcolor{purple!50}81} &
  \multicolumn{1}{l|}{\cellcolor{teal!20}68} &
  \multicolumn{1}{l|}{\cellcolor{orange!50}78} &
  \multicolumn{1}{l|}{\cellcolor{purple!50}86} &
  \multicolumn{1}{l|}{\cellcolor{teal!20}68} &
  \multicolumn{1}{l|}{\cellcolor{orange!50}75} &
  \cellcolor{purple!50}82   \\ \cline{2-9}  &
  Caffeinated item consumption$^{1,5,6}$ &
  \multicolumn{1}{l|}{\cellcolor{teal!20}69} &
  \multicolumn{1}{l|}{\cellcolor{orange!50}77} &
  \multicolumn{1}{l|}{\cellcolor{teal!20}65} &
  \multicolumn{1}{l|}{\cellcolor{orange!50}71} &
  \multicolumn{1}{l|}{\cellcolor{teal!20}66} &
  \multicolumn{1}{l|}{\cellcolor{teal!20}58} &
  \cellcolor{orange!50}72  \\ \cline{2-9}
 &
  Concentration$^{3,6}$ &
  \multicolumn{1}{l|}{\cellcolor{orange!50}73} &
  \multicolumn{1}{l|}{\cellcolor{orange!50}72} &
  \multicolumn{1}{l|}{\cellcolor{gray!45}49.5} &
  \multicolumn{1}{l|}{\cellcolor{gray!45}48} &
  \multicolumn{1}{l|}{\cellcolor{orange!50}70} &
  \multicolumn{1}{l|}{\cellcolor{teal!20}60} &
  \cellcolor{orange!50}72 \\ \cline{2-9}
 &
  Violence tolerance$^{1,4,6}$ &
  \multicolumn{1}{l|}{\cellcolor{orange!50}74} &
  \multicolumn{1}{l|}{\cellcolor{orange!50}70} &
  \multicolumn{1}{l|}{\cellcolor{gray!45}42} &
  \multicolumn{1}{l|}{\cellcolor{teal!20}67} &
  \multicolumn{1}{l|}{\cellcolor{orange!50}70} &
  \multicolumn{1}{l|}{\cellcolor{orange!50}70} &
  \cellcolor{orange!50}70 \\ \cline{2-9}
 &
  Introvert/Extrovert$^{\text{6}}$ &
  \multicolumn{1}{l|}{\cellcolor{teal!20}55} &
  \multicolumn{1}{l|}{\cellcolor{teal!20}65} &
  \multicolumn{1}{l|}{\cellcolor{gray!45}45} &
  \multicolumn{1}{l|}{\cellcolor{gray!45}45} &
  \multicolumn{1}{l|}{\cellcolor{teal!20}61} &
  \multicolumn{1}{l|}{\cellcolor{gray!45}47.5} &
  \cellcolor{teal!20}55 \\ \cline{2-9}
 &
  Organized/Unorganized$^{\text{6}}$ &
  \multicolumn{1}{l|}{\cellcolor{purple!50}88} &
  \multicolumn{1}{l|}{\cellcolor{purple!50}95} &
  \multicolumn{1}{l|}{\cellcolor{purple!50}90} &
  \multicolumn{1}{l|}{\cellcolor{purple!50}88} &
  \multicolumn{1}{l|}{\cellcolor{purple!50}92} &
  \multicolumn{1}{l|}{\cellcolor{orange!50}70} &
  \cellcolor{purple!50}80 \\ \cline{2-9}
 &
  Social media usage$^{4,6}$ &
  \multicolumn{1}{l|}{\cellcolor{orange!50}75} &
  \multicolumn{1}{l|}{\cellcolor{teal!20}60} &
  \multicolumn{1}{l|}{\cellcolor{purple!50}80} &
  \multicolumn{1}{l|}{\cellcolor{purple!50}90} &
  \multicolumn{1}{l|}{\cellcolor{orange!50}70} &
  \multicolumn{1}{l|}{\cellcolor{orange!50}71} &
  \cellcolor{purple!50}90 \\ \cline{2-9} &
      Openness$^{6}$ &
  \multicolumn{1}{l|}{\cellcolor{teal!20}68} &
  \multicolumn{1}{l|}{\cellcolor{orange!50}71} &
  \multicolumn{1}{l|}{\cellcolor{orange!50}70} &
  \multicolumn{1}{l|}{\cellcolor{teal!20}68} &
  \multicolumn{1}{l|}{\cellcolor{teal!20}67} &
  \multicolumn{1}{l|}{\cellcolor{teal!20}65} &
  \cellcolor{teal!20}62 \\ \cline{2-9} &

    Emotional Stability$^{6}$ &
  \multicolumn{1}{l|}{\cellcolor{orange!50}76} &
  \multicolumn{1}{l|}{\cellcolor{orange!50}74} &
  \multicolumn{1}{l|}{\cellcolor{teal!20}53} &
  \multicolumn{1}{l|}{\cellcolor{orange!50}74} &
  \multicolumn{1}{l|}{\cellcolor{orange!50}78} &
  \multicolumn{1}{l|}{\cellcolor{teal!20}66} &
  \cellcolor{orange!50}75  \\ \hline
\end{tabular}%
}
\end{table*}

\subsection{Research Questions}\label{subsec:rqs}
We evaluate the experimental results by answering the following Research Questions (RQs):

\begin{itemize}[leftmargin=*,topsep=0pt]
    \item{RQ1. \label{RQ1}How well users can be profiled using \textit{only} VR sensor data? (Section~\ref{subsec:eval_profiling_quant})} 

    \item{RQ2. \label{RQ2}What are the top features for profiling? (Section~\ref{subsec:eval_featureAnalysis})}

    \item{RQ3. \label{RQ3}How can user attributes be inferred from sensor data across different VR app groups? (Section~\ref{subsec:eval_appgroup})}

    \item{RQ4. \label{RQ4}How do combinations of different sensor groups expose personal attributes? (Section~\ref{subsec:eval_Privileged_Adversary})}
    
\end{itemize}

Moreover, we include an in-depth discussion in Section~\ref{Discussion} that addresses the following:

\begin{itemize}[leftmargin=*,topsep=0pt]
    \item{RQ5. \label{RQ5}How does revealing one or multiple attributes place users at risk across different threat scenarios? (Section~\ref{discussion:Risk})} 
    \item{RQ6. \label{RQ6}What are the design and regulatory implications of our findings, and how can these insights support service providers and lawmakers in designing safer VR experiences for users? (Sections~\ref{discussion:Lessons_for_Privacydefense} and \ref{Discussion:Compliance})}
\end{itemize}

\subsection{Quantifying Profiling Risks (RQ1)}  \label{subsec:eval_profiling_quant}
We evaluate the effectiveness of attribute inferences using VR sensor data by answering \textbf{RQ1}. Details are in Table~\ref{tab:f1score-bm} (BM), Appendix~\ref{app:Results} Tables~\ref{tab:f1score-fe} (FE), \ref{tab:f1score-eg} (EG), and \ref{tab:f1score-hj} (HJ).

Attributes from demographics, anthropometrics, and health, are inferred with moderately high to high F1 across all sensor groups. 
Demographic attributes, such as gender$^{\text{1,2,4,5,6}}$, age$^{\text{1,2,3,4,5,6,\ferpa}}$, and ethnicity$^{\text{4,5,6,\sens}}$, are inferred with moderately high (70--80\%) to high (80--100\%) F1 using BM, FE, and HJ. HJ and BM exhibit the strongest overall performance; \eg{} for gender$^{\text{1,2,4,5,6}}$, HJ achieves an F1 of 73--87\% and BM 65--94\%, and for age$^{\text{1,2,3,4,5,6,\ferpa}}$ and ethnicity$^{\text{4,5,6,\sens}}$, 71--85\% and 70--73\%, respectively, across most app-groups. FE provides high performance on ethnicity$^{\text{4,5,6,\sens}}$ (78--86\%) and age$^{\text{1,2,3,4,5,6,\ferpa}}$ (71--88\%), but moderate to moderately high for gender$^{\text{1,2,4,5,6}}$ (68--78\%). Alternatively, attribute inferences based on EG show low to moderate adversarial risks (49--60\%).

For anthropometrics, BM provides high F1 across app groups (70–100\%), particularly for height$^{\text{3,5,6}}$, weight$^{\text{3,4,6}}$, and reaction time$^{\text{3,5,6}}$, and moderate/moderate-high for BMI$^{\text{3,6}}$ (60–80\%). FE performs well for face length (65–80\%) but yields lower F1 for other attributes. HJ shows moderate to low performance, and EG yields the lowest. Health attributes show moderately high to high risk for BM and HJ (\eg{} 80\% for physical fitness$^{\text{1,4,5,6}}$ and stress$^{4,6}$, and 70--85\% for anxiety$^{4,7,8}$ and height phobia$^{4,6}$), depending on the app groups. FE performs better in stress$^{4,6}$ (85--94\%) and anxiety$^{4,7,8}$ (70--85\%), but moderate for height phobia$^{4,6}$ (55--73\%). 

BM provides high F1 for user behavior and interests; \eg{} for organized versus unorganized$^{\text{6}}$ (84--95\%) and emotional stability$^{\text{6}}$ (70--79\%). FE provides high F1 for organized behavior$^{\text{6}}$ (76--97\%) and openness$^{\text{6}}$ (73--84\%). %
For attributes such as violence tolerance and emotional stability, FE provides high to moderately high F1 (74--86\%), depending on the app groups. HJ demonstrates moderately high F1 for concentration$^{\text{3, 6}}$ (75--82\%), violence tolerance$^{\text{1, 4, 6}}$ (70--81\%), shooting experiences$^{1,4,6}$ (70--79\%), and moderate in others.

\begin{tcolorbox}[title=Key Takeaway 1]

\textbf{Observations:} BM and HJ provide high F1 across demographic, health and behavioral attributes, while FE for mental-health attributes. Overall, results show that a substantial set of private user attributes can be inferred with high F1 (80-100\%) from sensor data.

\textbf{Implications:} VR sensor data enables substantial user profiling. Profiling risk is sensor-dependent rather than uniform. Therefore, users privacy protections must be sensor-specific, with users control per sensor group.

\end{tcolorbox}

\subsection{Feature Analysis (RQ2)}\label{subsec:eval_featureAnalysis} 
Next, we analyze features for attributes with high and moderately high-risk, prioritizing those most likely to be exposed and most in need of mitigation. For BM (see Figures~\ref{fig:feature_BM_main} and \ref{fig:feature_analysis_bm}, Appendix~\ref{app:Results}), demographic (\eg{} gender$^{\text{1,2,4,5,6}}$, ethnicity$^{\text{4,5,6,\sens}}$) and health attributes (\eg{} BMI$^{\text{3,6}}$, physical fitness$^{\text{1,4,5,6}}$) that provide high F1 for social, archery, shooting, and rhythm have max standing height as the high importance (HI) feature. Additional features such as sitting/standing statistics, controller span (analogous to wingspan), and left–right head position also emerge as important. For physical fitness$^{\text{1,4,5,6}}$, head and controller rotations are identified as medium-high to high importance features, particularly in physically demanding app groups (\eg{} archery, rhythm). For mental health attributes such as stress$^{4,6}$ and anxiety$^{4,7,8}$, high-performing app groups (flight, social, archery) primarily rely on dynamic motion features—head and controller rotations and forward–backward movement—rather than static user-specific measurements. Similar trends are observed for HJ (Appendix~\ref{app:Results}, Figure~\ref{fig:feature_analysis_hj}). Demographic inference primarily depends on hand positional features, while mental health and behavioral attributes rely more on hand joint rotations (\eg{} height phobia$^{4,6}$). In contrast, physical fitness$^{\text{1,4,5,6}}$ is inferred using a combination of physical measurements and movement-based features, that encode physical biometrics.

For FE (see Figure~\ref{fig:feature_analysis_fe} in Appendix~\ref{app:Results}) we observe demographic attributes such as age$^{\text{1,2,3,4,5,6,\ferpa}}$ and gender$^{\text{1,2,4,5,6}}$ show feature importance relevant to app groups. App groups involving social interactions (\eg{} social, KW) predominantly contribute features related to positive or low-arousal negative emotions (\eg{} surprise, disgust, happy), compared to features linked to non-emotional expressions or high-arousal negative emotions. In contrast, app groups where users may experience negative emotions (\eg{} shooting) show high importance for features associated with negative emotions (fear, anger, or disgust). Across most app groups, important non-emotional features include jaw sideways, lip suck, and chin raiser, indicating older vs. younger users exhibit distinct facial feature around chin, jaw and lip. Behavioral and mental health attributes show variety of feature important for different app groups; \eg{} for social apps, disgust and surprise are top features for stress$^{4,6}$. %

\begin{tcolorbox}[title=Key Takeaway 2]

\textbf{Observations}: Demographics and anthropometric attributes stem from sensor measurements, whereas mental health and behavioral attributes stem from emotion- and movement-based features.

\textbf{Implications:} Since each attribute category is driven by a specific set of features (\eg{} measurements for demographic), suppressing those features can reduce profiling risk across multiple attributes.

\end{tcolorbox}

\subsection{Profiling Across App Groups (RQ3)}\label{subsec:eval_appgroup} 
Next, we examine the relationship between user attributes and app groups (\ie{} app activities and emotional states; Section~\ref{subsec:method_app_group}), drawing from Table~\ref{tab:f1score-bm} (BM) and Appendix~\ref{app:Results} Tables~\ref{tab:f1score-fe} (FE), \ref{tab:f1score-eg} (EG), and \ref{tab:f1score-hj} (HJ). Several app groups yield higher attack performance for specific attributes using BM, HJ, and FE. For example, social, shooting, and archery apps achieve high F1 for gender$^{\text{1,2,4,5,6}}$ (70--90\%), age$^{\text{1,2,3,4,5,6,\ferpa}}$, and ethnicity$^{\text{4,5,6,\sens}}$ (65--90\%) across sensor groups. For \textit{demographics}, attack performance varies by app group: using BM and HJ, gender$^{\text{1,2,4,5,6}}$ reaches high risk levels (80--100\%) in social, rhythm, and shooting/archery apps, but remains low in flight, KW, and IN. As shown in Section~\ref{subsec:eval_featureAnalysis}, body measurements such as height$^{\text{3,5,6}}$ and hand span are key for inferring gender. 
however, flight, IN, and KW involve seated or tilt-walking interactions (\eg{} gorilla-style), which obscure true measurements and reduce inference performance.

Alternatively, \textit{anthropometrics} such as height$^{\text{3,5,6}}$, weight$^{\text{3,4,6}}$, and BMI$^{\text{3,6}}$ achieve moderate to high F1 across most app groups using BM, with lower performance in flight, KW, and IN (44--69\%), where true physical measurements are less exposed. Physical fitness$^{\text{1,4,5,6}}$ and mental health attributes (anxiety$^{4,7,8}$, stress$^{4,6}$, height phobia$^{4,6}$) reach high to very high F1 (70--100\%) in social, flight (mental health only), archery, KW, and rhythm apps, where activity-driven body motions reveal health signals;\eg{} HJ infers height phobia with high F1 in flight (85\%) and KW (77\%). FE performs poorly for physical fitness (45--65\%) but moderately for mental health attributes, with limited variation across apps. For \textit{user interests and behaviors}, BM achieves moderately high performance (70--79\%) when aligned with app context; \eg{} shooting experience$^{1,4,6}$ reaches high F1 in shooting (81\%), rhythm (86\%), and archery (82\%) apps, where movements resemble shooting actions.

\begin{tcolorbox}[title=Key Takeaway 3]

\textbf{Observations:} App groups influence user profiling risk, as app-specific contexts and user interactions expose certain attributes more strongly than others.

\textbf{Implications:} Profiling risk is influenced by app groups: different apps influence different attribute exposures. Thus, defenses must consider \emph{which attributes are exposed in which apps} rather than applying a uniform policy across apps.

\end{tcolorbox}

\subsection{Multi-Sensor Adversary (RQ4)}\label{subsec:eval_Privileged_Adversary}We present the evaluation results of the multi-sensor adversary settings (see Section~\ref{subsubsec:ThreatModel_Adversaries}) to address \textbf{RQ4}. Our multi-sensor analysis is scoped as a proof-of-concept and evaluates a representative subset of sensor group combinations based on the realistic scenarios (details are below). We acknowledge that additional adversarial combinations are possible, and our methodology readily extends to explore them in future work.

\subsubsection{BM and FE}

Realistic adversaries for this setting include: (1) apps that use BM and FE together for functionality purposes %
(\eg{} social apps) with user-granted permissions, and (2) third-party apps (\eg{} ALVR~\cite{alvr}) that record both BM and FE without permission.
The performance for \textit{demographic} attributes, such as age$^{\text{1,2,3,4,5,6,\ferpa}}$ and gender$^{\text{1,2,4,5,6}}$ remains similar for some app groups (\eg{} social), but improves for others (\eg{} KW and IN)--raising their adversarial risk from medium-high to high (see Table~\ref{tab:f1score-bm-fe}).
As discussed in Sections~\ref{subsec:eval_profiling_quant} and \ref{subsec:eval_featureAnalysis}, BM loosely provides users' measurements (\eg{} height) for KW, IN, which are crucial to infer gender, combining BM with FE enhances accuracy in those app groups. 
For anthropometrics, performance remains unchanged, since facial features lack %
users' body measurements and vice versa.
We also observe an 8–30\% performance increase 
for height phobia, motion sickness, and interest/behavior attributes (\eg{} violence tolerance). 
Another observation is, attack improvements further depend on app groups: violence tolerance increases in shooting, KW, and archery apps with fear-inducing activities. %

\subsubsection{BM, FE, and EG} 
In this setting, the attack risk remains high, comparable to the BM+FE adversary with only a slight increase (2–5\%) in some groups (see Table~\ref{tab:f1score-eg-bm-fe}). Since EG is the weakest sensor group and BM+FE already yields strong performance, adding EG does not significantly improve results.

\begin{figure}[t!]
    \begin{subfigure}{.96\columnwidth} 
        \centering
        \includegraphics[width=.94\textwidth]{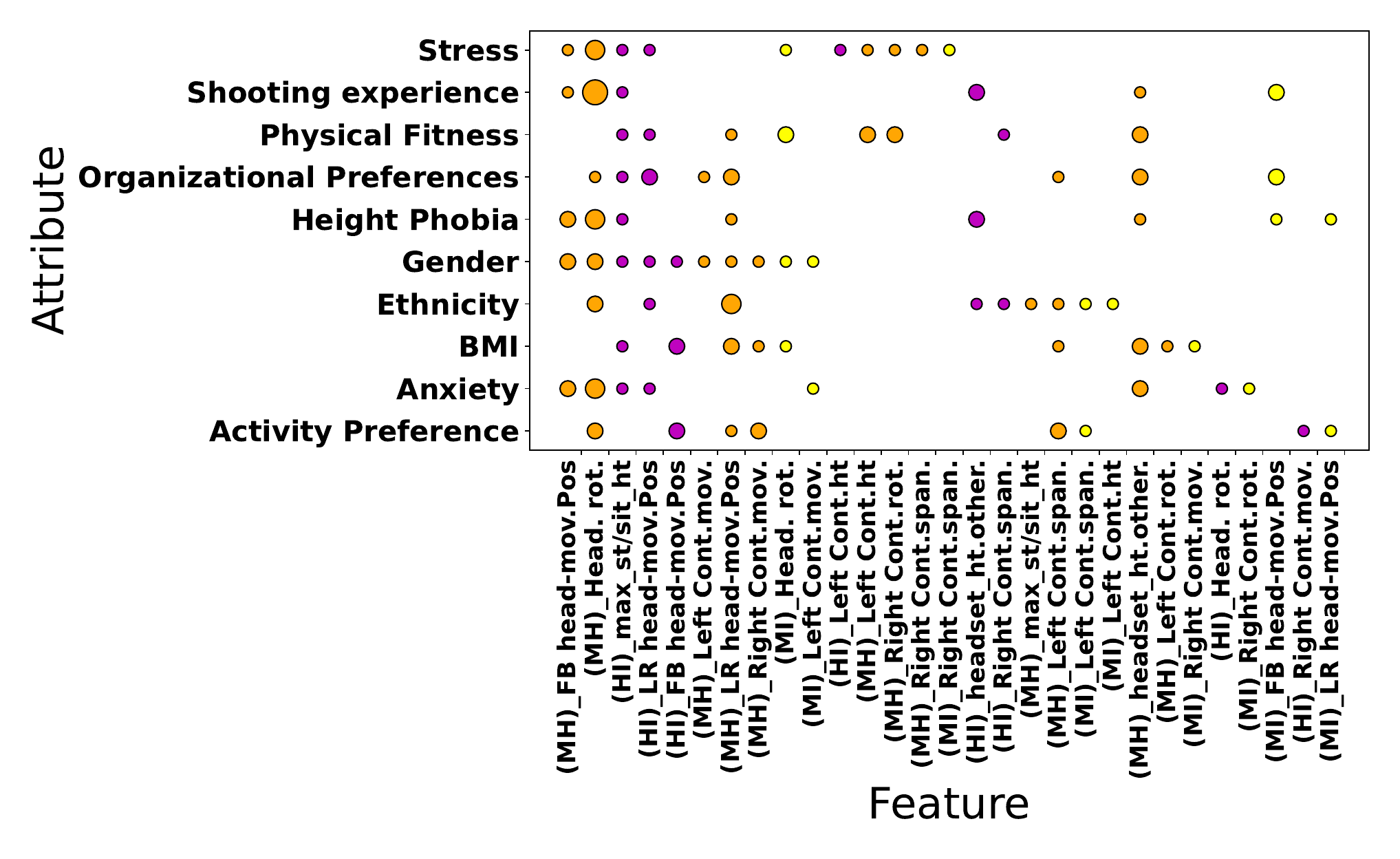}
        \caption{Social} \label{fig:feature_BM_Social}
    \end{subfigure}
    \begin{subfigure}{.94\columnwidth} 
        \centering
        \includegraphics[width=.95\textwidth]
         {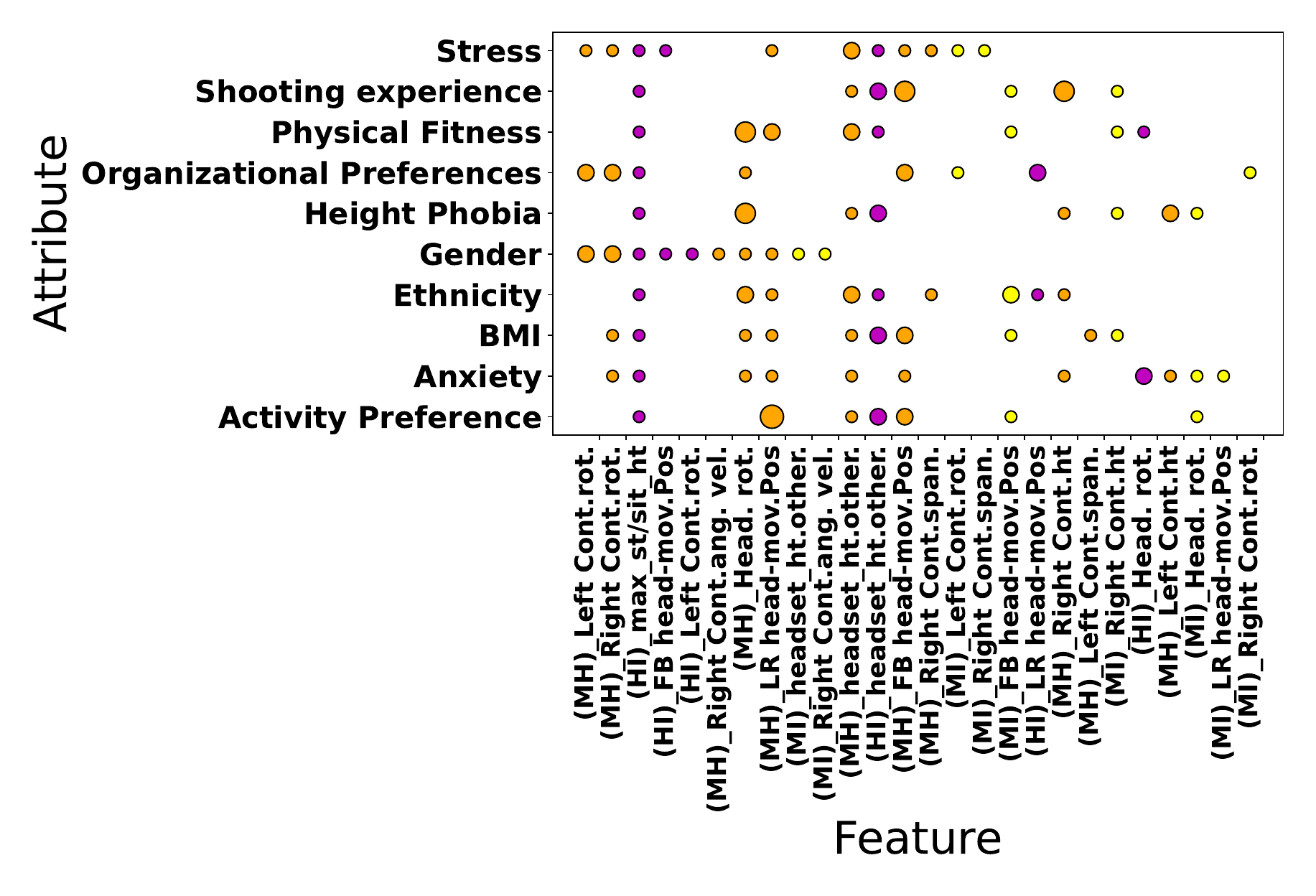}
        \caption{Archery}
        \label{fig:feature_BM_Arch}
    \end{subfigure}
    \caption{\textbf{Feature Analysis on BM for Social and Archery App Groups.} The Y-axis lists attributes and the X-axis shows top features. Color encodes feature importance: HI (high, pink), MH (medium-high, orange), MI (medium, yellow).
    }
    \label{fig:feature_BM_main}
\end{figure}

\begin{tcolorbox}[title=Key Takeaway 4]

\textbf{Observations:} Multi-sensor adversary (BM+FE) elevates risk for health and several demographics (\eg{} gender, age) from moderate to high (8–30\%), while anthropometrics remain unchanged. Adding EG yields marginal gains (2–5\%), indicating limited contribution.

\textbf{Implications:} Combining sensor groups amplifies profiling risk, highlighting the necessity of multi-sensor access protections to mitigate high-risk outcomes.

\end{tcolorbox}

\section{Discussion}\label{Discussion}

This section discusses the implications of our findings (see Section \ref{sec:evaluation}) for users' privacy risks (\ref{discussion:Risk}), guided by \textbf{RQ5}, and both design (\ref{discussion:Lessons_for_Privacydefense}) and regulatory (\ref{Discussion:Compliance}) implications, guided by \textbf{RQ6}. We also discuss limitations and future directions (\ref{limitationsand Future directions}) as well as ethical considerations (\ref{Discussion:Ethics}).

\subsection{Potential User Risks in Consumer VR Apps (RQ5)} \label{discussion:Risk}

\subsubsection{Feasibility of User Profiling} \label{discussion:profiling_feasibility}
We now revisit our central question: is it possible to profile users using \textit{only} sensor data from non-adversarial, consumer VR apps?
We demonstrate that BM, FE, and HJ enable moderately high to high F1 for multiple attributes (see Table~\ref{tab:f1score-bm} and Appendix Tables~\ref{tab:f1score-fe}-\ref{tab:f1score-eg-bm-fe}). Our feature analysis suggests that adversaries can infer one or more attributes even from a subset of sensor data: for example, in the social app group, adversaries can infer gender, physical fitness, and BMI using only height, left-right head movement, and other vertical headset statistics, which consistently rank as top features across these attributes (see Figure~\ref{fig:feature_BM_Social}).

\subsubsection{User Attributes vs. Threat Scenarios} \label{discussion:att_vs_threats}
Compared to prior works~\cite{nair2022metadata, Nair1kPersonal2023}, we study data collected from multiple consumer VR apps, capturing a more realistic VR ecosystem used by millions of users, including vulnerable populations~\cite{steamcharts,demandsage_vr_stats_2025}. 
Our findings highlight serious risks to users: a wide range of attributes, including sensitive demographic, health, and behavioral information were inferred with high F1 (70--100\%). Exposure of one or a combination of multiple attributes can pose various threats to users (see Section~\ref{subsec:method_profilingTaxonomy}). 
For example, height phobia$^{4,6}$, anxiety$^{4,6}$, and stress$^{4,6}$ achieve up to 81--90\% F1 using BM in several app groups (\eg{} flight, KW, archery). A user who shares BM with those apps could be exposed to safety threats (\eg{} virtual shock) with high %
risk, according to our risk level described in Section \ref{subsec:evaluation_metrics}, and as indicated by superscripts `4' and `6' (see Section \ref{subsubsec:ThreatModel_ThreatScenarios}).
Exposure of age$^{1,2,3,4,5,6}$ and gender$^{1,2,4,5,6}$ can put users at risk of targeted advertising (\eg{} especially if combined with user interests and behavioral data), identity theft (\eg{} if combined with other unique identifiers, such as their name and anthropometric data), and safety concerns, such as harassment or cyberbullying in VR for vulnerable groups, which can be more extreme when age and gender are known~\cite{hinduja2024metaverseSafety}. 

\subsubsection{Threats Involving Anthropometrics}\label{discussion:biometric_profiling} Anthropometric data poses privacy and security risks, as it can enable real-world or virtual identity theft: an adversary could infer a user's measurements, such as height$^{3,5,6}$, face length$^{3,5,6}$, and arm length$^{3,5,6}$, as well as attributes like age$^{1,2,3,4,5,6}$, from sensor data, enabling the replication of a user's digital identity in the virtual world. As discussed in Section~\ref{sec:evaluation}, our results show that users sharing BM, FE, or HJ data have moderately high to high adversarial risk (70--100\% F1) for these attacks.

\subsubsection{User Risks Across Multiple Apps}\label{discussion:acrossapprisks} The attacks, discussed in Section~\ref{discussion:profiling_feasibility}-\ref{discussion:biometric_profiling}, can be more extreme, as users can be identified across different apps~\cite{jarin2025behavr,acrossapp2025} and different settings within an app~\cite{jarin2025behavr}. Thus, an attacker can track users across apps and settings, aggregate inferred attributes, and launch profiling-based attacks across multiple apps.
 
\subsection{Design Implications (RQ6)}\label{discussion:Lessons_for_Privacydefense} The varying strengths of sensor data across different apps in inferring sensitive user attributes highlight the necessity of sensor- and app-specific privacy protections for users. We will discuss design implications next, considering both usability and risk mitigation tradeoffs. 

\subsubsection{User Awareness}\label{discussion:UserAwareness}
Users often share sensor data %
without recognizing the privacy risks~\cite{CHI24UsersAwarenessVR}  
as such sharing feels routine and harmless rather than an explicit or suspicious request that would raise concern. Moreover, marginalized users often conceal attributes (\eg{} gender, age, body measurements) to avoid potential harm, even while choosing avatars~\cite{PETS2025reveal}. 
We recommend that service providers (\eg{} Meta) raise awareness by clearly communicating risk levels across attributes \wrt{} sensor and app groups as shown in \sysname{}, such as through a simple risk dashboard or by sending quick alerts when multiple sensors are active together, since combining multiple sensor data raises profiling risk.

\subsubsection{Sensor Permission Design} \label{discussion:PermissionDesign} 
Another potential defense is to turn off the sharing of certain sensor groups based on associated risks. %
Users should be able to independently enable or disable each type of sensor data across VR platforms. Modern VR platforms (\eg{} Oculus) provide permission checks for FE, EG, and HJ ~\cite{USECRoesnerXRPermission2025,radway_permissionOculus23, metapolicy}, whereas others (\eg{} SteamVR) still lack equivalent controls for sensor data access (see Section \ref{subsec:method_dataCollectionPractices}). Platforms should also clearly disclose the purpose of collecting sensor data, ensuring it aligns with the functionality of the app group. Currently, these purposes are often vague~\cite{USECRoesnerXRPermission2025}; for example, the purpose of facial data collection in Meta Horizon Workstation, an IN app (see Section~\ref{subsec:method_dataCollectionPractices}). A major challenge is the lack of granularity in data-sharing choices, which can significantly limit usability. For example, if users do not share FE sensor data in social apps, it may reduce social interaction quality based on facial expression.
\subsubsection{Obfuscation-Based Design}\label{discussion:Obfuscation}
A more granular defense approach would be to obfuscate sensor data (\eg{} with local differential privacy (LDP)~\cite{dwork2006differential}), allowing users to share data without fully disabling sensors. Guided by \sysname{}’s feature analysis, obfuscation can target the most important features for inferring sensitive attributes, thereby reducing adversarial risks.
A more fine-grained approach is to target top features that affect multiple attributes. For example, for BM, the height feature is of high to medium-high importance for inferring gender and physical fitness. Obfuscating height can help lower the risk of exposing all three attributes and relevant threats (\eg{} safety). The main challenge is to find the best privacy-usability trade-off while obfuscating attributes. 

\subsubsection{Privacy Assistance Agent} \label{discussion:Obfuscation} VR platforms could integrate a privacy-assistance agent (\eg{} AI agent) to help users manage privacy decisions in real time. 
Guided by our findings, this agent may (i) provide permission nudges or suggestions and/or (ii) automatically predict and configure permission choices on the user’s behalf, as is done in other domains~\cite{wu2026towardsSP2026,MorelIF25}. The agent can analyze the risk level associated with each sensor group in each app and scenario (\eg{} private vs. public virtual space) and suggest or take actions, such as turning off sensor access or enabling privacy-preserving sharing (\eg{} recommending or selecting noise levels based on users’ privacy choices). This automation would reduce cognitive burden, enabling privacy and usability.

\subsection{Regulatory Implications and Recommendations (RQ6)} 
\label{Discussion:Compliance}

\subsubsection{Implications for Compliance} 
VR devices from different companies can vary widely in their sensor permission options (see Section~\ref{subsec:method_dataCollectionPractices}). %
Permissions are only meaningful if users have real choices, as opposed to forcing access to all sensors in order to use an app.
The CCPA definition of personal information~\cite{CCPA2018} (\ie{} foundation for our \taxonomy{}) classifies biometric information as ``sensitive personal information'', which affords it special protections.
Meta discloses that Meta and third parties are able to access ``abstracted'' sensor data, not raw sensor data~\cite{metapolicy}. We consider that ``abstracted'' sensor data aligns with CCPA's definition of biometric information, and we demonstrate its risks for user profiling.
Depending on VR providers' interpretation of ``abstracted'' sensor data, they may skirt associated protections for biometric information, endangering users' privacy.

\subsubsection{Regulatory Recommendations}
We recommend that lawmakers and regulators scrutinize VR providers' privacy policies regarding their definitions of sensor data to better align law with practice. We urge lawmakers and regulators to pay special attention to vulnerable attributes we have identified through the development of our \taxonomy{} and analysis of threat scenarios (see Sections~\ref{subsec:method_profilingTaxonomy} and \ref{subsubsec:ThreatModel_ThreatScenarios}). Regulations should mandate that VR providers only collect sensor data that is required for functionality. Further, regulations should require granular, opt-in sensor permissions and obfuscation-based privacy options for sensor data among VR devices. 
CCPA regulations already enforce strict privacy rights for vulnerable groups (\eg{} consumers under 16 years old) and for sensitive personal information~\cite{CCPA2018_law}, and thus we recommend similar protections be required for VR and other devices that are able to collect sensitive biometric data.

\subsection{Limitations and Future Directions}\label{limitationsand Future directions}

\subsubsection{User Study Size}
A limitation of our study is the number of participants (see Section~\ref{subsec:profiling_dataset}). Our goal was to evaluate profiling across multiple apps, sensors, and attributes, thus we relied on an in-person user study, resulting in a sample size comparable to prior in-person VR studies (\eg{}~\cite{nair2022metadata, TricomiNPCG23CanNotHide,miller2022combining,liebers2021understanding}) but with more app and sensor coverage. 

Recruitment was challenging, as participation required multi-hour gameplay across 10 apps and a follow-up survey. Some attributes in the survey were self-reported, which may introduce bias. Our attack classifier design (Section~\ref{subsec:model_training}) is guided by our threat model and participant distribution. Due to the limited participant pool, we could not exhaustively evaluate all adversarial capabilities; \eg{} the absence of participants under 18 prevents demonstrating child–adult age gating, even though such inference aligns some attackers objective. While these limitations may constrain generalizability, we hope our work offers new insights by expanding the problem space across sensors, apps, and attributes.  We hope our method and released artifact\footnote{\sysname{} source code is publicly available at \url{https://github.com/UCI-Networking-Group/VR-Profilens.git}.} enable future work with more users, additional apps, and new threat scenarios. %

\subsubsection{Design Defense Tools} 
As discussed in Sections~\ref{discussion:Lessons_for_Privacydefense} and \ref{Discussion:Compliance}, future directions include designing defenses, such as AI agents to assist users with privacy decisions, sensor data permissions, more granular privacy-preserving options by platforms, and the enforcement of data minimization through regulation. Future defenses should be evaluated through user studies to ensure acceptable privacy–usability tradeoffs.

\subsection{Ethical Considerations} \label{Discussion:Ethics}
We conducted an IRB-approved user study with careful attention to ethical considerations. 
We select apps to align with our IRB risk minimization protocol~\cite{irbminimalrisk}, such as excluding horror or violence genres that may cause psychological harm or distress to users. Participation in our study was voluntary, allowing participants to opt out at any time. Participants provided informed consent through a written consent form prior to participation. Our data was stored in a secure, password-protected database, and access was restricted to the lead researcher and only used for research purposes.

Personal attributes in our study were selected using our methodology and ethical considerations, with collection limited to attributes necessary for the study. Attributes were derived from prior VR literature~\cite{NairSurvey23,garrido2023sok,Nair1kPersonal2023} as part of our taxonomy development (see Section~\ref{subsec:method_profilingTaxonomy}), which identifies both explicit and implicit attributes, linked directly or indirectly to sensor/behavioral signals. 
Additionally, given the sensitivity of the survey content, all questions were optional and included a ``Prefer not to answer'' response. No personally identifiable information (\eg{} name, email) was collected. Responses were identified using random unique IDs. Our reported results are not linked to any individual, and our experiments do not cause harm to the participants.

\section{Conclusion}\label{sec:conclusion}
We present \sysname{}, a framework for identifying and analyzing user profiling risks in Virtual Reality using "abstracted" sensor data. To the best of our knowledge, \sysname{} is the first holistic demonstration of VR user profiling by (1) developing an expandable and systematic \taxonomy{}, (2) designing a methodology to examine profiling across sensor and app groups via a user study, (3) demonstrating profiling feasibility through empirical evaluation, and (4) providing both user-centered design and regulatory recommendations based on our findings. Additionally, our app groupings, threat model, and taxonomy are designed to be expandable, allowing new apps, emerging threats, and new attributes to be incorporated. Overall, our results demonstrated the adversarial feasibility of inferring user attributes %
using ``abstracted'' sensor data, underscoring the need for user-centered privacy protections and regulatory attention to improve users' privacy and safety in VR.

\section*{Acknowledgments}
This work was supported by the National Science Foundation under award numbers 1956393 and 1900654 and a gift from the Noyce Initiative. We also gratefully acknowledge that Ismat Jarin was partially supported by the ESET Women in Cybersecurity Scholarship, and Olivia Figueira was partially supported by the UCI ICS Steckler Family Endowed Fellowship and the ARCS and Danaher Foundation Scholar Award. We would like to thank Rahmadi Trimananda for his feedback on the initial version of this work. We thank the anonymous USEC reviewers for their constructive feedback, which helped improve this paper. The authors used OpenAI’s ChatGPT~\cite{chatgpt} for minor editing, including correcting typos, improving grammar, and paraphrasing.

\bibliographystyle{IEEEtran}
\bibliography{references}

\section*{Appendices}\label{sec:appendix}

\begin{table*}[t]
        \tiny
	\centering
	\caption{\textbf{List of \apps{} selected consumer VR apps ($a_1,..., a_{10}$) for \sysname{} study and corresponding app activities.} See Section~\ref{subsec:VR_apps} and Appendix~\ref{app:Apps} for app selection details.}
	\begin{tabular}{r | p{20mm}| p{90mm}}
	    \toprule
		\textbf{App No.} & \textbf{App Title} & \textbf{App Activities}
		\\
		\midrule

            $a_{1}$ & Rec Room                             & Explore welcome room or virtual recreation center; Users use bare hand for waiving or handshaking. \\ 
            $a_{2}$ & VRChat                               & Explore virtual scene by walking around; The user will wave or greet with bare hands. \\ 
            $a_3$ & DCS World Steam Edition               & Fly a military aircraft: the user first control the aircraft with controllers and then with bare hands. \\ 
            $a_{4}$ & X-Plane 11                           & Fly a civilian aircraft and interact with the virtual objects with the controllers and with bare hands. \\
            $a_5$ & Job Simulator                         & Explore office-worker simulation; The user is to interact with a virtual office objects with controllers and then with bare hands. \\
            $a_{6}$ & Tabletop Simulator                   & Move chess pieces: first with the controllers and then with bare hands. \\
            $a_7$ & Gorilla Tag                           & Perform gorilla movement (walk like gorilla to explore the environment): first with the controllers, then with bare hands.\\
            $a_8$ & Beat Saber                            & Cut objects with light-sabers: with the controllers and then with bare hands. \\ 
            $a_{9}$ & No Man's Sky                         & Explore an unknown planet by teleporting; the user interacts with a laser gun (shoot targets) with controllers and then with bare hands. \\
            $a_{10}$ & Elven Assassin                         & Shoot arrows to monsters: with the controllers and then with bare hands. \\ 
           
	    \bottomrule
	\end{tabular}
	\label{tab:list-of-apps}
\end{table*}

\subsection{Details on Methodology} \label{app:methodology}
In this appendix, we elaborate on our methodology, as introduced in Section~\ref{sec:methodology}, including the sensor data structure (\ref{app:sensor_data_structure}), VR apps we studied (\ref{app:Apps}), VR app groups (\ref{app:App groups}), sensor data collection practices (\ref{app:data_collection_practices}), and the \taxonomy{} (\ref{app:taxonomy_details}).

\subsubsection{Details on Sensor Data Structure} \label{app:sensor_data_structure}
Among the sensor groups, Body Motion (BM) includes the position, rotation, angular and linear velocity of the two controllers, and only the position and rotation of the headset~\cite{openxrcoordinate}. This group has been studied in prior works~\cite{miller2020personal, nair2023unique, Nair1kPersonal2023,jarin2025behavr}. \Eyedata{} (EG) includes the position and rotation of both left and right eyes (7 values per eye), as specified by OpenXR~\cite{eyegazepositionrotationunity,eyegazepositionrotationkhronos}. \Handdata{} (HJ) consists of 26 articulated joints per hand, as defined by the OpenXR \texttt{XrHandJointEXT} structure~\cite{openxrhandjointsconvention} and include the position and rotation of each joint %
~\cite{openxrhandtracking}. Finally, \Facedata{} (FE) includes 64 facial expression elements tracked by the OpenXR standard, following the \texttt{XrFaceExpressionFB} structure~\cite{openxrfacetracking}.
These elements map to 31 Facial Action Units (AUs) in the FACS system~\cite{Ekman1978FACS}, each representing a facial muscle movement. 

\begin{table*}[t!]
  \tiny
  \centering %
  \caption{\textbf{Grouping apps ($a_1, ..., a_{10}$) listed in in Table \ref{tab:list-of-apps} based on their similarity of activities and emotional states (arousal/valence \cite{bouchard2011emotions_valence}).} {\em Sensor Groups:}  BM, EG, HJ, FE. {\em Emotional States:} LA = low arousal, HA = high arousal, PV = positive valence, NV = negative valence. Important sensors is based on our app grouping, where each sensor is relevant with app specific activities.}
  
  \begin{tabular}{l p{7mm} p{12mm} p{19mm} p{48mm} p{20mm}}
    \toprule
    \textbf{App Groups} & \textbf{App No.} & \textbf{Important Sensors}& \textbf{Scope}& \textbf{App-Specific Activities}  & \textbf{Arousal/Valence} 
    \\
    \midrule
      Social & $a_{1}$, $a_{2}$ & BM, EG, FE, HJ & Social Media & Walking, waving, socializing and sightseeing & LA/PV, HA/PV  \\
      Flight Simulation & $a_3$, $a_{4}$ & BM, HJ& Train./Education & Holding onto the airplane control stick, interacting with control panel/buttons in an airplane cockpit & LA/NV, HA/NV, LA/PV \\ %
      Interactive Navigation & $a_{5}$, $a_{6}$  & BM,HJ & Office/Entertainment & Grabbing, moving objects \ie{} short time interaction with objects & Neutral, LA/PV, LA/NV \\
      Knuckle-walking & $a_7$ & BM, HJ, FE &Social Media & Walking using an open fist like a gorilla & LA/PV, HA/PV, LA/NV \\ %
      Rhythm & $a_8$ & BM, HJ & Entertainment & Dancing-like moves and cutting objects in quick pace & All  \\ %
      Shooting \& Archery & $a_{9}, a_{10}$ & BM, HJ &Train./Education. & Grabbing/holding a gun/arrow, aiming+shooting objects  & LA/NV, HA/NV  \\ %
    \bottomrule
  \end{tabular}
  \label{tab:Apps-Grouping}
\end{table*}

\subsubsection{More about VR Apps} \label{app:Apps}
The \sysname{} study is based on \apps{} apps selected from the SteamVR store~\cite{steamvrstore}, as listed in Table \ref{tab:list-of-apps} and discussed in Section \ref{subsec:VR_apps}. Starting from the top 100 apps listed under ``Most Played VR Games'' on Steam~\cite{apprankings}, we exclude apps based on user discomfort, such as horror or violent genres, in accordance with 45 CFR § 46.111(a)(1) to minimize participant risk~\cite{irbminimalrisk}.

\begin{table*}[htbp]
\centering
\scriptsize
\caption{\textbf{Attributes Obtained from \taxonomy{} for our \sysname{} Study.} The \textbf{Class:Statistics} column represents each class name along with its corresponding statistics. If an attribute is continuous in nature, we use regression and mark the \textbf{Class:Statistics} field as N/A. If two attributes are highly correlated and yield the same response, we mark the response as N/P. The \textbf{Selection} column indicates whether an attribute is selected (\checkmark) or filtered out (\xmark), with the corresponding reason provided for exclusion.} \label{tab:Attributes_and_statistics}

\begin{tabular}{|p{1cm}|p{4.2cm}|p{4.3cm}|p{3.42cm}|}
\hline
\textbf{Category} & \textbf{Attribute} & \textbf{Class: Statistics} & \textbf{Selection} \\
\hline
\multirow{7}{*}{\shortstack{Demo-\\graphics}}
& Sex/Gender$^{1,2,4,5,6}$ & male: 55\%, female: 45\% & \checkmark \\
\cline{2-4}
& Age / Date of Birth$^{1,2,3,4,5,6}$ & $<30:65\%$, $\geq$30: 35\%, no-res.: 10\% & \checkmark \\
\cline{2-4}
& Religion$^{4,5,6}$ &religious:25\%,non-religious:65\%,no-res:10\% & \xmark : does not map \\
\cline{2-4}
& Marital Status$^{1,2,5,6}$ & married:35\%, unmarried:65\%, no-res.:10\% & \checkmark \\
\cline{2-4}
& Ethnicity/National Origin$^{4,5,6}$ & Asian:65\%, others:35\% & \checkmark \\
\cline{2-4}
& Race$^{4,5,6}$ & Asian:65\%, others:35\% & \xmark : Same response as Ethnicity \\
\hline
\multirow{3}{*}{\shortstack{Personal\\History}}
& Education (Highest Level)$^{1,5,6}$ &$<$graduate:0\%, graduate:100\%  & \xmark : Non-Discriminative Response  \\
\cline{2-4}
& Academic Interests$^{1,5,6}$ & CS/EECS:100\% & \xmark : Non-Discriminative Response \\
\cline{2-4}
& Income (USD)$^{1,5,6}$ & $\leq$40k:50\%,$>$40k:50\% & \xmark\\
\hline
\multirow{10}{*}{\shortstack{Anthro-\\pometrics}}
& Hand shape/length$^{3,5,6}$ & N/A & \checkmark \\
\cline{2-4}
& Face length$^{3,5,6}$ & N/A & \checkmark \\
\cline{2-4}
& Height$^{3,5,6}$ & N/A & \checkmark \\
\cline{2-4}
& Arms Length$^{3,5,6}$ & N/A & \checkmark \\
\cline{2-4}
& IPD$^{3,5,6}$ & respond:50\%, non-res:50\% & \xmark : Limited Response by users \\
\cline{2-4}
& Wingspan$^{3,5,6}$ & respond:40\%, non-res:60\% & \xmark : Limited Response by users \\
\cline{2-4}
& Weight$^{3,4,6}$ & N/A & \checkmark \\
\cline{2-4}
& BMI$^{3,6}$ & normal: 55\%, overweight/obese: 45\%& \checkmark \\
\hline
\multirow{11}{*}{Health}
& Physical Fitness$^{1,4,5,6}$ & fit:45\%, unfit:45\%, no-res.:10\% & \checkmark \\
\cline{2-4}
& Illness (COVID)$^{1,5,6}$ & yes: 5\%, no: 80\%, no-res.:15\%& \xmark : Limited Response by users\\
\cline{2-4}
& Color blindness$^{1,5,6}$ & yes:0\%, no:90\%, no-res.:10\% & \xmark : Non-Discriminative Response \\
\cline{2-4}
& Close / Distance vision and lenses$^{1,5,6}$ & yes:35\%, No:55\% no-res.:5\%& \checkmark \\
\cline{2-4}
& Mental disability$^{4,5,6}$ & yes:0\%, no:100\% & \xmark : Non-Discriminative Response \\
\cline{2-4}
& Physical disability$^{4,5,6}$ & yes:0\%, no:100\% & \xmark : Non-Discriminative Response \\
\cline{2-4}
& Anxiety$^{4,6}$ & yes:45\%, no:55\% & \checkmark \\
\cline{2-4}
& Stress$^{4,6}$ & yes: 65\%, no:20\%, no-res.:15\%& \checkmark \\
\cline{2-4}
& Height phobia$^{4,6}$ & yes:45\%, no:50\%, non-res.:5\% & \checkmark \\
\cline{2-4}
& Motion sickness$^{4,6}$ & yes:30\%, no:65\%, no-res.:5\% & \checkmark \\
\cline{2-4}
& Sleepiness$^{5,6}$ & yes: 0\%, no: 100\% & \xmark : Non-Discriminative Response \\
\hline
\multirow{17}{*}{\shortstack{Interests \&\\Behaviors}}
& Political orientation$^{1,4,5,6}$ & respond:50\%, no-res.:50\% & \xmark : Limited Response \\
\cline{2-4}
& Musical instrument experience$^{1,5,6}$ & N/P & \xmark \\
\cline{2-4}
& Dance experience$^{1,5,6}$ & N/P & \xmark \\
\cline{2-4}
& VR experience$^{6}$ &experienced:45\%, inexperienced:55\% & \checkmark \\
\cline{2-4}
& Athletics/sports experience$^{1,4,5,6}$ & N/P & \xmark : high correlation w/ another class \\
\cline{2-4}
& Shooting experience$^{1,4,6}$ &yes:45\%, no:55\% & \checkmark \\
\cline{2-4}
& Violence tolerance$^{1,4,6}$ & yes: 45\%, no: 45\%, no-res.:10\%& \checkmark \\
\cline{2-4}
& Alcohol Consumption$^{1,5,6}$ & yes:30\%,no/occasionally:60\%, no-res.:10\% & \checkmark \\
\cline{2-4}
& Caffeinated item consumption$^{1,5,6}$ & high: 35\%, moderate:60\%, no-res.:5\% & \checkmark \\
\cline{2-4}
& Social media usage$^{4,6}$ & Active:60\%, Inactive:40\% & \checkmark \\
\cline{2-4}
& Attention / Concentration$^{5,6}$ & good: 65\%, poor: 35\%& \checkmark \\
\cline{2-4}
& Handedness$^{5,6}$ & left:0\%, right:90\%, no-res.:10\% & \xmark : Non-Discriminative Response \\
\cline{2-4}
& Problem Solving Abilities$^{6}$ &confident:70\%, moderate:30\% & \checkmark \\
\cline{2-4}
& Working Preferences (Remote Working)$^{6}$ & N/P & \xmark : high correlation w/ another class \\
\cline{2-4}
& Organizational Preferences$^{6}$ & organized: 80\%,unorganized:20\% & \checkmark \\
\cline{2-4}
& Activity Preference (Indoor/Outdoor)$^{6}$ &indoor:50\%,outdoor:50\% & \checkmark \\
\cline{2-4}
& Introvert/Extrovert (Extraversion)$^{6}$ & Introvert:45\%, Extrovert:50\%, no-res.:5\% & \checkmark \\
\cline{2-4}
& Openness$^{6}$ & open:60\%, neutral:40\% & \checkmark \\
\cline{2-4}
& Conscientiousness$^{6}$ & N/P & \xmark:high correlation w/ another class \\
\cline{2-4}
& Emotional stability$^{6}$ & stable:50\%, unstable:45\%, no-res.0\%& \checkmark \\
\hline
\end{tabular}
\end{table*}

\begin{table*}[t!]
\centering
\tiny
\caption{\textbf{Feature Interpretations of Body Motion (BM) Sensor Group.} Based on the OpenXR data structures~\cite{openxrcoordinate}.}
\begin{tabular}{|p{5cm}|p{8cm}|}
\hline
\textbf{Feature Interpretation} & \textbf{Actual Feature Name} \\
\hline
Head rotation & Quatx Headset, Quaty  Headset, Quatz Headset, Quatw Headset \\
\hline
LR (Left-Right) head-movement Pos & Position.x Max Headset, Position.x Min Headset, Position.x Mean Headset, Position.x Std Headset, Position.x Median Headset \\
\hline
Max standing/sitting height & Position.y Max Headset \\
\hline
FB (Forward-Backward) head-movement Pos & Position.z Headset \\
\hline
Headset height other stats. & Position.y (Min, Mean, Std, Median) Headset \\
\hline
Left Controller rotation & Quatx Left Controller, Quaty Left Controller, Quatz Left Controller, Quatw Left Controller \\
\hline
Left Controller span & Position.x Left Controller \\
\hline
Left Controller height & Position.y Left Controller \\
\hline
Left Controller movement & Position.z Left Controller \\
\hline
Left Controller linear velocity & Lin0 Left Controller, Lin1 Left Controller, Lin2 Left Controller \\
\hline
Left Controller angular velocity & Ang0 Left Controller, Ang1 Left Controller, Ang2 Left Controller \\
\hline
Right Controller rotation & Quatx Right Controller, Quaty Right Controller, Quatz Right Controller, Quatw Right Controller \\
\hline
Right Controller span & Position.x Max Right Controller, Position.x Min Right Controller, Position.x Mean Right Controller, Position.x Std Right Controller, Position.x Median Right Controller \\
\hline
Right Controller height & Position.y Right Controller \\
\hline
Right Controller movement & Position.z Right Controller \\
\hline
Right Controller linear velocity & Lin0 Right Controller, Lin1 Right Controller, Lin2 Right Controller\\
\hline
Right Controller angular velocity & Ang0 Right Controller, Ang1 Right Controller, Ang2 Right Controller \\
\hline
\end{tabular}
\label{tab:feature_interpretation_BM}
\end{table*}

\subsubsection{Details on App groups} \label{app:App groups}
We define app groups based on similarities in activities, which mostly influence the BM and HJ sensor groups, and emotional (valence-arousal~\cite{suhaimi2018modeling_valence}) states, which mostly influence the FE sensor group as discussed in Section \ref{subsec:method_app_group} and listed in Table \ref{tab:Apps-Grouping}. Details are as follows: 

\parheading{Social App Group.} Social apps focus primarily on facilitating real-life social experiences through an enhanced sense of ownership over an avatar within a 3D environment, distinguishing them from traditional 2D text/image-based social media. Social apps facilitate various forms of social interactions~\cite{SocialVRIEEEVR21} and contribute to social engagements ~\cite{SocialVRCHI23}. Users engage in various activities within these apps, including walking, waving at friends, handshakes, conversations, and interacting with each other (\eg{} hangouts). The emotional state associated with social apps is typically positive, as users are expected to be relaxed and happy while using them. The social apps in our study are $a_{1}$, RecRoom, and $a_{2}$, VRChat.

\parheading{Flight Simulation App Group.} Flight simulation primarily resembles flying an airplane, helicopter, or other flying vehicle. These apps can be used broadly in air force training ~\cite{VR_airforce22}, military education~\cite{military_applicationVR13}, and entertainment purposes. The app-specific activities can include but are not limited to holding a joystick, interacting with a control-panel and buttons, and looking at the surrounding environment to avoid possible crashes. The arousal/valence mostly tends to be negative (fear/stress) depending on the user, or there can be low arousal/positive valence (surprise). The flight apps in our study are $a_{3}$, DCS World Steam Edition, and $a_{4}$, X-Plane 11.

\parheading{Interactive Navigation App Group.} Interactive navigation apps refer to apps in which the user has frequent interaction with virtual objects using their hands, controllers, and/or eyes. The app-specific activities include but are not limited to grabbing, pressing/moving objects, interacting with virtual keyboard/virtual office objects, cooking, and playing chess. The arousal/valence should be neutral as users tends to concentrate on activities. The interactive navigation apps in our study are $a_{5}$, Job Simulator, and $a_6$, Tabletop Simulator.

\parheading{Knuckle-Walking App Group.} This group includes apps where players imitate gorilla-like locomotion, using hand and knuckle positions to walk and jump through the environment. Regarding the valence/arousal, it is similar to the social app group, as the app is multiplayer and people explore the environment and go on adventures together. However, this kind of app includes climbing and jumping from tall structures, and thus can induce negative arousal/valence as well. The knuckle-walking app in our study is $a_7$, Gorilla Tag.

\parheading{Shooting App Group.} The shooting group refers to apps that involve simulated shooting activities. This typically includes apps in which players engage in activities such as aiming at and hitting targets with firearms. The arousal/valence state is most likely to be negative, as it involves looking for targets or enemies and shooting at them, \eg{} $a_9$, No Man's Sky.

\parheading{Archery App Group.} This app group simulates archery experiences, where users shoot at targets with bows and arrows, track targets with head movements, and looking around the environment. It has similarity as the shooting group, however, it requires users to hold a virtual bow and arrow, that is different than holding a virtual gun, resulting in distinct HJ/BM characteristics. The arousal/valence state is most likely to be negative as it involves shooting at targets/enemies, similarly to the shooting group, \eg{} $a_{10}$, Elven Assassin.

\subsubsection{Details on Sensor Data Collection Practices} \label{app:data_collection_practices} 
We examine Oculus Quest~\cite{metastore} and SteamVR~\cite{steamvr} apps data collection practices (introduced in Section~\ref{subsec:method_dataCollectionPractices}). We selected 20 apps based on popularity and relevance to our app groups, then manually inspected their sensor data collection and privacy policies to determine stated purposes of use.
We find that data-collection permissions vary across platforms. SteamVR~\cite{steamvr} apps require no runtime permissions and offer no disclosure of sensor data collection in their privacy policies. For Oculus Store apps, social apps collect data from all four sensor groups to support realistic avatars, and their policies state that this data is used for app functionality under Meta’s standard policy.
Interactive navigation apps disclose collecting hand joint and body motion data, aligning with app-specific activities (see Section \ref{subsec:method_app_group}). However, apps like Virtual Desktop and Meta Horizon Workstation also mention facial and eye tracking data collection, which aligns less with purposes of their app groups. Notably, some interactive navigation apps (\eg{} Job Simulator, Lost Recipes) either lack privacy disclosures or state that they collect data for advertising and share it with third parties. Flight and shooting apps collect both body-motion and voice data, whereas rhythm apps collect only body-motion. Specifically, apps such as Shuttle Commander and Pavlov VR disclose that they use collected data for marketing purposes. Shuttle Commander specifies that this usage is restricted to first-party marketing only. 
\begin{table*}[t!]
\centering
\tiny
\caption{\textbf{Feature Interpretations of Facial Expression (FE) Sensor Group.} Based on the OpenXR~\cite{openxrfacetracking} feature indices and Facial Action Coding System (FACS)~\cite{facsexplained} system.}
\begin{tabularx}{.984\linewidth}{|p{2.7cm}|p{3.7cm}|p{3.7cm}|p{6cm}|}
\hline
\textbf{Feature Interpretation} & \textbf{Element Number} & \textbf{Feature Interpretation} & \textbf{Element Number} \\
\hline
Happy            & [5], [6], [33], [34]                          & Surprise         & [23], [24], [25], [58], [59], [60], [61] \\
\hline
Anger            & [1], [2], [60], [61], [29], [30], [49], [50]  & Contempt         & [33], [11], [12]                       \\
\hline
Disgust          & [56], [57], [31], [32], [52], [53]            & Fear             & [1], [2], [23], [24], [25], [29], [30], [31], [32], [43], [44], [58], [59], [60], [61] \\
\hline
Sadness          & [23], [24], [1], [2], [31], [32]              & NEFE: Cheek Puff & [3], [4]                              \\
\hline
NEFE: Cheek Suck & [7], [8]                                     & NEFE: Chin Raiser & [9], [10]                            \\
\hline
NEFE: Eyes Closed & [13], [14]                                  & NEFE: Eyes Look Down & [15], [16]                        \\
\hline
NEFE: Eyes Look Left & [17], [18]                              & NEFE: Eyes Look Right & [19], [20]                       \\
\hline
NEFE: Eyes Look Up & [21], [22]                                 & NEFE: Jaw Sideways & [26], [27]                         \\
\hline
NEFE: Jaw Thrust & [28]                                       & NEFE: Lip Funneler & [35], [36], [37], [38]               \\
\hline
NEFE: Lip Pressor & [39], [40]                                 & NEFE: Lip Pucker & [41], [42]                          \\
\hline
NEFE: Lip Suck   & [45], [46], [47], [48]                      & NEFE: Lips Toward & [51]                                \\
\hline
NEFE: Mouth Stretch & [54], [55]                              & NEFE: Upper Lip Raiser & [62], [63]                 \\
\hline
\end{tabularx}
\label{tab:feature_interpretation_FE}
\end{table*}

\subsubsection{Details on the \taxonomy{}} \label{app:taxonomy_details}

In this section, we discuss further details regarding how we extracted and organized attributes and categories from the CCPA definition of personal information and included them in our taxonomy, as discussed in Section~\ref{subsec:taxonomy_legal}. Some of the categories within the CCPA definition include references to other CCPA definitions or legal texts, such as sensitive personal information, biometric information, personally identifiable information defined in the Family Educational Rights and Privacy Act (FERPA)\footnote{20 U.S.C. Sec. 1232g; 34 C.F.R. Part 99}, and personal information described in subdivision (e) of Section 1798.80. For such cases, we extracted all the attributes from those separate definitions and either moved them to other more specific categories due to contextual similarities (\eg{} sensitive personal information in the CCPA includes precise geolocation, which we moved to the existing geolocation category) or removed them because the attribute already existed in another category (\eg{} name and address appear in both the FERPA definition and the identifiers category, so we kept them only in the identifiers category as it is more specific). For all other categories in the CCPA definition of personal information that did not reference other definitions, we included %
the categories and their attributes as stated in the legal text.

\begin{table*}[ht!]
  \centering
  \tiny
  \caption{\textbf{Feature Interpretations of Hand Joint (HJ) Sensor Group.} Based on the list of 26 joints in the \texttt{handData} data structure per OpenXR convention~\cite{openxrhandtracking}.}
  \begin{tabularx}{\textwidth}{|X|X|X|X|}
    \hline
    \textbf{Feature Interpretation} & \textbf{Joint No. \& Type} & \textbf{Feature Interpretation} & \textbf{Joint No. \& Type} \\
    \hline
    Palm (rotation/position)               & [1] rotation/position     & Wrist (rotation/position)              & [2] rotation/position     \\
    \hline
    Thumb Metacarpal (rotation/position)   & [3] rotation/position     & Thumb Proximal (rotation/position)     & [4] rotation/position     \\
    \hline
    Thumb Distal (rotation/position)       & [5] rotation/position     & Thumb Tip (rotation/position)          & [6] rotation/position     \\
    \hline
    Index Metacarpal (rotation/position)   & [7] rotation/position     & Index Proximal (rotation/position)     & [8] rotation/position     \\
    \hline
    Index Intermediate (rotation/position) & [9] rotation/position     & Index Distal (rotation/position)       & [10] rotation/position    \\
    \hline
    Index Tip (rotation/position)          & [11] rotation/position    & Middle Metacarpal (rotation/position)  & [12] rotation/position    \\
    \hline
    Middle Proximal (rotation/position)    & [13] rotation/position    & Middle Intermediate (rotation/position)& [14] rotation/position    \\
    \hline
    Middle Distal (rotation/position)      & [15] rotation/position    & Middle Tip (rotation/position)         & [16] rotation/position    \\
    \hline
    Ring Metacarpal (rotation/position)    & [17] rotation/position    & Ring Proximal (rotation/position)      & [18] rotation/position    \\
    \hline
    Ring Intermediate (rotation/position)  & [19] rotation/position    & Ring Distal (rotation/position)        & [20] rotation/position    \\
    \hline
    Ring Tip (rotation/position)           & [21] rotation/position    & Little Metacarpal (rotation/position)  & [22] rotation/position    \\
    \hline
    Little Proximal (rotation/position)    & [23] rotation/position    & Little Intermediate (rotation/position)& [24] rotation/position    \\
    \hline
    Little Distal (rotation/position)      & [25] rotation/position    & Little Tip (rotation/position)         & [26] rotation/position    \\
    \hline
  \end{tabularx}
  \label{tab:feature_interpretation_hj}
\end{table*}

\begin{table*}[!t]
\centering
\tiny 
\renewcommand{\arraystretch}{0.8}
\caption{\textbf{User Profiling Using FE across 7 App Groups.} The color code, designed to be color-blind friendly~\cite{colorBlindFriendly}, represents four risk levels based on F1: High/Very High Risk (F1=80-100\%,
purple), Moderate-High Risk (F1=70-80\%, orange), Moderate Risk (F1=50-70\%, light-blue), and Low Risk ($F1<50\%$, gray) as per~\ref{subsec:evaluation_metrics}. Attribute superscripts indicate associated threats according to our threat scenarios (see Section~\ref{subsubsec:ThreatModel_ThreatScenarios}) and taxonomy (see Section~\ref{subsec:taxonomy_VR_UserProfiling}).
}
\label{tab:f1score-fe}
\resizebox{\textwidth}{!}{%
\begin{tabular}{|l|l|lllllll|}
\hline
\multicolumn{1}{|c|}{\multirow{2}{*}{\textbf{Attribute Groups}}} &
  \multicolumn{1}{c|}{\multirow{2}{*}{\textbf{Attributes}}} &
  \multicolumn{7}{c|}{\textbf{App Groups}} \\ \cline{3-9}
\multicolumn{1}{|c|}{} &
  \multicolumn{1}{c|}{} &
  \multicolumn{1}{c|}{Social} &
  \multicolumn{1}{c|}{Flight} &
  \multicolumn{1}{c|}{Shoot} &
  \multicolumn{1}{c|}{Rhythm} &
  \multicolumn{1}{c|}{IN} &
  \multicolumn{1}{c|}{KW} &
  \multicolumn{1}{c|}{Archery} \\ \hline
\multirow{5}{*}{Demographics} &
  Gender$^{\text{\text{1,2,4,5,6}}}$ &
  \multicolumn{1}{l|}{\cellcolor{teal!20}68} &
  \multicolumn{1}{l|}{\cellcolor{orange!50}78} &
  \multicolumn{1}{l|}{\cellcolor{orange!50}70} &
  \multicolumn{1}{l|}{\cellcolor{orange!50}70} &
  \multicolumn{1}{l|}{\cellcolor{orange!50}72} &
  \multicolumn{1}{l|}{\cellcolor{teal!20}65} &
  \cellcolor{orange!50}71 \\ \cline{2-9}
 &
 Age$^{\text{\text{1,2,3,4,5,6,\ferpa}}}$ &
  \multicolumn{1}{l|}{\cellcolor{purple!50}88} &
  \multicolumn{1}{l|}{\cellcolor{teal!20}65} &
  \multicolumn{1}{l|}{\cellcolor{purple!50}80} &
  \multicolumn{1}{l|}{\cellcolor{orange!50}71} &
  \multicolumn{1}{l|}{\cellcolor{orange!50}78} &
  \multicolumn{1}{l|}{\cellcolor{orange!50}73} &
  \cellcolor{orange!50}75 \\ \cline{2-9}
 &
  Ethnicity$^{\text{\text{4,5,6,\sens}}}$ &
  \multicolumn{1}{l|}{\cellcolor{purple!50}80} &
  \multicolumn{1}{l|}{\cellcolor{orange!50}78} &
  \multicolumn{1}{l|}{\cellcolor{purple!50}86} &
  \multicolumn{1}{l|}{\cellcolor{purple!50}82} &
  \multicolumn{1}{l|}{\cellcolor{purple!50}85} &
  \multicolumn{1}{l|}{\cellcolor{purple!50}85} &
  \cellcolor{purple!50}80 \\ \cline{2-9}
 &
  Marital status$^{\text{\text{1,2,5,6}}}$ &
  \multicolumn{1}{l|}{\cellcolor{gray!45}49} &
  \multicolumn{1}{l|}{\cellcolor{gray!45}47} &
  \multicolumn{1}{l|}{\cellcolor{gray!45}43} &
  \multicolumn{1}{l|}{\cellcolor{teal!20}52} &
  \multicolumn{1}{l|}{\cellcolor{teal!20}55} &
  \multicolumn{1}{l|}{\cellcolor{teal!20}55} &
  \cellcolor{gray!45}40  \\ \hline

\multirow{6}{*}{Anthropometrics} &
 Height$^{\text{\text{3,5,6}}}$ &
  \multicolumn{1}{l|}{\cellcolor{teal!20}60} &
  \multicolumn{1}{l|}{\cellcolor{gray!45}48} &
  \multicolumn{1}{l|}{\cellcolor{gray!45}45} &
  \multicolumn{1}{l|}{\cellcolor{teal!20}57} &
  \multicolumn{1}{l|}{\cellcolor{gray!45}32} &
  \multicolumn{1}{l|}{\cellcolor{gray!45}30} &
  \cellcolor{teal!20}68 \\ \cline{2-9}
 &
  Reaction Time$^{\text{\text{3,5,6}}}$ &
  \multicolumn{1}{l|}{\cellcolor{orange!50}70} &
  \multicolumn{1}{l|}{\cellcolor{teal!20}60} &
  \multicolumn{1}{l|}{\cellcolor{orange!50}70} &
  \multicolumn{1}{l|}{\cellcolor{purple!50}90} &
  \multicolumn{1}{l|}{\cellcolor{purple!50}80} &
  \multicolumn{1}{l|}{\cellcolor{purple!50}80} &
  \cellcolor{purple!50}80
   \\ \cline{2-9}
 &
  Face Length$^{\text{\text{3,5,6}}}$ &
  \multicolumn{1}{l|}{\cellcolor{teal!20}65} &
  \multicolumn{1}{l|}{\cellcolor{teal!20}56} &
  \multicolumn{1}{l|}{\cellcolor{orange!50}70} &
  \multicolumn{1}{l|}{\cellcolor{orange!50}75} &
  \multicolumn{1}{l|}{\cellcolor{purple!50}80} &
  \multicolumn{1}{l|}{\cellcolor{orange!50}71} &
  \cellcolor{orange!50}75 \\ \cline{2-9}
  &
    Arm Length$^{\text{\text{3,5,6}}}$ &
    \multicolumn{1}{l|}{\cellcolor{teal!20}52} &
    \multicolumn{1}{l|}{\cellcolor{teal!20}54} &
    \multicolumn{1}{l|}{\cellcolor{teal!20}52} &
    \multicolumn{1}{l|}{\cellcolor{teal!20}50} &
    \multicolumn{1}{l|}{\cellcolor{gray!45}37} &
    \multicolumn{1}{l|}{\cellcolor{gray!45}46} &
    \cellcolor{teal!20}52 \\ \cline{2-9}
 &
  Weight$^{\text{\text{3,4,6}}}$ &
  \multicolumn{1}{l|}{\cellcolor{gray!45}39} &
  \multicolumn{1}{l|}{\cellcolor{gray!45}46} &
  \multicolumn{1}{l|}{\cellcolor{gray!45}45} &
  \multicolumn{1}{l|}{\cellcolor{gray!45}30} &
  \multicolumn{1}{l|}{\cellcolor{gray!45}45} &
  \multicolumn{1}{l|}{\cellcolor{gray!45}40} &
  \cellcolor{gray!45}35 \\ \cline{2-9}
 &
  BMI$^{\text{\text{3,6}}}$ &
  \multicolumn{1}{l|}{\cellcolor{teal!20}68} &
  \multicolumn{1}{l|}{\cellcolor{teal!20}65} &
  \multicolumn{1}{l|}{\cellcolor{orange!50}70} &
  \multicolumn{1}{l|}{\cellcolor{teal!20}66} &
  \multicolumn{1}{l|}{\cellcolor{orange!50}70} &
  \multicolumn{1}{l|}{\cellcolor{orange!50}79} &
  \cellcolor{teal!20}65 \\ \hline
\multirow{6}{*}{Health} &
  Close / Distance vision and lenses$^{\text{\text{1,5,6}}}$ &
  \multicolumn{1}{l|}{\cellcolor{teal!20}64} &
  \multicolumn{1}{l|}{\cellcolor{teal!20}59} &
  \multicolumn{1}{l|}{\cellcolor{teal!20}54} &
  \multicolumn{1}{l|}{\cellcolor{teal!20}63} &
  \multicolumn{1}{l|}{\cellcolor{teal!20}68} &
  \multicolumn{1}{l|}{\cellcolor{teal!20}69} &
  \cellcolor{teal!20}59
   \\ \cline{2-9}
 &
  Physical fitness$^{\text{\text{1,4,5,6}}}$ &
  \multicolumn{1}{l|}{\cellcolor{teal!20}55} &
  \multicolumn{1}{l|}{\cellcolor{teal!20}63} &
  \multicolumn{1}{l|}{\cellcolor{teal!20}65} &
  \multicolumn{1}{l|}{\cellcolor{gray!45}45} &
  \multicolumn{1}{l|}{\cellcolor{teal!20}52} &
  \multicolumn{1}{l|}{\cellcolor{teal!20}61} &
  \cellcolor{teal!20}65 \\ \cline{2-9}
 &
  Anxiety$^{\text{4,7,8}}$ &
  \multicolumn{1}{l|}{\cellcolor{orange!50}78} &
  \multicolumn{1}{l|}{\cellcolor{orange!50}71} &
  \multicolumn{1}{l|}{\cellcolor{orange!50}70} &
  \multicolumn{1}{l|}{\cellcolor{purple!50}82} &
  \multicolumn{1}{l|}{\cellcolor{orange!50}78} &
  \multicolumn{1}{l|}{\cellcolor{orange!50}70} &
  \cellcolor{teal!20}65 \\ \cline{2-9}
 &
  Stress$^{\text{4,6}}$ &
  \multicolumn{1}{l|}{\cellcolor{purple!50}85} &
  \multicolumn{1}{l|}{\cellcolor{purple!50}85} &
  \multicolumn{1}{l|}{\cellcolor{purple!50}90} &
  \multicolumn{1}{l|}{\cellcolor{purple!50}82} &
  \multicolumn{1}{l|}{\cellcolor{purple!50}90} &
  \multicolumn{1}{l|}{\cellcolor{purple!50}94} &
  \cellcolor{purple!50}90 \\ \cline{2-9}
 &
  Height phobia$^{\text{4,6}}$  &
  \multicolumn{1}{l|}{\cellcolor{teal!20}61.5} &
  \multicolumn{1}{l|}{\cellcolor{teal!20}65} &
  \multicolumn{1}{l|}{\cellcolor{orange!50}73} &
  \multicolumn{1}{l|}{\cellcolor{teal!20}62} &
  \multicolumn{1}{l|}{\cellcolor{teal!20}59} &
  \multicolumn{1}{l|}{\cellcolor{teal!20}57} &
  \cellcolor{teal!20}55 \\ \cline{2-9}
 &
  Motion sickness$^{\text{4,6}}$ &
  \multicolumn{1}{l|}{\cellcolor{gray!45}45} &
  \multicolumn{1}{l|}{\cellcolor{teal!20}50} &
  \multicolumn{1}{l|}{\cellcolor{gray!45}47} &
  \multicolumn{1}{l|}{\cellcolor{gray!45}45} &
  \multicolumn{1}{l|}{\cellcolor{gray!45}45} &
  \multicolumn{1}{l|}{\cellcolor{teal!20}55} &
  \cellcolor{gray!45}40 \\ \hline
\multirow{12}{*}{User Interests \& Behaviors} & 

Problem Solving Abilities$^{\text{6}}$ &  
  \multicolumn{1}{l|}{\cellcolor{gray!45}48} &
  \multicolumn{1}{l|}{\cellcolor{gray!45}48} &
  \multicolumn{1}{l|}{\cellcolor{gray!45}42} &
  \multicolumn{1}{l|}{\cellcolor{gray!45}40} &
  \multicolumn{1}{l|}{\cellcolor{gray!45}49} &
  \multicolumn{1}{l|}{\cellcolor{gray!45}44} &
  \cellcolor{teal!20}53    \\ \cline{2-9}  &    
Alcohol consumption$^{\text{1,5,6}}$ &  
\multicolumn{1}{l|}{\cellcolor{teal!20}68} &
  \multicolumn{1}{l|}{\cellcolor{teal!20}69} &
  \multicolumn{1}{l|}{\cellcolor{teal!20}67} &
  \multicolumn{1}{l|}{\cellcolor{orange!50}70} &
  \multicolumn{1}{l|}{\cellcolor{orange!50}78} &
  \multicolumn{1}{l|}{\cellcolor{teal!20}67} &
  \cellcolor{teal!20}62    \\ \cline{2-9} &    
VR Experience$^{\text{5,6}}$ &   
  \multicolumn{1}{l|}{\cellcolor{teal!20}67} &
  \multicolumn{1}{l|}{\cellcolor{teal!20}65} &
  \multicolumn{1}{l|}{\cellcolor{orange!50}70} &
  \multicolumn{1}{l|}{\cellcolor{orange!50}73} &
  \multicolumn{1}{l|}{\cellcolor{teal!20}65} &
  \multicolumn{1}{l|}{\cellcolor{orange!50}76} &
  \cellcolor{teal!20}66    \\ \cline{2-9}   &   
Activity preference$^{\text{6}}$ &   
  \multicolumn{1}{l|}{\cellcolor{teal!20}53} &
  \multicolumn{1}{l|}{\cellcolor{gray!45}48} &
  \multicolumn{1}{l|}{\cellcolor{teal!20}63} &
  \multicolumn{1}{l|}{\cellcolor{gray!45}43} &
  \multicolumn{1}{l|}{\cellcolor{teal!20}50} &
  \multicolumn{1}{l|}{\cellcolor{gray!45}32} &
  \cellcolor{teal!20}58   \\ \cline{2-9}        &    
Shooting Experiences$^{\text{1,4,6}}$ &     \multicolumn{1}{l|}{\cellcolor{teal!20}66} &
  \multicolumn{1}{l|}{\cellcolor{orange!50}74} &
  \multicolumn{1}{l|}{\cellcolor{orange!50}76} &
  \multicolumn{1}{l|}{\cellcolor{teal!20}69} &
  \multicolumn{1}{l|}{\cellcolor{orange!50}73} &
  \multicolumn{1}{l|}{\cellcolor{orange!50}79} &
  \cellcolor{orange!50}75   \\ \cline{2-9}  &
  
  Caffeinated item consumption$^{\text{1,5,6}}$ &
  \multicolumn{1}{l|}{\cellcolor{orange!50}75} &
  \multicolumn{1}{l|}{\cellcolor{teal!20}66} &
  \multicolumn{1}{l|}{\cellcolor{teal!20}64} &
  \multicolumn{1}{l|}{\cellcolor{orange!50}77} &
  \multicolumn{1}{l|}{\cellcolor{orange!50}75} &
  \multicolumn{1}{l|}{\cellcolor{orange!50}79} &
  \cellcolor{teal!20}69
   \\ \cline{2-9}
 &
  Concentration$^{\text{3,6}}$ &
  \multicolumn{1}{l|}{\cellcolor{gray!45}46} &
  \multicolumn{1}{l|}{\cellcolor{gray!45}45} &
  \multicolumn{1}{l|}{\cellcolor{teal!20}55} &
  \multicolumn{1}{l|}{\cellcolor{gray!45}40} &
  \multicolumn{1}{l|}{\cellcolor{gray!45}45} &
  \multicolumn{1}{l|}{\cellcolor{teal!20}55} &
  \cellcolor{gray!45}45 \\ \cline{2-9}
 &
  Violence tolerance$^{\text{1,4,6}}$ &
  \multicolumn{1}{l|}{\cellcolor{teal!20}68} &
  \multicolumn{1}{l|}{\cellcolor{orange!50}70} &
  \multicolumn{1}{l|}{\cellcolor{orange!50}75} &
  \multicolumn{1}{l|}{\cellcolor{teal!20}57} &
  \multicolumn{1}{l|}{\cellcolor{teal!20}67} &
  \multicolumn{1}{l|}{\cellcolor{orange!50}72} &
  \cellcolor{teal!20}62 \\ \cline{2-9}
 &
  Introvert/Extrovert$^{\text{\text{6}}}$ &
  \multicolumn{1}{l|}{\cellcolor{orange!50}70} &
  \multicolumn{1}{l|}{\cellcolor{teal!20}60} &
  \multicolumn{1}{l|}{\cellcolor{teal!20}62} &
  \multicolumn{1}{l|}{\cellcolor{teal!20}58} &
  \multicolumn{1}{l|}{\cellcolor{teal!20}65} &
  \multicolumn{1}{l|}{\cellcolor{teal!20}60} &
  \cellcolor{teal!20}56 \\ \cline{2-9}
 &
  Organized/Unorganized$^{\text{\text{6}}}$ &
  \multicolumn{1}{l|}{\cellcolor{purple!50}82} &
  \multicolumn{1}{l|}{\cellcolor{purple!50}90} &
  \multicolumn{1}{l|}{\cellcolor{purple!50}81} &
  \multicolumn{1}{l|}{\cellcolor{purple!50}97} &
  \multicolumn{1}{l|}{\cellcolor{purple!50}92} &
  \multicolumn{1}{l|}{\cellcolor{orange!50}76} &
  \cellcolor{purple!50}80 \\ \cline{2-9}
 &
  Social media usage$^{\text{4,6}}$ &
  \multicolumn{1}{l|}{\cellcolor{gray!45}43} &
  \multicolumn{1}{l|}{\cellcolor{gray!45}49} &
  \multicolumn{1}{l|}{\cellcolor{orange!50}71} &
  \multicolumn{1}{l|}{\cellcolor{teal!20}56} &
  \multicolumn{1}{l|}{\cellcolor{gray!45}49} &
  \multicolumn{1}{l|}{\cellcolor{teal!20}62} &
  \cellcolor{teal!20}61 \\ \cline{2-9}
 &
     Openness$^{\text{6}}$ &
  \multicolumn{1}{l|}{\cellcolor{purple!50}80} &
  \multicolumn{1}{l|}{\cellcolor{orange!50}73} &
  \multicolumn{1}{l|}{\cellcolor{orange!50}74} &
  \multicolumn{1}{l|}{\cellcolor{purple!50}84} &
  \multicolumn{1}{l|}{\cellcolor{orange!50}73} &
  \multicolumn{1}{l|}{\cellcolor{purple!50}81} &
  \cellcolor{purple!50}80 \\ \cline{2-9} &
  
  Emotional stability$^{\text{6}}$ &
  \multicolumn{1}{l|}{\cellcolor{teal!20}68} &
  \multicolumn{1}{l|}{\cellcolor{orange!50}71} &
  \multicolumn{1}{l|}{\cellcolor{purple!50}86} &
  \multicolumn{1}{l|}{\cellcolor{teal!20}65} &
  \multicolumn{1}{l|}{\cellcolor{orange!50}76} &
  \multicolumn{1}{l|}{\cellcolor{teal!20}68} &
  \cellcolor{teal!20}69
   \\ \hline
\end{tabular}%
}
\end{table*}

\subsection{Details on Experimental Setup}\label{app:experimental_setup}
We elaborate on our experimental setup discussed in Section \ref{sec: experimental_setup}, including sensor data collection (\ref{app:dataset}), survey protocol (\ref{app:survey}), and attributes and their statistics (\ref{app:Attributes_and_statistics}) here.
\begin{table*}[!t]
\centering
\renewcommand{\arraystretch}{0.78}
\tiny 
\caption{\textbf{User Profiling Using EG Sensor Data Across 7 App Groups.} (Color code representation similar as Table \ref{tab:f1score-fe}).
}
\label{tab:f1score-eg}
\resizebox{\textwidth}{!}{%
\begin{tabular}{|l|l|lllllll|}
\hline
\multicolumn{1}{|c|}{\multirow{2}{*}{\textbf{Attribute Groups}}} &
  \multicolumn{1}{c|}{\multirow{2}{*}{\textbf{Attributes}}} &
  \multicolumn{7}{c|}{\textbf{App Groups}} \\ \cline{3-9}
\multicolumn{1}{|c|}{} &
  \multicolumn{1}{c|}{} &
  \multicolumn{1}{c|}{Social} &
  \multicolumn{1}{c|}{Flight} &
  \multicolumn{1}{c|}{Shoot} &
  \multicolumn{1}{c|}{Rhythm} &
  \multicolumn{1}{c|}{IN} &
  \multicolumn{1}{c|}{KW} &
  \multicolumn{1}{c|}{Archery} \\ \hline
\multirow{5}{*}{Demographics} &
  Gender$^{\text{\text{1,2,4,5,6}}}$ &
  \multicolumn{1}{l|}{\cellcolor{teal!20}63} &
  \multicolumn{1}{l|}{\cellcolor{teal!20}62} &
  \multicolumn{1}{l|}{\cellcolor{teal!20}62} &
  \multicolumn{1}{l|}{\cellcolor{teal!20}66} &
  \multicolumn{1}{l|}{\cellcolor{teal!20}55} &
  \multicolumn{1}{l|}{\cellcolor{teal!20}61} &
  \cellcolor{teal!20}58 \\ \cline{2-9}
 &
 Age$^{\text{\text{1,2,3,4,5,6,\ferpa}}}$&
  \multicolumn{1}{l|}{\cellcolor{teal!20}62} &
  \multicolumn{1}{l|}{\cellcolor{teal!20}52} &
  \multicolumn{1}{l|}{\cellcolor{gray!45}49} &
  \multicolumn{1}{l|}{\cellcolor{teal!20}58} &
  \multicolumn{1}{l|}{\cellcolor{teal!20}63} &
  \multicolumn{1}{l|}{\cellcolor{gray!45}39} &
  \cellcolor{teal!20}62 \\ \cline{2-9}
 &
  Ethnicity$^{\text{\text{4,5,6,\sens}}}$ &
  \multicolumn{1}{l|}{\cellcolor{gray!45}42} &
  \multicolumn{1}{l|}{\cellcolor{teal!20}53} &
  \multicolumn{1}{l|}{\cellcolor{teal!20}69} &
  \multicolumn{1}{l|}{\cellcolor{teal!20}58} &
  \multicolumn{1}{l|}{\cellcolor{teal!20}62} &
  \multicolumn{1}{l|}{\cellcolor{teal!20}61} &
  \cellcolor{gray!45}47 \\ \cline{2-9}
 &
  Marital status$^{\text{\text{1,2,5,6}}}$ &
  \multicolumn{1}{l|}{\cellcolor{teal!20}58} &
  \multicolumn{1}{l|}{\cellcolor{gray!45}42} &
  \multicolumn{1}{l|}{\cellcolor{teal!20}58} &
  \multicolumn{1}{l|}{\cellcolor{gray!45}46} &
  \multicolumn{1}{l|}{\cellcolor{gray!45}48} &
  \multicolumn{1}{l|}{\cellcolor{teal!20}59} &
  \cellcolor{teal!20}60 
   \\  \hline
\multirow{6}{*}{Anthropometrics} &
 Height$^{\text{\text{3,5,6}}}$ &
  \multicolumn{1}{l|}{\cellcolor{teal!20}56} &
  \multicolumn{1}{l|}{\cellcolor{teal!20}53} &
  \multicolumn{1}{l|}{\cellcolor{teal!20}56} &
  \multicolumn{1}{l|}{\cellcolor{teal!20}59} &
  \multicolumn{1}{l|}{\cellcolor{teal!20}58} &
  \multicolumn{1}{l|}{\cellcolor{teal!20}56} &
  \cellcolor{teal!20}56 \\ \cline{2-9}
 &
  Reaction Time$^{\text{\text{3,5,6}}}$ &
  \multicolumn{1}{l|}{\cellcolor{gray!45}45} &
  \multicolumn{1}{l|}{\cellcolor{teal!20}50} &
  \multicolumn{1}{l|}{\cellcolor{gray!45}40} &
  \multicolumn{1}{l|}{\cellcolor{teal!20}56} &
  \multicolumn{1}{l|}{\cellcolor{gray!45}44} &
  \multicolumn{1}{l|}{\cellcolor{teal!20}57} &
  \cellcolor{teal!20}52 \\ \cline{2-9}
 &
  Face Length$^{\text{\text{3,5,6}}}$ &
  \multicolumn{1}{l|}{\cellcolor{gray!45}39} &
  \multicolumn{1}{l|}{\cellcolor{gray!45}31} &
  \multicolumn{1}{l|}{\cellcolor{gray!45}49} &
  \multicolumn{1}{l|}{\cellcolor{gray!45}16} &
  \multicolumn{1}{l|}{\cellcolor{gray!45}45} &
  \multicolumn{1}{l|}{\cellcolor{gray!45}38} &
  \cellcolor{gray!45}49 \\ \cline{2-9}
  &
    Arm Length$^{\text{\text{3,5,6}}}$ &
\multicolumn{1}{l|}{\cellcolor{teal!20}53} &
\multicolumn{1}{l|}{\cellcolor{teal!20}58} &
\multicolumn{1}{l|}{\cellcolor{teal!20}54} &
\multicolumn{1}{l|}{\cellcolor{gray!45}40} &
\multicolumn{1}{l|}{\cellcolor{gray!45}50} &
\multicolumn{1}{l|}{\cellcolor{teal!20}59} &
\cellcolor{teal!20}54 \\ \cline{2-9}
 &
  Weight$^{\text{\text{3,4,6}}}$&
  \multicolumn{1}{l|}{\cellcolor{teal!20}56} &
  \multicolumn{1}{l|}{\cellcolor{teal!20}53} &
  \multicolumn{1}{l|}{\cellcolor{teal!20}56} &
  \multicolumn{1}{l|}{\cellcolor{teal!20}51} &
  \multicolumn{1}{l|}{\cellcolor{teal!20}58} &
  \multicolumn{1}{l|}{\cellcolor{teal!20}56} &
  \cellcolor{teal!20}56 \\ \cline{2-9}
 &
  BMI$^{\text{\text{3,6}}}$ &
  \multicolumn{1}{l|}{\cellcolor{gray!45}49} &
  \multicolumn{1}{l|}{\cellcolor{teal!20}59} &
  \multicolumn{1}{l|}{\cellcolor{gray!45}47} &
  \multicolumn{1}{l|}{\cellcolor{teal!20}60} &
  \multicolumn{1}{l|}{\cellcolor{teal!20}52} &
  \multicolumn{1}{l|}{\cellcolor{gray!45}49} &
  \cellcolor{teal!20}55 \\ \hline
\multirow{6}{*}{Health} &
  Close / Distance vision and lenses$^{\text{1,5,6}}$ &
  \multicolumn{1}{l|}{\cellcolor{teal!20}53} &
  \multicolumn{1}{l|}{\cellcolor{teal!20}55} &
  \multicolumn{1}{l|}{\cellcolor{gray!45}48} &
  \multicolumn{1}{l|}{\cellcolor{gray!45}45} &
  \multicolumn{1}{l|}{\cellcolor{teal!20}68} &
  \multicolumn{1}{l|}{\cellcolor{teal!20}57} &
  \cellcolor{gray!45}48
   \\ \cline{2-9}
 &
  Physical fitness$^{\text{\text{1,4,5,6}}}$ &
  \multicolumn{1}{l|}{\cellcolor{teal!20}53} &
  \multicolumn{1}{l|}{\cellcolor{teal!20}57} &
  \multicolumn{1}{l|}{\cellcolor{teal!20}52} &
  \multicolumn{1}{l|}{\cellcolor{teal!20}67} &
  \multicolumn{1}{l|}{\cellcolor{teal!20}48} &
  \multicolumn{1}{l|}{\cellcolor{teal!20}58} &
  \cellcolor{gray!45}38 \\ \cline{2-9}
 &
  Anxiety$^{\text{4,7,8}}$ &
  \multicolumn{1}{l|}{\cellcolor{teal!20}57} &
  \multicolumn{1}{l|}{\cellcolor{teal!20}65} &
  \multicolumn{1}{l|}{\cellcolor{teal!20}51} &
  \multicolumn{1}{l|}{\cellcolor{teal!20}66} &
  \multicolumn{1}{l|}{\cellcolor{teal!20}60} &
  \multicolumn{1}{l|}{\cellcolor{teal!20}57} &
  \cellcolor{teal!20}54 \\ \cline{2-9}
 &
  Stress$^{\text{4,6}}$ &
  \multicolumn{1}{l|}{\cellcolor{gray!45}46} &
  \multicolumn{1}{l|}{\cellcolor{gray!45}48} &
  \multicolumn{1}{l|}{\cellcolor{teal!20}51} &
  \multicolumn{1}{l|}{\cellcolor{teal!20}64} &
  \multicolumn{1}{l|}{\cellcolor{teal!20}55} &
  \multicolumn{1}{l|}{\cellcolor{teal!20}69} &
  \cellcolor{teal!20}65 \\ \cline{2-9}
 &
  Height phobia$^{\text{4,6}}$  &
  \multicolumn{1}{l|}{\cellcolor{gray!45}49} &
  \multicolumn{1}{l|}{\cellcolor{teal!20}54} &
  \multicolumn{1}{l|}{\cellcolor{gray!45}37} &
  \multicolumn{1}{l|}{\cellcolor{teal!20}62} &
  \multicolumn{1}{l|}{\cellcolor{teal!20}55} &
  \multicolumn{1}{l|}{\cellcolor{teal!20}57} &
  \cellcolor{gray!45}49 \\ \cline{2-9}
 &
  Motion sickness$^{\text{4,6}}$ &
  \multicolumn{1}{l|}{\cellcolor{gray!45}47} &
  \multicolumn{1}{l|}{\cellcolor{gray!45}48} &
  \multicolumn{1}{l|}{\cellcolor{gray!45}45} &
  \multicolumn{1}{l|}{\cellcolor{gray!45}42} &
  \multicolumn{1}{l|}{\cellcolor{teal!20}54} &
  \multicolumn{1}{l|}{\cellcolor{teal!20}54} &
  \cellcolor{gray!45}50 \\ \hline
\multirow{12}{*}{User Interests \& Behaviors} &   
Problem Solving Abilities$^{\text{6}}$ &   
  \multicolumn{1}{l|}{\cellcolor{gray!45}42} &
  \multicolumn{1}{l|}{\cellcolor{gray!45}44} &
  \multicolumn{1}{l|}{\cellcolor{gray!45}41} &
  \multicolumn{1}{l|}{\cellcolor{teal!20}53} &
  \multicolumn{1}{l|}{\cellcolor{gray!45}48} &
  \multicolumn{1}{l|}{\cellcolor{gray!45}44} &
  \cellcolor{gray!45}43   \\ \cline{2-9}  &    
Alcohol consumption$^{\text{1,5,6}}$ &
\multicolumn{1}{l|}{\cellcolor{teal!20}52} &
  \multicolumn{1}{l|}{\cellcolor{teal!20}50} &
  \multicolumn{1}{l|}{\cellcolor{teal!20}52} &
  \multicolumn{1}{l|}{\cellcolor{teal!20}63} &
  \multicolumn{1}{l|}{\cellcolor{gray!45}45} &
  \multicolumn{1}{l|}{\cellcolor{gray!45}43} &
  \cellcolor{gray!45}40    \\ \cline{2-9} &    
VR Experience$^{\text{5,6}}$ &   \multicolumn{1}{l|}{\cellcolor{gray!45}38} &
  \multicolumn{1}{l|}{\cellcolor{teal!20}54} &
  \multicolumn{1}{l|}{\cellcolor{gray!45}48} &
  \multicolumn{1}{l|}{\cellcolor{teal!20}59} &
  \multicolumn{1}{l|}{\cellcolor{teal!20}49} &
  \multicolumn{1}{l|}{\cellcolor{teal!20}64} &
  \cellcolor{teal!20}52    \\ \cline{2-9}   &    
Activity preference$^{\text{6}}$ &   
\multicolumn{1}{l|}{\cellcolor{gray!45}44} &
  \multicolumn{1}{l|}{\cellcolor{teal!20}59} &
  \multicolumn{1}{l|}{\cellcolor{gray!45}47} &
  \multicolumn{1}{l|}{\cellcolor{teal!20}61} &
  \multicolumn{1}{l|}{\cellcolor{teal!20}55} &
  \multicolumn{1}{l|}{\cellcolor{gray!45}45} &
  \cellcolor{gray!45}33   \\ \cline{2-9}        &   
Shooting Experiences$^{\text{1,4,6}}$ &   
\multicolumn{1}{l|}{\cellcolor{teal!20}55} &
  \multicolumn{1}{l|}{\cellcolor{teal!20}65} &
  \multicolumn{1}{l|}{\cellcolor{teal!20}60} &
  \multicolumn{1}{l|}{\cellcolor{teal!20}57} &
  \multicolumn{1}{l|}{\cellcolor{gray!45}47} &
  \multicolumn{1}{l|}{\cellcolor{teal!20}57} &
  \cellcolor{teal!20}56   \\ \cline{2-9}  &
  Caffeinated item consumption$^{\text{1,5,6}}$ &
  \multicolumn{1}{l|}{\cellcolor{teal!20}59} &
  \multicolumn{1}{l|}{\cellcolor{gray!45}41} &
  \multicolumn{1}{l|}{\cellcolor{gray!45}39} &
  \multicolumn{1}{l|}{\cellcolor{gray!45}49} &
  \multicolumn{1}{l|}{\cellcolor{gray!45}45} &
  \multicolumn{1}{l|}{\cellcolor{teal!20}58} &
  \cellcolor{teal!20}52 
   \\ \cline{2-9}
 &
  Concentration$^{\text{3,6}}$ &
  \multicolumn{1}{l|}{\cellcolor{teal!20}66} &
  \multicolumn{1}{l|}{\cellcolor{gray!45}42} &
  \multicolumn{1}{l|}{\cellcolor{gray!45}42} &
  \multicolumn{1}{l|}{\cellcolor{gray!45}42} &
  \multicolumn{1}{l|}{\cellcolor{gray!45}46} &
  \multicolumn{1}{l|}{\cellcolor{gray!45}43} &
  \cellcolor{gray!45}43 \\ \cline{2-9}
 &
  Violence tolerance$^{\text{1,4,6}}$ &
  \multicolumn{1}{l|}{\cellcolor{teal!20}67} &
  \multicolumn{1}{l|}{\cellcolor{teal!20}56} &
  \multicolumn{1}{l|}{\cellcolor{teal!20}57} &
  \multicolumn{1}{l|}{\cellcolor{teal!20}60} &
  \multicolumn{1}{l|}{\cellcolor{teal!20}58} &
  \multicolumn{1}{l|}{\cellcolor{orange!50}71} &
  \cellcolor{teal!20}63 \\ \cline{2-9}
 &
  Introvert/Extrovert$^{\text{\text{6}}}$ &
  \multicolumn{1}{l|}{\cellcolor{teal!20}60} &
  \multicolumn{1}{l|}{\cellcolor{teal!20}53} &
  \multicolumn{1}{l|}{\cellcolor{gray!45}45} &
  \multicolumn{1}{l|}{\cellcolor{teal!20}53} &
  \multicolumn{1}{l|}{\cellcolor{orange!50}70} &
  \multicolumn{1}{l|}{\cellcolor{teal!20}57} &
  \cellcolor{teal!20}55 \\ \cline{2-9}
 &
  Organized/Unorganized$^{\text{\text{6}}}$ &
  \multicolumn{1}{l|}{\cellcolor{teal!20}52} &
  \multicolumn{1}{l|}{\cellcolor{gray!45}48} &
  \multicolumn{1}{l|}{\cellcolor{teal!20}61} &
  \multicolumn{1}{l|}{\cellcolor{teal!20}66} &
  \multicolumn{1}{l|}{\cellcolor{teal!20}50} &
  \multicolumn{1}{l|}{\cellcolor{teal!20}57} &
  \cellcolor{teal!20}55 \\ \cline{2-9}
 &
  Social media usage$^{\text{4,6}}$ &
  \multicolumn{1}{l|}{\cellcolor{gray!45}48} &
  \multicolumn{1}{l|}{\cellcolor{gray!45}46} &
  \multicolumn{1}{l|}{\cellcolor{gray!45}42} &
  \multicolumn{1}{l|}{\cellcolor{gray!45}46} &
  \multicolumn{1}{l|}{\cellcolor{gray!45}48} &
  \multicolumn{1}{l|}{\cellcolor{gray!45}48} &
  \cellcolor{gray!45}41 \\ \cline{2-9}
 &
     Openness$^{\text{6}}$ &
  \multicolumn{1}{l|}{\cellcolor{teal!20}62} &
  \multicolumn{1}{l|}{\cellcolor{teal!20}58} &
  \multicolumn{1}{l|}{\cellcolor{teal!20}60} &
  \multicolumn{1}{l|}{\cellcolor{teal!20}56} &
  \multicolumn{1}{l|}{\cellcolor{teal!20}52} &
  \multicolumn{1}{l|}{\cellcolor{gray!45}47} &
  \cellcolor{gray!45}36 \\ \cline{2-9} &
  
  Emotional stability$^{\text{6}}$ &
  \multicolumn{1}{l|}{\cellcolor{gray!45}26} &
  \multicolumn{1}{l|}{\cellcolor{gray!45}39} &
  \multicolumn{1}{l|}{\cellcolor{teal!20}64} &
  \multicolumn{1}{l|}{\cellcolor{teal!20}62} &
  \multicolumn{1}{l|}{\cellcolor{gray!45}25} &
  \multicolumn{1}{l|}{\cellcolor{teal!20}55} &
  \cellcolor{teal!20}62 
   \\ \hline
\end{tabular}%
}
\vspace{2mm}

\centering
\renewcommand{\arraystretch}{0.78}
\tiny 
\caption{\textbf{User Profiling Using HJ Sensor Data Across 7 App Groups.} (Color code representation similar as Table \ref{tab:f1score-fe}). %
}
\label{tab:f1score-hj}
\resizebox{\textwidth}{!}{%
\begin{tabular}{|l|l|lllllll|}
\hline
\multicolumn{1}{|c|}{\multirow{2}{*}{\textbf{Attribute Groups}}} &
  \multicolumn{1}{c|}{\multirow{2}{*}{\textbf{Attributes}}} &
  \multicolumn{7}{c|}{\textbf{App Groups}} \\ \cline{3-9}
\multicolumn{1}{|c|}{} &
  \multicolumn{1}{c|}{} &
  \multicolumn{1}{c|}{Social} &
  \multicolumn{1}{c|}{Flight} &
  \multicolumn{1}{c|}{Shoot} &
  \multicolumn{1}{c|}{Rhythm} &
  \multicolumn{1}{c|}{IN} &
  \multicolumn{1}{c|}{KW} &
  \multicolumn{1}{c|}{Archery} \\ \hline
\multirow{5}{*}{Demographics} &
  Gender$^{\text{\text{1,2,4,5,6}}}$ &
  \multicolumn{1}{l|}{\cellcolor{orange!50}73} &
  \multicolumn{1}{l|}{\cellcolor{orange!50}75} &
  \multicolumn{1}{l|}{\cellcolor{teal!20}61} &
  \multicolumn{1}{l|}{\cellcolor{orange!50}75} &
  \multicolumn{1}{l|}{\cellcolor{orange!50}75} &
  \multicolumn{1}{l|}{\cellcolor{purple!50}87} &
  \cellcolor{teal!20}67 \\ \cline{2-9}
 &
 Age$^{\text{\text{1,2,3,4,5,6,\ferpa}}}$&
  \multicolumn{1}{l|}{\cellcolor{teal!20}63} &
  \multicolumn{1}{l|}{\cellcolor{orange!50}73} &
  \multicolumn{1}{l|}{\cellcolor{purple!50}83} &
  \multicolumn{1}{l|}{\cellcolor{orange!50}73} &
  \multicolumn{1}{l|}{\cellcolor{orange!50}75} &
  \multicolumn{1}{l|}{\cellcolor{orange!50}75} &
  \cellcolor{orange!50}70 \\ \cline{2-9}
 &
  Ethnicity$^{\text{\text{4,5,6,\sens}}}$ &
  \multicolumn{1}{l|}{\cellcolor{orange!50}73} &
  \multicolumn{1}{l|}{\cellcolor{purple!50}80} &
  \multicolumn{1}{l|}{\cellcolor{orange!50}78} &
  \multicolumn{1}{l|}{\cellcolor{teal!20}61} &
  \multicolumn{1}{l|}{\cellcolor{orange!50}71} &
  \multicolumn{1}{l|}{\cellcolor{orange!50}78} &
  \cellcolor{orange!50}70 \\ \cline{2-9}
 &
  Marital status$^{\text{\text{1,2,5,6}}}$ &
  \multicolumn{1}{l|}{\cellcolor{orange!50}78} &
  \multicolumn{1}{l|}{\cellcolor{teal!20}67} &
  \multicolumn{1}{l|}{\cellcolor{purple!50}82} &
  \multicolumn{1}{l|}{\cellcolor{teal!20}52} &
  \multicolumn{1}{l|}{\cellcolor{teal!20}64} &
  \multicolumn{1}{l|}{\cellcolor{orange!50}79} &
  \cellcolor{teal!20}68 
   \\   \hline
\multirow{6}{*}{Anthropometrics} &
 Height$^{\text{\text{3,5,6}}}$ &
  \multicolumn{1}{l|}{\cellcolor{teal!20}60} &
  \multicolumn{1}{l|}{\cellcolor{teal!20}51} &
  \multicolumn{1}{l|}{\cellcolor{orange!50}76} &
  \multicolumn{1}{l|}{\cellcolor{teal!20}55} &
  \multicolumn{1}{l|}{\cellcolor{teal!20}55} &
  \multicolumn{1}{l|}{\cellcolor{teal!20}68} &
  \cellcolor{teal!20}55 \\ \cline{2-9}
 &
  Reaction Time$^{\text{\text{3,5,6}}}$ &
  \multicolumn{1}{l|}{\cellcolor{purple!50}90} &
  \multicolumn{1}{l|}{\cellcolor{purple!50}98} &
  \multicolumn{1}{l|}{\cellcolor{purple!50}91} &
  \multicolumn{1}{l|}{\cellcolor{purple!50}89} &
  \multicolumn{1}{l|}{\cellcolor{purple!50}93} &
  \multicolumn{1}{l|}{\cellcolor{purple!50}86} &
  \cellcolor{purple!50}94
   \\ \cline{2-9}
 &
  Face Length$^{\text{\text{3,5,6}}}$ &
\multicolumn{1}{l|}{\cellcolor{teal!20}57} &
\multicolumn{1}{l|}{\cellcolor{teal!20}55} &
\multicolumn{1}{l|}{\cellcolor{teal!20}57} &
\multicolumn{1}{l|}{\cellcolor{gray!45}44} &
\multicolumn{1}{l|}{\cellcolor{gray!45}20} &
\multicolumn{1}{l|}{\cellcolor{gray!45}36} &
\cellcolor{gray!45}38 \\ \cline{2-9}
  &
    Arm Length$^{\text{\text{3,5,6}}}$ &
\multicolumn{1}{l|}{\cellcolor{gray!45}49} &
\multicolumn{1}{l|}{\cellcolor{gray!45}38} &
\multicolumn{1}{l|}{\cellcolor{teal!20}56} &
\multicolumn{1}{l|}{\cellcolor{orange!50}71} &
\multicolumn{1}{l|}{\cellcolor{teal!20}56} &
\multicolumn{1}{l|}{\cellcolor{teal!20}67} &
\cellcolor{gray!45}50 \\ \cline{2-9}
 &
  Weight$^{\text{\text{3,4,6}}}$&
  \multicolumn{1}{l|}{\cellcolor{teal!20}62} &
  \multicolumn{1}{l|}{\cellcolor{teal!20}58} &
  \multicolumn{1}{l|}{\cellcolor{teal!20}56} &
  \multicolumn{1}{l|}{\cellcolor{teal!20}55} &
  \multicolumn{1}{l|}{\cellcolor{teal!20}62} &
  \multicolumn{1}{l|}{\cellcolor{orange!50}74} &
  \cellcolor{teal!20}68 \\ \cline{2-9}
 &
  BMI$^{\text{\text{3,6}}}$ &
  \multicolumn{1}{l|}{\cellcolor{teal!20}63} &
  \multicolumn{1}{l|}{\cellcolor{orange!50}74} &
  \multicolumn{1}{l|}{\cellcolor{teal!20}67} &
  \multicolumn{1}{l|}{\cellcolor{teal!20}66} &
  \multicolumn{1}{l|}{\cellcolor{orange!50}77} &
  \multicolumn{1}{l|}{\cellcolor{teal!20}61} &
  \cellcolor{teal!20}56 \\ \hline
\multirow{6}{*}{Health} &
  Close / Distance vision and lenses$^{\text{1,5,6}}$ &
  \multicolumn{1}{l|}{\cellcolor{orange!50}71} &
  \multicolumn{1}{l|}{\cellcolor{teal!20}68} &
  \multicolumn{1}{l|}{\cellcolor{teal!20}65} &
  \multicolumn{1}{l|}{\cellcolor{teal!20}67} &
  \multicolumn{1}{l|}{\cellcolor{orange!50}72} &
  \multicolumn{1}{l|}{\cellcolor{orange!50}73} &
  \cellcolor{orange!50}76
   \\ \cline{2-9}
 &
  Physical fitness$^{\text{\text{1,4,5,6}}}$ &
  \multicolumn{1}{l|}{\cellcolor{orange!50}79} &
  \multicolumn{1}{l|}{\cellcolor{purple!50}82} &
  \multicolumn{1}{l|}{\cellcolor{orange!50}72} &
  \multicolumn{1}{l|}{\cellcolor{teal!20}63} &
  \multicolumn{1}{l|}{\cellcolor{orange!50}75} &
  \multicolumn{1}{l|}{\cellcolor{orange!50}77} &
  \cellcolor{orange!50}76 \\ \cline{2-9}
 &
  Anxiety$^{\text{4,6}}$ &
  \multicolumn{1}{l|}{\cellcolor{purple!50}80} &
  \multicolumn{1}{l|}{\cellcolor{purple!50}82} &
  \multicolumn{1}{l|}{\cellcolor{teal!20}54} &
  \multicolumn{1}{l|}{\cellcolor{orange!50}78} &
  \multicolumn{1}{l|}{\cellcolor{orange!50}77} &
  \multicolumn{1}{l|}{\cellcolor{orange!50}75} &
  \cellcolor{purple!50}81 \\ \cline{2-9}
 &
  Stress$^{\text{4,6}}$ &
  \multicolumn{1}{l|}{\cellcolor{purple!50}85} &
  \multicolumn{1}{l|}{\cellcolor{purple!50}90} &
  \multicolumn{1}{l|}{\cellcolor{orange!50}79} &
  \multicolumn{1}{l|}{\cellcolor{orange!50}75} &
  \multicolumn{1}{l|}{\cellcolor{purple!50}85} &
  \multicolumn{1}{l|}{\cellcolor{orange!50}72} &
  \cellcolor{purple!50}88 \\ \cline{2-9}
 &
  Height phobia$^{\text{4,6}}$  &
  \multicolumn{1}{l|}{\cellcolor{teal!20}68} &
  \multicolumn{1}{l|}{\cellcolor{purple!50}85} &
  \multicolumn{1}{l|}{\cellcolor{teal!20}65} &
  \multicolumn{1}{l|}{\cellcolor{teal!20}67} &
  \multicolumn{1}{l|}{\cellcolor{teal!20}65} &
  \multicolumn{1}{l|}{\cellcolor{orange!50}77} &
  \cellcolor{orange!50}70 \\ \cline{2-9}
 &
  Motion sickness$^{\text{4,6}}$ &
  \multicolumn{1}{l|}{\cellcolor{orange!50}78} &
  \multicolumn{1}{l|}{\cellcolor{purple!50}80} &
  \multicolumn{1}{l|}{\cellcolor{orange!50}73} &
  \multicolumn{1}{l|}{\cellcolor{teal!20}53} &
  \multicolumn{1}{l|}{\cellcolor{orange!50}70} &
  \multicolumn{1}{l|}{\cellcolor{purple!50}89} &
  \cellcolor{orange!50}75 \\ \hline
\multirow{12}{*}{User Interests \& Behaviors} &
  Problem Solving Abilities$^{\text{6}}$ &
  \multicolumn{1}{l|}{\cellcolor{teal!20}62} &
  \multicolumn{1}{l|}{\cellcolor{orange!50}76} &
  \multicolumn{1}{l|}{\cellcolor{orange!50}76} &
  \multicolumn{1}{l|}{\cellcolor{teal!20}59} &
  \multicolumn{1}{l|}{\cellcolor{orange!50}75} &
  \multicolumn{1}{l|}{\cellcolor{teal!20}69} &
  \cellcolor{teal!20}82
   \\ \cline{2-9}
 &
   Alcohol consumption$^{\text{1,5,6}}$ &
  \multicolumn{1}{l|}{\cellcolor{orange!50}63} &
  \multicolumn{1}{l|}{\cellcolor{orange!50}72} &
  \multicolumn{1}{l|}{\cellcolor{teal!20}63} &
  \multicolumn{1}{l|}{\cellcolor{teal!20}63} &
  \multicolumn{1}{l|}{\cellcolor{orange!50}72} &
  \multicolumn{1}{l|}{\cellcolor{teal!20}69} &
  \cellcolor{teal!20}65
   \\ \cline{2-9} &
   VR Experience$^{\text{5,6}}$ &
  \multicolumn{1}{l|}{\cellcolor{teal!20}66} &
  \multicolumn{1}{l|}{\cellcolor{purple!50}69} &
  \multicolumn{1}{l|}{\cellcolor{purple!50}82} &
  \multicolumn{1}{l|}{\cellcolor{teal!20}55} &
  \multicolumn{1}{l|}{\cellcolor{teal!20}62} &
  \multicolumn{1}{l|}{\cellcolor{orange!50}76} &
  \cellcolor{orange!50}81
   \\ \cline{2-9}
  &
   Activity preference$^{\text{6}}$ &
  \multicolumn{1}{l|}{\cellcolor{teal!20}66} &
  \multicolumn{1}{l|}{\cellcolor{orange!50}76} &
  \multicolumn{1}{l|}{\cellcolor{orange!50}76} &
  \multicolumn{1}{l|}{\cellcolor{teal!20}68} &
  \multicolumn{1}{l|}{\cellcolor{teal!20}70} &
  \multicolumn{1}{l|}{\cellcolor{teal!20}68} &
  \cellcolor{teal!20}63
   \\ \cline{2-9} 

     &
   Shooting Experiences$^{\text{1,4,6}}$ &
  \multicolumn{1}{l|}{\cellcolor{orange!50}76} &
  \multicolumn{1}{l|}{\cellcolor{orange!50}72} &
  \multicolumn{1}{l|}{\cellcolor{orange!50}77} &
  \multicolumn{1}{l|}{\cellcolor{teal!20}69} &
  \multicolumn{1}{l|}{\cellcolor{orange!50}74} &
  \multicolumn{1}{l|}{\cellcolor{purple!50}83} &
  \cellcolor{orange!50}71
   \\ \cline{2-9}
 &
 Caffeinated item consumption$^{\text{1,5,6}}$ &
  \multicolumn{1}{l|}{\cellcolor{teal!20}71} &
  \multicolumn{1}{l|}{\cellcolor{teal!20}71} &
  \multicolumn{1}{l|}{\cellcolor{purple!50}85} &
  \multicolumn{1}{l|}{\cellcolor{teal!20}61} &
  \multicolumn{1}{l|}{\cellcolor{orange!50}74} &
  \multicolumn{1}{l|}{\cellcolor{orange!50}73} &
  \cellcolor{teal!20}68 \\
\cline{2-9} &

  Concentration$^{\text{3,6}}$ &
  \multicolumn{1}{l|}{\cellcolor{orange!50}75} &
  \multicolumn{1}{l|}{\cellcolor{purple!50}80} &
  \multicolumn{1}{l|}{\cellcolor{teal!20}69} &
  \multicolumn{1}{l|}{\cellcolor{teal!20}59} &
  \multicolumn{1}{l|}{\cellcolor{orange!50}76} &
  \multicolumn{1}{l|}{\cellcolor{orange!50}71} &
  \cellcolor{orange!50}79 \\ \cline{2-9}
 &
  Violence tolerance$^{\text{1,4,6}}$ &
  \multicolumn{1}{l|}{\cellcolor{orange!50}70} &
  \multicolumn{1}{l|}{\cellcolor{purple!50}81} &
  \multicolumn{1}{l|}{\cellcolor{teal!20}55} &
  \multicolumn{1}{l|}{\cellcolor{orange!50}77} &
  \multicolumn{1}{l|}{\cellcolor{teal!20}66} &
  \multicolumn{1}{l|}{\cellcolor{orange!50}71} &
  \cellcolor{orange!50}78 \\ \cline{2-9}
 &
  Introvert/Extrovert$^{\text{\text{6}}}$ &
  \multicolumn{1}{l|}{\cellcolor{orange!50}72} &
  \multicolumn{1}{l|}{\cellcolor{teal!20}65} &
  \multicolumn{1}{l|}{\cellcolor{teal!20}52} &
  \multicolumn{1}{l|}{\cellcolor{teal!20}63} &
  \multicolumn{1}{l|}{\cellcolor{orange!50}74} &
  \multicolumn{1}{l|}{\cellcolor{orange!50}71} &
  \cellcolor{teal!20}65 \\ \cline{2-9}
 &
  Organized/Unorganized$^{\text{\text{6}}}$ &
  \multicolumn{1}{l|}{\cellcolor{purple!50}81} &
  \multicolumn{1}{l|}{\cellcolor{purple!50}86} &
  \multicolumn{1}{l|}{\cellcolor{purple!50}90} &
  \multicolumn{1}{l|}{\cellcolor{teal!20}67} &
  \multicolumn{1}{l|}{\cellcolor{purple!50}88} &
  \multicolumn{1}{l|}{\cellcolor{orange!50}79} &
  \cellcolor{purple!50}89 \\ \cline{2-9}
 &
  Social media usage$^{\text{4,6}}$ &
  \multicolumn{1}{l|}{\cellcolor{orange!50}72} &
  \multicolumn{1}{l|}{\cellcolor{teal!20}62} &
  \multicolumn{1}{l|}{\cellcolor{orange!50}78} &
  \multicolumn{1}{l|}{\cellcolor{teal!20}68} &
  \multicolumn{1}{l|}{\cellcolor{orange!50}78} &
  \multicolumn{1}{l|}{\cellcolor{orange!50}73} &
  \cellcolor{orange!50}78 \\ \cline{2-9}
 &
     Openness$^{\text{6}}$ &
  \multicolumn{1}{l|}{\cellcolor{orange!50}73} &
  \multicolumn{1}{l|}{\cellcolor{orange!50}72} &
  \multicolumn{1}{l|}{\cellcolor{teal!20}68} &
  \multicolumn{1}{l|}{\cellcolor{purple!50}84} &
  \multicolumn{1}{l|}{\cellcolor{orange!50}74} &
  \multicolumn{1}{l|}{\cellcolor{orange!50}70} &
  \cellcolor{teal!20}61 \\ \cline{2-9} &

  Emotional stability$^{\text{6}}$ &
  \multicolumn{1}{l|}{\cellcolor{orange!50}75} &
  \multicolumn{1}{l|}{\cellcolor{orange!50}77} &
  \multicolumn{1}{l|}{\cellcolor{teal!20}66} &
  \multicolumn{1}{l|}{\cellcolor{orange!50}73} &
  \multicolumn{1}{l|}{\cellcolor{orange!50}76} &
  \multicolumn{1}{l|}{\cellcolor{orange!50}70} &
  \cellcolor{orange!50}76
   \\ \hline
\end{tabular}%
}
\end{table*}

\begin{table*}[!t]
\centering
\renewcommand{\arraystretch}{0.78}
\tiny  
\caption{\textbf{User Profiling for Multi-Sensor (BM \& FE) Adversary Across 7 App Groups.} %
}
\label{tab:f1score-bm-fe}
\resizebox{\textwidth}{!}{%
\begin{tabular}{|l|l|lllllll|}
\hline
\multicolumn{1}{|c|}{\multirow{2}{*}{\textbf{Attribute Groups}}} &
  \multicolumn{1}{c|}{\multirow{2}{*}{\textbf{Attributes}}} &
  \multicolumn{7}{c|}{\textbf{App Groups}} \\ \cline{3-9}
\multicolumn{1}{|c|}{} &
  \multicolumn{1}{c|}{} &
  \multicolumn{1}{c|}{Social} &
  \multicolumn{1}{c|}{Flight} &
  \multicolumn{1}{c|}{Shoot} &
  \multicolumn{1}{c|}{Rhythm} &
  \multicolumn{1}{c|}{IN} &
  \multicolumn{1}{c|}{KW} &
  \multicolumn{1}{c|}{Archery} \\ \hline
\multirow{5}{*}{Demographics} &
  Gender$^{\text{1,2,4,5,6}}$ &
  \multicolumn{1}{l|}{\cellcolor{purple!50}90} &
  \multicolumn{1}{l|}{\cellcolor{orange!50}75} &
  \multicolumn{1}{l|}{\cellcolor{purple!50}91} &
  \multicolumn{1}{l|}{\cellcolor{purple!50}90} &
  \multicolumn{1}{l|}{\cellcolor{purple!50}80} &
  \multicolumn{1}{l|}{\cellcolor{purple!50}81} &
  \cellcolor{purple!50}86 \\ \cline{2-9}
 &
 Age$^{\text{1,2,3,4,5,6,\ferpa}}$&
  \multicolumn{1}{l|}{\cellcolor{orange!50}78} &
  \multicolumn{1}{l|}{\cellcolor{orange!50}71} &
  \multicolumn{1}{l|}{\cellcolor{purple!50}84} &
  \multicolumn{1}{l|}{\cellcolor{purple!50}80} &
  \multicolumn{1}{l|}{\cellcolor{purple!50}80} &
  \multicolumn{1}{l|}{\cellcolor{orange!50}74} &
  \cellcolor{orange!50}71 \\ \cline{2-9}
 &
  Ethnicity$^{\text{4,5,6,\sens}}$ &
  \multicolumn{1}{l|}{\cellcolor{purple!50}80} &
  \multicolumn{1}{l|}{\cellcolor{orange!50}75} &
  \multicolumn{1}{l|}{\cellcolor{purple!50}83} &
  \multicolumn{1}{l|}{\cellcolor{orange!50}72} &
  \multicolumn{1}{l|}{\cellcolor{purple!50}85} &
  \multicolumn{1}{l|}{\cellcolor{purple!50}80} &
  \cellcolor{orange!50}78 \\ \cline{2-9}
 &
  Marital status$^{\text{1,2,5,6}}$ &
  \multicolumn{1}{l|}{\cellcolor{teal!20}57} &
  \multicolumn{1}{l|}{\cellcolor{teal!20}54} &
  \multicolumn{1}{l|}{\cellcolor{orange!50}70} &
  \multicolumn{1}{l|}{\cellcolor{teal!20}63} &
  \multicolumn{1}{l|}{\cellcolor{teal!20}67} &
  \multicolumn{1}{l|}{\cellcolor{teal!20}65} &
  \cellcolor{teal!20}69  \\  \hline
  
\multirow{6}{*}{Anthropometrics} &
 Height$^{\text{3,5,6}}$ &
  \multicolumn{1}{l|}{\cellcolor{purple!50}80} &
  \multicolumn{1}{l|}{\cellcolor{gray!45}49} &
  \multicolumn{1}{l|}{\cellcolor{purple!50}90} &
  \multicolumn{1}{l|}{\cellcolor{purple!50}100} &
  \multicolumn{1}{l|}{\cellcolor{teal!20}52} &
  \multicolumn{1}{l|}{\cellcolor{orange!50}70} &
  \cellcolor{purple!50}90 \\ \cline{2-9}
 &
  Reaction Time$^{\text{3,5,6}}$ &
  \multicolumn{1}{l|}{\cellcolor{purple!50}90} &
  \multicolumn{1}{l|}{\cellcolor{purple!50}90} &
  \multicolumn{1}{l|}{\cellcolor{purple!50}89} &
  \multicolumn{1}{l|}{\cellcolor{purple!50}90} &
  \multicolumn{1}{l|}{\cellcolor{purple!50}89} &
  \multicolumn{1}{l|}{\cellcolor{purple!50}89} &
  \cellcolor{purple!50}90
   \\ \cline{2-9}
 &
  Face Length$^{\text{3,5,6}}$ &
  \multicolumn{1}{l|}{\cellcolor{teal!20}65} &
  \multicolumn{1}{l|}{\cellcolor{teal!20}56} &
  \multicolumn{1}{l|}{\cellcolor{orange!50}70} &
  \multicolumn{1}{l|}{\cellcolor{orange!50}75} &
  \multicolumn{1}{l|}{\cellcolor{purple!50}80} &
  \multicolumn{1}{l|}{\cellcolor{orange!50}71} &
  \cellcolor{orange!50}75 \\ \cline{2-9}
  &
    Arm Length$^{\text{3,5,6}}$ &
\multicolumn{1}{l|}{\cellcolor{teal!20}68} &
\multicolumn{1}{l|}{\cellcolor{gray!45}39} &
\multicolumn{1}{l|}{\cellcolor{teal!20}60} &
\multicolumn{1}{l|}{\cellcolor{teal!20}65} &
\multicolumn{1}{l|}{\cellcolor{teal!20}65} &
\multicolumn{1}{l|}{\cellcolor{teal!20}61} &
\cellcolor{orange!50}75 \\ \cline{2-9}
 &
    Weight$^{\text{3,4,6}}$&
  \multicolumn{1}{l|}{\cellcolor{orange!50}75} &
  \multicolumn{1}{l|}{\cellcolor{gray!45}48} &
  \multicolumn{1}{l|}{\cellcolor{gray!45}40} &
  \multicolumn{1}{l|}{\cellcolor{orange!50}72} &
  \multicolumn{1}{l|}{\cellcolor{orange!50}73} &
  \multicolumn{1}{l|}{\cellcolor{orange!50}70} &
  \cellcolor{orange!50}75 \\ \cline{2-9}
 &
  BMI$^{\text{3,6}}$ &
  \multicolumn{1}{l|}{\cellcolor{orange!50}72} &
  \multicolumn{1}{l|}{\cellcolor{orange!50}73} &
  \multicolumn{1}{l|}{\cellcolor{orange!50}78} &
  \multicolumn{1}{l|}{\cellcolor{teal!20}68} &
  \multicolumn{1}{l|}{\cellcolor{orange!50}70} &
  \multicolumn{1}{l|}{\cellcolor{orange!50}76} &
  \cellcolor{orange!50} 70 \\ \hline
\multirow{6}{*}{Health} &
  Close / Distance vision and lenses$^{\text{1,5,6}}$ &
  \multicolumn{1}{l|}{\cellcolor{teal!20}73} &
  \multicolumn{1}{l|}{\cellcolor{teal!20}67} &
  \multicolumn{1}{l|}{\cellcolor{teal!20}56} &
  \multicolumn{1}{l|}{\cellcolor{teal!20}69} &
  \multicolumn{1}{l|}{\cellcolor{teal!20}68} &
  \multicolumn{1}{l|}{\cellcolor{teal!20}67} &
  \cellcolor{teal!20}64
   \\ \cline{2-9}
 &
  Physical fitness$^{\text{1,4,5,6}}$ &
  \multicolumn{1}{l|}{\cellcolor{purple!50}89} &
  \multicolumn{1}{l|}{\cellcolor{orange!50}76} &
  \multicolumn{1}{l|}{\cellcolor{purple!50}87} &
  \multicolumn{1}{l|}{\cellcolor{purple!50}80} &
  \multicolumn{1}{l|}{\cellcolor{orange!50}78} &
  \multicolumn{1}{l|}{\cellcolor{purple!50}80} &
  \cellcolor{purple!50}95 \\ \cline{2-9}
 &
  Anxiety$^{\text{4,7,8}}$ &
  \multicolumn{1}{l|}{\cellcolor{purple!50}83} &
  \multicolumn{1}{l|}{\cellcolor{orange!50}75} &
  \multicolumn{1}{l|}{\cellcolor{teal!20}63} &
  \multicolumn{1}{l|}{\cellcolor{purple!50}80} &
  \multicolumn{1}{l|}{\cellcolor{teal!20}65} &
  \multicolumn{1}{l|}{\cellcolor{purple!50}80} &
  \cellcolor{teal!20}66 \\ \cline{2-9}
 &
  Stress$^{\text{4,6}}$ &
  \multicolumn{1}{l|}{\cellcolor{purple!50}84} &
  \multicolumn{1}{l|}{\cellcolor{purple!50}87} &
  \multicolumn{1}{l|}{\cellcolor{purple!50}88} &
  \multicolumn{1}{l|}{\cellcolor{purple!50}93} &
  \multicolumn{1}{l|}{\cellcolor{purple!50}88} &
  \multicolumn{1}{l|}{\cellcolor{purple!50}85} &
  \cellcolor{purple!50}95 \\ \cline{2-9}
 &
  Height phobia$^{\text{4,6}}$  &
  \multicolumn{1}{l|}{\cellcolor{orange!50}78} &
  \multicolumn{1}{l|}{\cellcolor{purple!50}80} &
  \multicolumn{1}{l|}{\cellcolor{orange!50}76} &
  \multicolumn{1}{l|}{\cellcolor{teal!20}63} &
  \multicolumn{1}{l|}{\cellcolor{teal!20}68} &
  \multicolumn{1}{l|}{\cellcolor{orange!50}77} &
  \cellcolor{orange!50}75 \\ \cline{2-9}
 &
  Motion sickness$^{\text{4,6}}$ &
  \multicolumn{1}{l|}{\cellcolor{orange!50}75} &
  \multicolumn{1}{l|}{\cellcolor{teal!20}68} &
  \multicolumn{1}{l|}{\cellcolor{orange!50}77} &
  \multicolumn{1}{l|}{\cellcolor{teal!20}60} &
  \multicolumn{1}{l|}{\cellcolor{teal!20}68} &
  \multicolumn{1}{l|}{\cellcolor{purple!50}80} &
  \cellcolor{gray!45}56 \\ \hline
\multirow{12}{*}{User Interests \& Behaviors} &   

Problem Solving Abilities$^{\text{6}}$ &     \multicolumn{1}{l|}{\cellcolor{gray!45}48} &
  \multicolumn{1}{l|}{\cellcolor{gray!45}48} &
  \multicolumn{1}{l|}{\cellcolor{gray!45}42} &
  \multicolumn{1}{l|}{\cellcolor{gray!45}48} &
  \multicolumn{1}{l|}{\cellcolor{gray!45}49} &
  \multicolumn{1}{l|}{\cellcolor{purple!50}84} &
  \cellcolor{purple!50}81     \\ \cline{2-9}  &  
  
Alcohol consumption$^{\text{1,5,6}}$ &   
  \multicolumn{1}{l|}{\cellcolor{teal!20}62} &
  \multicolumn{1}{l|}{\cellcolor{orange!50}74} &
  \multicolumn{1}{l|}{\cellcolor{teal!20}68} &
  \multicolumn{1}{l|}{\cellcolor{teal!20}63} &
  \multicolumn{1}{l|}{\cellcolor{orange!50}74} &
  \multicolumn{1}{l|}{\cellcolor{teal!20}69} &
  \cellcolor{teal!20}65   \\ \cline{2-9} &
  
VR Experience$^{\text{5,6}}$ &  
  \multicolumn{1}{l|}{\cellcolor{orange!50}74} &
  \multicolumn{1}{l|}{\cellcolor{orange!50}70} &
  \multicolumn{1}{l|}{\cellcolor{teal!20}64} &
  \multicolumn{1}{l|}{\cellcolor{orange!50}72} &
  \multicolumn{1}{l|}{\cellcolor{teal!20}67} &
  \multicolumn{1}{l|}{\cellcolor{orange!50}75} &
  \cellcolor{orange!50}72   \\ \cline{2-9}   &    
Activity preference$^{\text{6}}$ &   
  \multicolumn{1}{l|}{\cellcolor{teal!20}54} &
  \multicolumn{1}{l|}{\cellcolor{gray!45}48} &
  \multicolumn{1}{l|}{\cellcolor{teal!20}63} &
  \multicolumn{1}{l|}{\cellcolor{gray!45}48} &
  \multicolumn{1}{l|}{\cellcolor{orange!50}74} &
  \multicolumn{1}{l|}{\cellcolor{teal!20}52} &
  \cellcolor{teal!20}66    \\ \cline{2-9}        &    

Shooting Experiences$^{\text{1,4,6}}$ &     \multicolumn{1}{l|}{\cellcolor{teal!20}68} &
  \multicolumn{1}{l|}{\cellcolor{orange!50}75} &
  \multicolumn{1}{l|}{\cellcolor{orange!50}76} &
  \multicolumn{1}{l|}{\cellcolor{teal!20}68} &
  \multicolumn{1}{l|}{\cellcolor{orange!50}77} &
  \multicolumn{1}{l|}{\cellcolor{orange!50}78} &
  \cellcolor{purple!50}86   \\ \cline{2-9}  &

  Caffeinated item consumption$^{\text{1,5,6}}$ &
  \multicolumn{1}{l|}{\cellcolor{orange!50}74} &
  \multicolumn{1}{l|}{\cellcolor{orange!50}76} &
  \multicolumn{1}{l|}{\cellcolor{teal!20}61} &
  \multicolumn{1}{l|}{\cellcolor{orange!50}70} &
  \multicolumn{1}{l|}{\cellcolor{orange!50}78} &
  \multicolumn{1}{l|}{\cellcolor{orange!50}72} &
  \cellcolor{orange!50}75
   \\ \cline{2-9}
 &
  Concentration$^{\text{5,6}}$ &
  \multicolumn{1}{l|}{\cellcolor{orange!50}75} &
  \multicolumn{1}{l|}{\cellcolor{teal!20}65} &
  \multicolumn{1}{l|}{\cellcolor{orange!50}74} &
  \multicolumn{1}{l|}{\cellcolor{teal!20}57} &
  \multicolumn{1}{l|}{\cellcolor{purple!50}80} &
  \multicolumn{1}{l|}{\cellcolor{teal!20}69} &
  \cellcolor{purple!50}85 \\ \cline{2-9}
 &
  Violence tolerance$^{\text{1,4,6}}$ &
  \multicolumn{1}{l|}{\cellcolor{purple!50}84} &
  \multicolumn{1}{l|}{\cellcolor{teal!20}66} &
  \multicolumn{1}{l|}{\cellcolor{orange!50}77} &
  \multicolumn{1}{l|}{\cellcolor{teal!20}60} &
  \multicolumn{1}{l|}{\cellcolor{orange!50}73} &
  \multicolumn{1}{l|}{\cellcolor{orange!50}75} &
  \cellcolor{orange!50}70 \\  \cline{2-9}
 &
  Introvert/Extrovert$^{\text{6}}$ &
  \multicolumn{1}{l|}{\cellcolor{orange!50}75} &
  \multicolumn{1}{l|}{\cellcolor{teal!20}67} &
  \multicolumn{1}{l|}{\cellcolor{teal!20}56} &
  \multicolumn{1}{l|}{\cellcolor{orange!50}75} &
  \multicolumn{1}{l|}{\cellcolor{orange!50}75} &
  \multicolumn{1}{l|}{\cellcolor{orange!50}70} &
  \cellcolor{teal!20}62 \\ \cline{2-9}
 &
  Organized/Unorganized$^{\text{6}}$ &
  \multicolumn{1}{l|}{\cellcolor{purple!50}86} &
  \multicolumn{1}{l|}{\cellcolor{purple!50}91} &
  \multicolumn{1}{l|}{\cellcolor{purple!50}91} &
  \multicolumn{1}{l|}{\cellcolor{purple!50}95} &
  \multicolumn{1}{l|}{\cellcolor{purple!50}91} &
  \multicolumn{1}{l|}{\cellcolor{orange!50}74} &
  \cellcolor{purple!50}95 \\ \cline{2-9}
 &
  Social media usage$^{\text{4,6}}$ &
  \multicolumn{1}{l|}{\cellcolor{purple!50}90} &
  \multicolumn{1}{l|}{\cellcolor{teal!20}69} &
  \multicolumn{1}{l|}{\cellcolor{purple!50}81} &
  \multicolumn{1}{l|}{\cellcolor{purple!50}82} &
  \multicolumn{1}{l|}{\cellcolor{orange!50}79} &
  \multicolumn{1}{l|}{\cellcolor{orange!50}83} &
  \cellcolor{purple!50}86 \\ \cline{2-9}
 &
     Openness$^{\text{6}}$ &
  \multicolumn{1}{l|}{\cellcolor{purple!50}83} &
  \multicolumn{1}{l|}{\cellcolor{orange!50}73} &
  \multicolumn{1}{l|}{\cellcolor{orange!50}72} &
  \multicolumn{1}{l|}{\cellcolor{purple!50}84} &
  \multicolumn{1}{l|}{\cellcolor{purple!50}81} &
  \multicolumn{1}{l|}{\cellcolor{orange!50}77} &
  \cellcolor{teal!20}72 \\ \cline{2-9} &
  
  Emotional stability$^{\text{6}}$ &
  \multicolumn{1}{l|}{\cellcolor{orange!50}82} &
  \multicolumn{1}{l|}{\cellcolor{orange!50}77} &
  \multicolumn{1}{l|}{\cellcolor{purple!50}85} &
  \multicolumn{1}{l|}{\cellcolor{purple!50}81} &
  \multicolumn{1}{l|}{\cellcolor{purple!50}83} &
  \multicolumn{1}{l|}{\cellcolor{purple!50}88} &
  \cellcolor{orange!50}75
   \\ \hline
\end{tabular}%
}

\vspace{2mm}

\centering
\renewcommand{\arraystretch}{0.78}
\tiny
\caption{\textbf{User Profiling for Multi-Sensor (BM, FE \& EG) Adversary Across 7 App Groups.} %
}
\label{tab:f1score-eg-bm-fe}
\resizebox{\textwidth}{!}{%
\begin{tabular}{|l|l|lllllll|}
\hline
\multicolumn{1}{|c|}{\multirow{2}{*}{\textbf{Attribute Groups}}} &
  \multicolumn{1}{c|}{\multirow{2}{*}{\textbf{Attributes}}} &
  \multicolumn{7}{c|}{\textbf{App Groups}} \\ \cline{3-9}
\multicolumn{1}{|c|}{} &
  \multicolumn{1}{c|}{} &
  \multicolumn{1}{c|}{Social} &
  \multicolumn{1}{c|}{Flight} &
  \multicolumn{1}{c|}{Shoot} &
  \multicolumn{1}{c|}{Rhythm} &
  \multicolumn{1}{c|}{IN} &
  \multicolumn{1}{c|}{KW} &
  \multicolumn{1}{c|}{Archery} \\ \hline
\multirow{5}{*}{Demographics} &
    Gender$^{\text{1,2,4,5,6}}$ &
  \multicolumn{1}{l|}{\cellcolor{purple!50}90} &
  \multicolumn{1}{l|}{\cellcolor{orange!50}75} &
  \multicolumn{1}{l|}{\cellcolor{purple!50}91} &
  \multicolumn{1}{l|}{\cellcolor{purple!50}90} &
  \multicolumn{1}{l|}{\cellcolor{purple!50}84} &
  \multicolumn{1}{l|}{\cellcolor{purple!50}82} &
  \cellcolor{purple!50}85  \\ \cline{2-9}
 &
 Age$^{\text{1,2,3,4,5,6,\ferpa}}$&
  \multicolumn{1}{l|}{\cellcolor{purple!50}80} &
  \multicolumn{1}{l|}{\cellcolor{orange!50}70} &
  \multicolumn{1}{l|}{\cellcolor{purple!50}85} &
  \multicolumn{1}{l|}{\cellcolor{purple!50}80} &
  \multicolumn{1}{l|}{\cellcolor{purple!50}80} &
  \multicolumn{1}{l|}{\cellcolor{orange!50}77} &
  \cellcolor{orange!50}71  \\ \cline{2-9}
 &
  Ethnicity$^{\text{4,5,6,\sens}}$ &
   \multicolumn{1}{l|}{\cellcolor{purple!50}80} &
  \multicolumn{1}{l|}{\cellcolor{orange!50}75} &
  \multicolumn{1}{l|}{\cellcolor{purple!50}83} &
  \multicolumn{1}{l|}{\cellcolor{orange!50}75} &
  \multicolumn{1}{l|}{\cellcolor{purple!50}85} &
  \multicolumn{1}{l|}{\cellcolor{purple!50}80} &
  \cellcolor{purple!50}81 \\ \cline{2-9}
 &
Marital status$^{\text{1,2,5,6}}$ &
\multicolumn{1}{l|}{\cellcolor{teal!20}58} & %
\multicolumn{1}{l|}{\cellcolor{teal!20}53} & %
\multicolumn{1}{l|}{\cellcolor{orange!50}71}  & %
\multicolumn{1}{l|}{\cellcolor{teal!20}69} & %
\multicolumn{1}{l|}{\cellcolor{teal!20}65} & %
\multicolumn{1}{l|}{\cellcolor{orange!50}70}  & %
\cellcolor{teal!20}64  \\  \hline

\multirow{6}{*}{Anthropometrics} &
 Height$^{\text{3,5,6}}$ &
  \multicolumn{1}{l|}{\cellcolor{purple!50}80} &
  \multicolumn{1}{l|}{\cellcolor{gray!45}49} &
  \multicolumn{1}{l|}{\cellcolor{purple!50}90} &
  \multicolumn{1}{l|}{\cellcolor{purple!50}100} &
  \multicolumn{1}{l|}{\cellcolor{teal!20}52} &
  \multicolumn{1}{l|}{\cellcolor{orange!50}70} &
  \cellcolor{purple!50}90 \\ \cline{2-9}
 &
  Reaction Time$^{\text{3,5,6}}$ &
  \multicolumn{1}{l|}{\cellcolor{purple!50}90} &
  \multicolumn{1}{l|}{\cellcolor{purple!50}95} &
  \multicolumn{1}{l|}{\cellcolor{purple!50}99} &
  \multicolumn{1}{l|}{\cellcolor{purple!50}100} &
  \multicolumn{1}{l|}{\cellcolor{purple!50}100} &
  \multicolumn{1}{l|}{\cellcolor{purple!50}89} &
  \cellcolor{purple!50}100
   \\ \cline{2-9}
 &
  Face Length$^{\text{3,5,6}}$ &
  \multicolumn{1}{l|}{\cellcolor{teal!20}65} &
  \multicolumn{1}{l|}{\cellcolor{teal!20}56} &
  \multicolumn{1}{l|}{\cellcolor{orange!50}70} &
  \multicolumn{1}{l|}{\cellcolor{orange!50}75} &
  \multicolumn{1}{l|}{\cellcolor{purple!50}80} &
  \multicolumn{1}{l|}{\cellcolor{orange!50}71} &
  \cellcolor{orange!50}75 \\ \cline{2-9}
  &
    Arm Length$^{\text{3,5,6}}$ &
\multicolumn{1}{l|}{\cellcolor{teal!20}68} &
\multicolumn{1}{l|}{\cellcolor{gray!45}39} &
\multicolumn{1}{l|}{\cellcolor{teal!20}60} &
\multicolumn{1}{l|}{\cellcolor{teal!20}65} &
\multicolumn{1}{l|}{\cellcolor{teal!20}65} &
\multicolumn{1}{l|}{\cellcolor{teal!20}61} &
\cellcolor{orange!50}75 \\ \cline{2-9}
 &
    Weight$^{\text{3,4,6}}$&
  \multicolumn{1}{l|}{\cellcolor{orange!50}75} &
  \multicolumn{1}{l|}{\cellcolor{gray!45}48} &
  \multicolumn{1}{l|}{\cellcolor{gray!45}40} &
  \multicolumn{1}{l|}{\cellcolor{orange!50}72} &
  \multicolumn{1}{l|}{\cellcolor{orange!50}73} &
  \multicolumn{1}{l|}{\cellcolor{orange!50}70} &
  \cellcolor{orange!50}75 \\ \cline{2-9}
 &
  BMI$^{\text{3,6}}$ &
  \multicolumn{1}{l|}{\cellcolor{orange!50}72} &
  \multicolumn{1}{l|}{\cellcolor{orange!50}73} &
  \multicolumn{1}{l|}{\cellcolor{orange!50}78} &
  \multicolumn{1}{l|}{\cellcolor{teal!20}68} &
  \multicolumn{1}{l|}{\cellcolor{orange!50}70} &
  \multicolumn{1}{l|}{\cellcolor{orange!50}76} &
  \cellcolor{orange!50} 70 \\ \hline

\multirow{6}{*}{Health} &
  Close / Distance vision and lenses$^{\text{1,5,6}}$ &
  \multicolumn{1}{l|}{\cellcolor{orange!50}73} &
  \multicolumn{1}{l|}{\cellcolor{teal!20}62} &
  \multicolumn{1}{l|}{\cellcolor{teal!20}56} &
  \multicolumn{1}{l|}{\cellcolor{teal!20}67} &
  \multicolumn{1}{l|}{\cellcolor{teal!20}68} &
  \multicolumn{1}{l|}{\cellcolor{teal!20}67} &
  \cellcolor{teal!20}64 
   \\ \cline{2-9} &
  Physical fitness$^{\text{1,4,5,6}}$ &
  \multicolumn{1}{l|}{\cellcolor{teal!20}53} &
  \multicolumn{1}{l|}{\cellcolor{teal!20}57} &
  \multicolumn{1}{l|}{\cellcolor{teal!20}52} &
  \multicolumn{1}{l|}{\cellcolor{teal!20}67} &
  \multicolumn{1}{l|}{\cellcolor{teal!20}48} &
  \multicolumn{1}{l|}{\cellcolor{teal!20}58} &
  \cellcolor{gray!45}38 \\ \cline{2-9}
 &
  Anxiety$^{\text{4,7,8}}$ &
  \multicolumn{1}{l|}{\cellcolor{orange!50}76} &
  \multicolumn{1}{l|}{\cellcolor{orange!50}79} &
  \multicolumn{1}{l|}{\cellcolor{orange!50}76} &
  \multicolumn{1}{l|}{\cellcolor{teal!20}68} &
  \multicolumn{1}{l|}{\cellcolor{teal!20}66} &
  \multicolumn{1}{l|}{\cellcolor{orange!50}77} &
  \cellcolor{orange!50}73  \\ \cline{2-9}
 &
  Stress$^{\text{4,6}}$ &
  \multicolumn{1}{l|}{\cellcolor{purple!50}89} &
  \multicolumn{1}{l|}{\cellcolor{purple!50}86} &
  \multicolumn{1}{l|}{\cellcolor{purple!50}93} &
  \multicolumn{1}{l|}{\cellcolor{purple!50}84} &
  \multicolumn{1}{l|}{\cellcolor{purple!50}91} &
  \multicolumn{1}{l|}{\cellcolor{purple!50}86} &
  \cellcolor{purple!50}91\\ \cline{2-9}
 &
  Height phobia$^{\text{4,6}}$  &
  \multicolumn{1}{l|}{\cellcolor{orange!50}70} &
  \multicolumn{1}{l|}{\cellcolor{orange!50}75} &
  \multicolumn{1}{l|}{\cellcolor{orange!50}71} &
  \multicolumn{1}{l|}{\cellcolor{teal!20}69} &
  \multicolumn{1}{l|}{\cellcolor{teal!20}66} &
  \multicolumn{1}{l|}{\cellcolor{teal!20}66} &
  \cellcolor{orange!50}78 \\ \cline{2-9}
 &
  Motion sickness$^{\text{4,6}}$ &
  \multicolumn{1}{l|}{\cellcolor{teal!20}59} &
  \multicolumn{1}{l|}{\cellcolor{teal!20}56} &
  \multicolumn{1}{l|}{\cellcolor{orange!50}73} &
  \multicolumn{1}{l|}{\cellcolor{gray!45}42} &
  \multicolumn{1}{l|}{\cellcolor{teal!20}64} &
  \multicolumn{1}{l|}{\cellcolor{purple!50}80} &
  \cellcolor{gray!45}40 \\ \hline
\multirow{12}{*}{User Interests \& Behaviors} &  

Problem Solving Abilities$^{\text{6}}$ &   
  \multicolumn{1}{l|}{\cellcolor{teal!20}81} &
  \multicolumn{1}{l|}{\cellcolor{teal!20}61} &
  \multicolumn{1}{l|}{\cellcolor{teal!20}66} &
  \multicolumn{1}{l|}{\cellcolor{teal!20}65} &
  \multicolumn{1}{l|}{\cellcolor{teal!20}72} &
  \multicolumn{1}{l|}{\cellcolor{teal!20}52} &
  \cellcolor{teal!20}63    \\ \cline{2-9}  &    
Alcohol consumption$^{\text{1,5,6}}$ &   
  \multicolumn{1}{l|}{\cellcolor{teal!20}63} &
  \multicolumn{1}{l|}{\cellcolor{orange!50}73} &
  \multicolumn{1}{l|}{\cellcolor{teal!20}68} &
  \multicolumn{1}{l|}{\cellcolor{teal!20}63} &
  \multicolumn{1}{l|}{\cellcolor{orange!50}72} &
  \multicolumn{1}{l|}{\cellcolor{teal!20}69} &
  \cellcolor{teal!20}65    \\ \cline{2-9} &    
VR Experience$^{\text{5,6}}$ &  
  \multicolumn{1}{l|}{\cellcolor{orange!50}74} &
  \multicolumn{1}{l|}{\cellcolor{orange!50}70} &
  \multicolumn{1}{l|}{\cellcolor{teal!20}64} &
  \multicolumn{1}{l|}{\cellcolor{orange!50}72} &
  \multicolumn{1}{l|}{\cellcolor{teal!20}68} &
  \multicolumn{1}{l|}{\cellcolor{orange!50}73} &
  \cellcolor{orange!50}74    \\ \cline{2-9}   & 
  
  Activity preference$^{\text{6}}$ &   
  \multicolumn{1}{l|}{\cellcolor{teal!20}61} &
  \multicolumn{1}{l|}{\cellcolor{teal!20}67} &
  \multicolumn{1}{l|}{\cellcolor{teal!20}63} &
  \multicolumn{1}{l|}{\cellcolor{teal!20}62} &
  \multicolumn{1}{l|}{\cellcolor{orange!50}70} &
  \multicolumn{1}{l|}{\cellcolor{teal!20}55} &
  \cellcolor{teal!20}67   \\ \cline{2-9}        &    
Shooting Experiences$^{\text{1,4,6}}$ &   
  \multicolumn{1}{l|}{\cellcolor{orange!50}73} &
  \multicolumn{1}{l|}{\cellcolor{orange!50}76} &
  \multicolumn{1}{l|}{\cellcolor{orange!50}73} &
  \multicolumn{1}{l|}{\cellcolor{orange!50}79} &
  \multicolumn{1}{l|}{\cellcolor{orange!50}77} &
  \multicolumn{1}{l|}{\cellcolor{purple!50}86} &
  \cellcolor{orange!50}75   \\ \cline{2-9}  &
  
  Caffeinated item consumption$^{\text{1,5,6}}$ &
  \multicolumn{1}{l|}{\cellcolor{orange!50}75} &
  \multicolumn{1}{l|}{\cellcolor{orange!50}75} &
  \multicolumn{1}{l|}{\cellcolor{teal!20}61} &
  \multicolumn{1}{l|}{\cellcolor{orange!50}71} &
  \multicolumn{1}{l|}{\cellcolor{orange!50}79} &
  \multicolumn{1}{l|}{\cellcolor{orange!50}72} &
  \cellcolor{orange!50}75
   \\ \cline{2-9}
 &
  Concentration$^{\text{3,6}}$ &
  \multicolumn{1}{l|}{\cellcolor{teal!20}68} &
  \multicolumn{1}{l|}{\cellcolor{teal!20}67} &
  \multicolumn{1}{l|}{\cellcolor{orange!50}72} &
  \multicolumn{1}{l|}{\cellcolor{teal!20}60} &
  \multicolumn{1}{l|}{\cellcolor{teal!20}68} &
  \multicolumn{1}{l|}{\cellcolor{orange!50}77} &
  \cellcolor{purple!50}85 \\ \cline{2-9}
 &
  Violence tolerance$^{\text{1,4,6}}$ &
  \multicolumn{1}{l|}{\cellcolor{teal!20}69} &
  \multicolumn{1}{l|}{\cellcolor{teal!20}68} &
  \multicolumn{1}{l|}{\cellcolor{teal!20}58} &
  \multicolumn{1}{l|}{\cellcolor{teal!20}57} &
  \multicolumn{1}{l|}{\cellcolor{orange!50}70} &
  \multicolumn{1}{l|}{\cellcolor{orange!50}75} &
  \cellcolor{teal!20}68 \\ \cline{2-9}
 &
  Introvert/Extrovert$^{\text{6}}$ &
  \multicolumn{1}{l|}{\cellcolor{teal!20}63} &
  \multicolumn{1}{l|}{\cellcolor{teal!20}62} &
  \multicolumn{1}{l|}{\cellcolor{teal!20}53} &
  \multicolumn{1}{l|}{\cellcolor{orange!50}72} &
  \multicolumn{1}{l|}{\cellcolor{teal!20}68} &
  \multicolumn{1}{l|}{\cellcolor{teal!20}67} &
  \cellcolor{teal!20}60 \\ \cline{2-9}
 &
  Organized/Unorganized$^{\text{6}}$ &
  \multicolumn{1}{l|}{\cellcolor{purple!50}90} &
  \multicolumn{1}{l|}{\cellcolor{purple!50}92} &
  \multicolumn{1}{l|}{\cellcolor{purple!50}91} &
  \multicolumn{1}{l|}{\cellcolor{purple!50}99} &
  \multicolumn{1}{l|}{\cellcolor{purple!50}89} &
  \multicolumn{1}{l|}{\cellcolor{gray!45}47} &
  \cellcolor{purple!50}92 \\ \cline{2-9}
 &
  Social media usage$^{\text{4,6}}$ &
  \multicolumn{1}{l|}{\cellcolor{purple!50}82} &
  \multicolumn{1}{l|}{\cellcolor{teal!20}65} &
  \multicolumn{1}{l|}{\cellcolor{purple!50}92} &
  \multicolumn{1}{l|}{\cellcolor{purple!50}80} &
  \multicolumn{1}{l|}{\cellcolor{orange!50}77} &
  \multicolumn{1}{l|}{\cellcolor{purple!50}82} &
  \cellcolor{purple!50}89 \\ \cline{2-9}
 &
     Openness$^{\text{6}}$ &
 \multicolumn{1}{l|}{\cellcolor{purple!50}84} &
\multicolumn{1}{l|}{\cellcolor{orange!50}70} &
\multicolumn{1}{l|}{\cellcolor{purple!50}84} &
\multicolumn{1}{l|}{\cellcolor{purple!50}84} &
\multicolumn{1}{l|}{\cellcolor{purple!50}82} &
\multicolumn{1}{l|}{\cellcolor{purple!50}80} &
\cellcolor{orange!50}73 \\ \cline{2-9} &
  
  Emotional stability$^{\text{6}}$ &
  \multicolumn{1}{l|}{\cellcolor{purple!50}80} &
  \multicolumn{1}{l|}{\cellcolor{orange!50}80} &
  \multicolumn{1}{l|}{\cellcolor{purple!50}84} &
  \multicolumn{1}{l|}{\cellcolor{purple!50}83} &
  \multicolumn{1}{l|}{\cellcolor{purple!50}82} &
  \multicolumn{1}{l|}{\cellcolor{purple!50}88} &
  \cellcolor{purple!50}75
   \\ \hline
\end{tabular}%
}
\end{table*}

\subsubsection{Sensor Data Collection} \label{app:dataset} In the sensor data collection phase, each participant wore the \vrdevice{} and interacted with all \apps{} apps, first using controllers and then using bare hands. During app interaction, all sensor data (BM with controller, HJ without controller, FE, EG) are being collected using the set-up from ~\cite{jarin2025behavr}. This data collection setup employs the Meta VR device, called \vrdevice{}, and instruments parts of \alvr{}'s source code that receives sensor data from the \vrdevice{}. Note that \alvr{}~\cite{alvr} is an open-source software that can run VR apps on a PC, and the sensor data sent from \vrdevice{} are received by \alvr{} as time series.

Our data collection process was based on real-world scenarios, minimizing control and bias created by a lab environment. For example, in social apps ($a_{1}, a_{2}$) users engage in multi-user environments (\ie{} public rooms) where they can naturally interact with other users (\eg{} waving, talking) and explore app environments (\eg{} walking like humans or gorillas). Social apps support more immersive and realistic interactions than 2D platforms, so we focus on natural user activities driven by their in-app social environments.

\subsubsection{Survey Protocol} \label{app:survey}
Details on demographics, personal history, and anthropometric are discussed in Section \ref{subsec:survey}. Here, we focus %
on health and user interests/behavior as follows:

    \parheading{\em Health.} Attributes were collected through a structured post-session survey response. Attributes such as chronic illness, motion sickness, stress, height phobia were derived from these responses, \eg{} participants rated their physical fitness level and reported whether they experienced stress during our study.

   \parheading{\em User Interests \& Behavior.} Participants self-reported their interests, experiences, and personality traits. These included political orientation, prior VR experiences, among others; \eg{} to gauge social media use, we asked how frequently participants engage with social media (\eg{} Instagram).

\subsubsection{Final Attributes} 
\label{app:Attributes_and_statistics}
 The demographic distributions of the participants are as follows: female is 9 (45\%), male is 11 (55\%). The age ranges is between 20-40 with a median age of 26 and mean age of $\sim28$. The participants’ heights range from 154–190 cm, with a median height of 174 cm and a mean of $\sim172$ cm, weights range from 57–118 kg, with a median weight of 73.5 kg and a mean of $\sim74.6$ kg. Among them, 11 (55\%) of users have prior VR experiences, 9 (45\%) was trained during our study by the authors. More details about user attributes and their statistics %
are shown in Table~\ref{tab:Attributes_and_statistics}.

\subsubsection{Classification} \label{app:classifier}
As discussed in Section~\ref{subsec:model_training}, our attack classifier is designed to demonstrate adversarial capability, while aligning with the distribution of participant survey responses. Our attack classifier design is guided by adversarial goals defined in our threat model (see Section~\ref{subsec:Threat_Model}), with each attribute mapped to relevant threat scenarios as indicated by its superscripts in the taxonomy. For example, \textit{Sex/Gender} (Men 55\%, Women 45\%; superscripts $^{1,2,4,5,6}$) is modeled as a binary classifier aligned with targeted advertising and profiling threats, where coarse gender inference is sufficient for advertiser-driven decision making (superscripts $^{1,2}$). Similarly, \textit{Ethnicity/National Origin} (Asian 65\%, Others 35\%; superscripts $^{4,5,6}$) is modeled as Asian vs.\ Others, reflecting plausible adversarial goals such as discriminatory targeting or safety-and-harm risks (\eg{} Asian hate crime\cite{asian_hate} or harassment), corresponding to the safety and harm threat scenario (see Section~\ref{subsubsec:ThreatModel_ThreatScenarios}).

Classifier granularity is further constrained by the participant distribution. For \textit{age}, while adversaries may aim to infer child vs.\ adult (\eg{} for age gating or regulatory evasion), our cohort contains no participants under 18. We therefore demonstrate a coarse age split ($<30$ vs.\ $\geq30$), which is commonly used in advertising taxonomies \cite{iab_data_transparency} and remains meaningful for targeted advertising threats. These design choices reflect proof-of-concept demonstrations rather than exhaustive evaluation of all adversarial goals.

\subsection{Data Processing and Feature Engineering} \label{app:data_processing}
\subsubsection{Sensor Data Processing}\label{subsec:time_series_abs} Sensor data is received as a time series, segmented into fixed 1-second intervals, referred to as blocks. We summarize the information in the time series of each block with a vector of five statistics, \ie{} maximum, minimum, mean, standard deviation, and median. This summarization was originally proposed in~\cite{miller2020personal} for \bodydata{} and was also used in~\cite{nair2023unique, jarin2025behavr}.

\subsubsection{Feature Engineering} \label{subsec:feature_engineering}
Each sensor group comprises multiple features, defined according to OpenXR standards \cite{openxrcoordinate} and further processed through our data processing pipeline (see Section \ref{subsec:time_series_abs}). There are 33 BM sensor readings, including 3 position and 4 rotation from controllers and the headset, and an additional 3 linear and 3 angular velocity readings for each controller. After the data processing %
, each sensor reading yields 5 statistics, resulting in 165 BM features per block. Similarly for EG, there are 7 readings (3 position and 4 rotation) for each eye, %
providing 46 features. For HJ, there are 182 readings per hand that describe 3 position and 4 rotation readings from each of 26 joints ~\cite{openxrhandtracking}. Finally, we have 364 sensor readings for 2 hands and 1820 features after data processing step.
FE comprises 64 readings %
~\cite{openxrfacetracking} to capture facial expression and emotions. We refer to each sensor reading as an ``element'' %
(See Appendix~\ref{app:feature_analysis}, Table~\ref{tab:feature_interpretation_FE}). After data processing, we obtain 320 features per block.

\subsubsection{Feature Analysis and Interpretation}
\label{app:feature_analysis}
Here, we elaborate on our feature analysis and interpretation approaches, as discussed in Sections~\ref{subsec:data_processing} and ~\ref{subsec:feature_analysis}.
While prior works have applied ad hoc feature sets to support their evaluations, there is no standardized or reproducible process for transforming sensor group features into semantically coherent descriptors. Moreover, previous studies often focused on a single attribute (\eg{} identification \cite{jarin2025behavr, nair2023unique, miller2020personal}) or did not provide any detailed analysis of the feature space~\cite{Nair1kPersonal2023}. Although such approaches may be suitable for domains with limited dimensionality, drawing conclusions across multiple dimensions (attribute vs. app vs. sensors) requires a more systematic and automated method.
For BM, we recall our 165 features from Section ~\ref{subsec:feature_engineering}. We reorganized these 165 attributes into 17 interpretable features that are more suitable to capture users' app-specific activities and characteristics within the context of the BM. For example, the maximum positional features from the headset can be interpreted as the user's height. Detailed descriptions of raw features to interpretable features are described in Table~\ref{tab:feature_interpretation_BM}.
For the FE,
we reorganized these 320 features into 24 interpretable features that are more suitable to capture users' app-specific emotion/valence state as well as facial expression within the context of the FE sensor group (see Table~\ref{tab:feature_interpretation_FE}). 
For the HJ, we recall our 1820 features from Section~\ref{subsec:feature_engineering}. We reorganized these 1820 features into 52 interpretable features as described in Table \ref{tab:feature_interpretation_hj}.
\subsection{Evaluations}  \label{app:evaluation}
\subsubsection{More Detailed on Risk Assessments} \label{app:Risk Assessments}
As described in Section~\ref{subsec:evaluation_metrics}, we rank user attributes by profiling risk using the F1 score, which jointly captures precision and recall and reflects adversarial inference strength. Our risk mapping follows industry ML evaluation practices (\eg{} Encord~\cite{encord}, Arize~\cite{arize}) and government risk assessment frameworks (\eg{} NIST SP~800-30 and the NIST AI RMF~\cite{NIST}). Consistent across these sources, F1 scores of 80--100\% indicate consistently reliable inference and are classified as High/Very High Risk, while F1 scores below 50\% reflect unreliable inference and are classified as Low Risk. Although F1 scores above 70\% are often considered strong in general ML practice~\cite{ML_F1}, industry and NIST frameworks reserve the highest risk classification for performance above 80\%. To avoid conflating emerging and consistently exploitable inference, we split the intermediate range: 50--70\% is classified as Moderate Risk, and 70--80\% as Moderately High Risk, capturing meaningful differences in adversarial capability.

\subsubsection{Results} \label{app:Results}
\begin{figure*}[ht!]
    \centering
    \begin{subfigure}{0.5\textwidth}
        \centering
        \includegraphics[width=\linewidth]{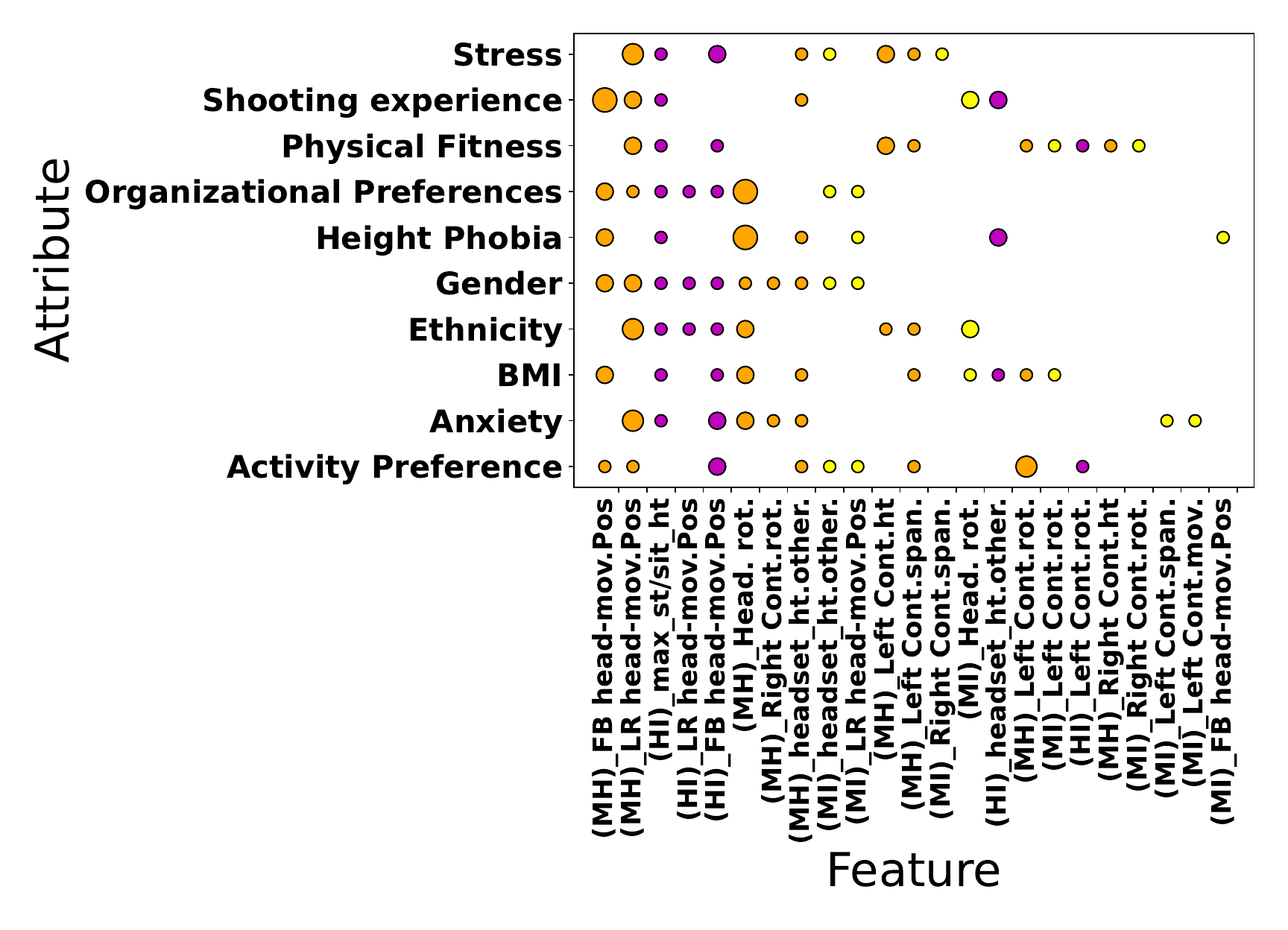}
        \caption{Shooting}
        \label{fig:AppAdv_AccBlockPerUser_motion}
    \end{subfigure}%
    \begin{subfigure}{0.5\textwidth}
        \centering
        \includegraphics[width=.98\linewidth]{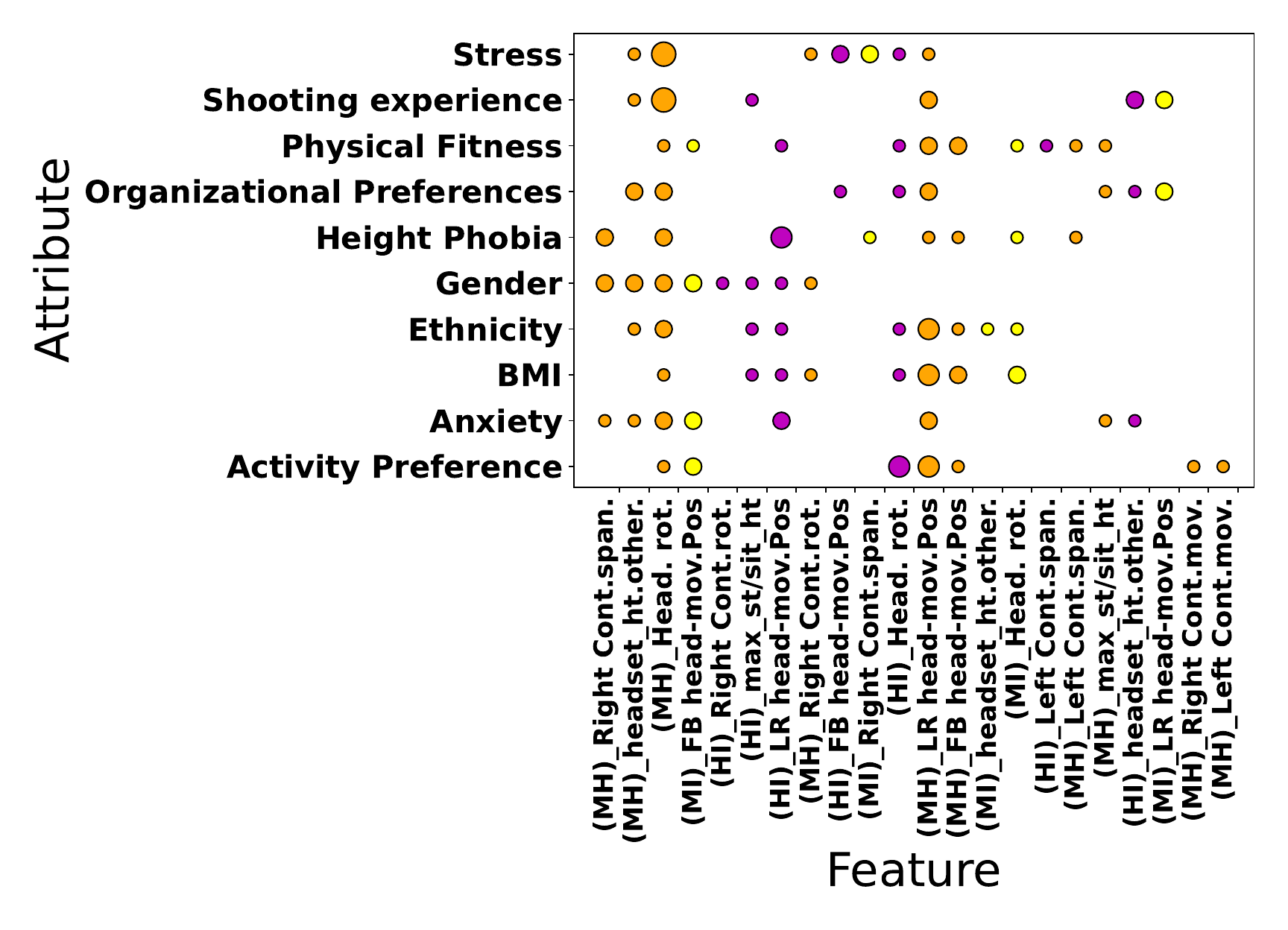}
        \caption{Flight Simulation}
        \label{fig:AppAdv_AccBlockPerUser_Eye}
    \end{subfigure}
    \begin{subfigure}{0.5\textwidth}
        \centering
        \includegraphics[width=\linewidth]{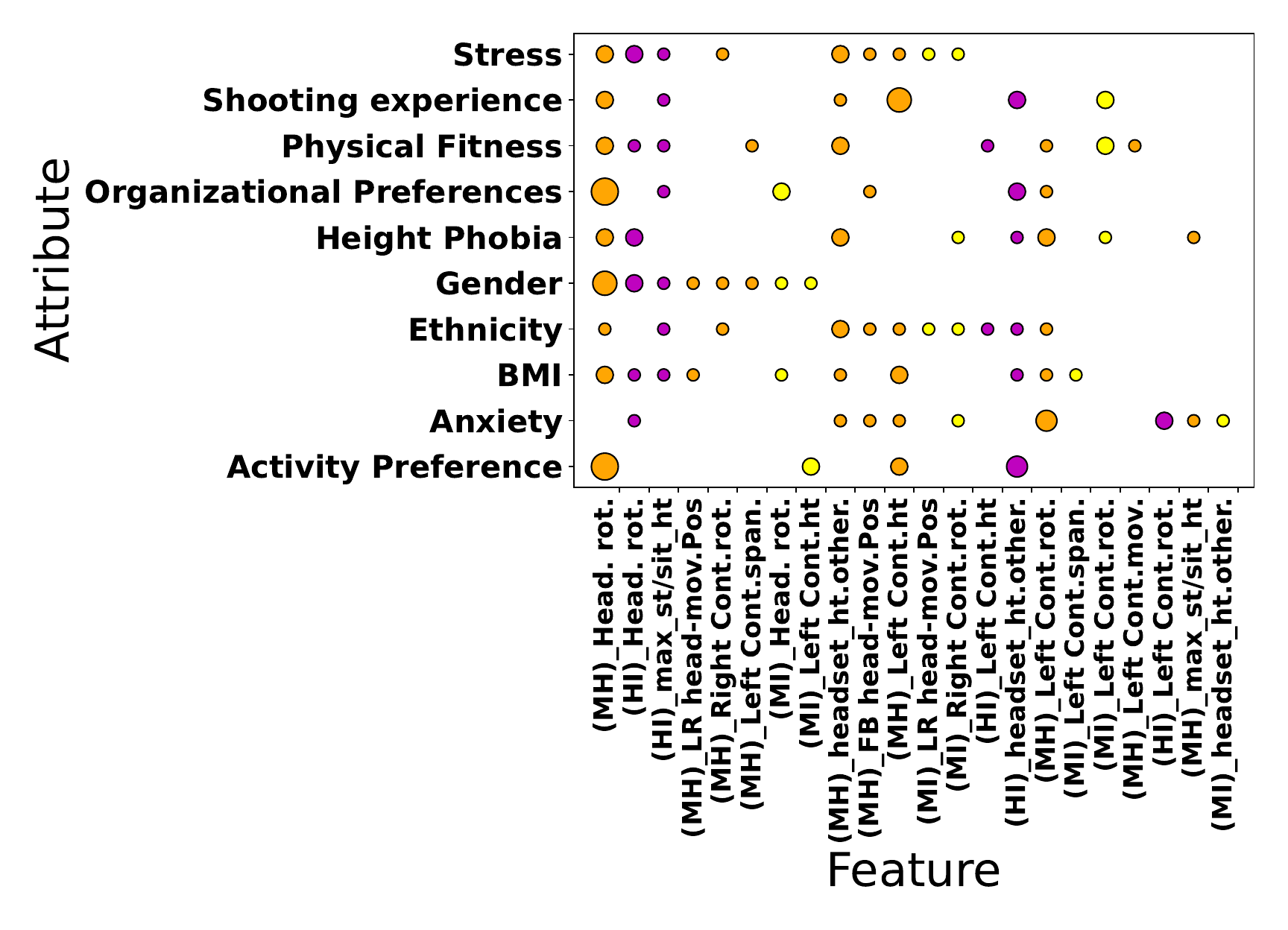}
        \caption{Interactive Navigation}
        \label{fig:AppAdv_AccBlockPerUser_hand}
    \end{subfigure}\hfill
    \begin{subfigure}{0.5\textwidth}
        \centering
        \includegraphics[width=\linewidth]{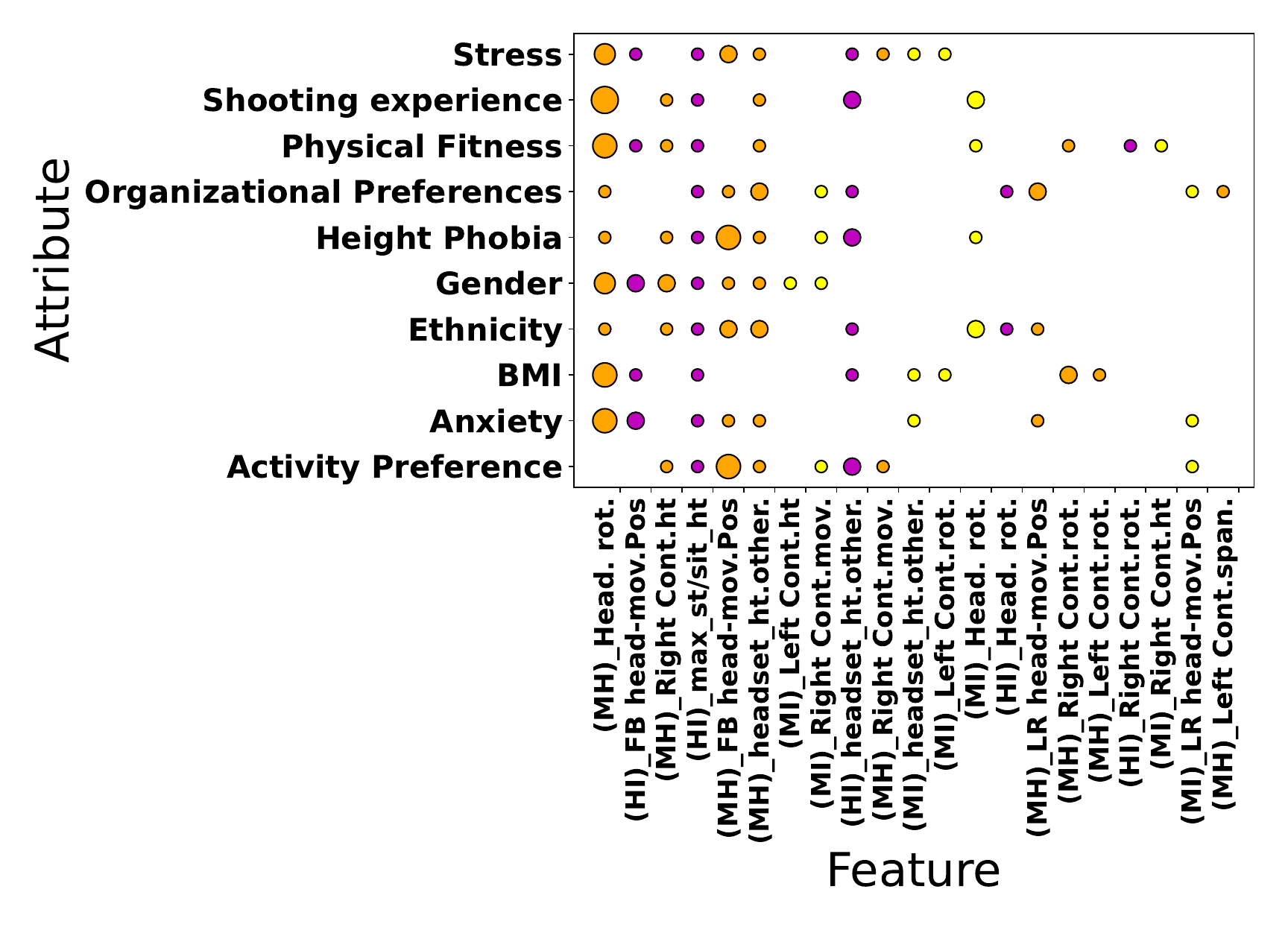}
        \caption{Rhythm}
        \label{fig:AppAdv_AccBlockPerUser_face}
    \end{subfigure}%
    \label{fig:BM_feature}
        \begin{subfigure}{0.5\textwidth}
        \centering
        \includegraphics[width=.92\linewidth]{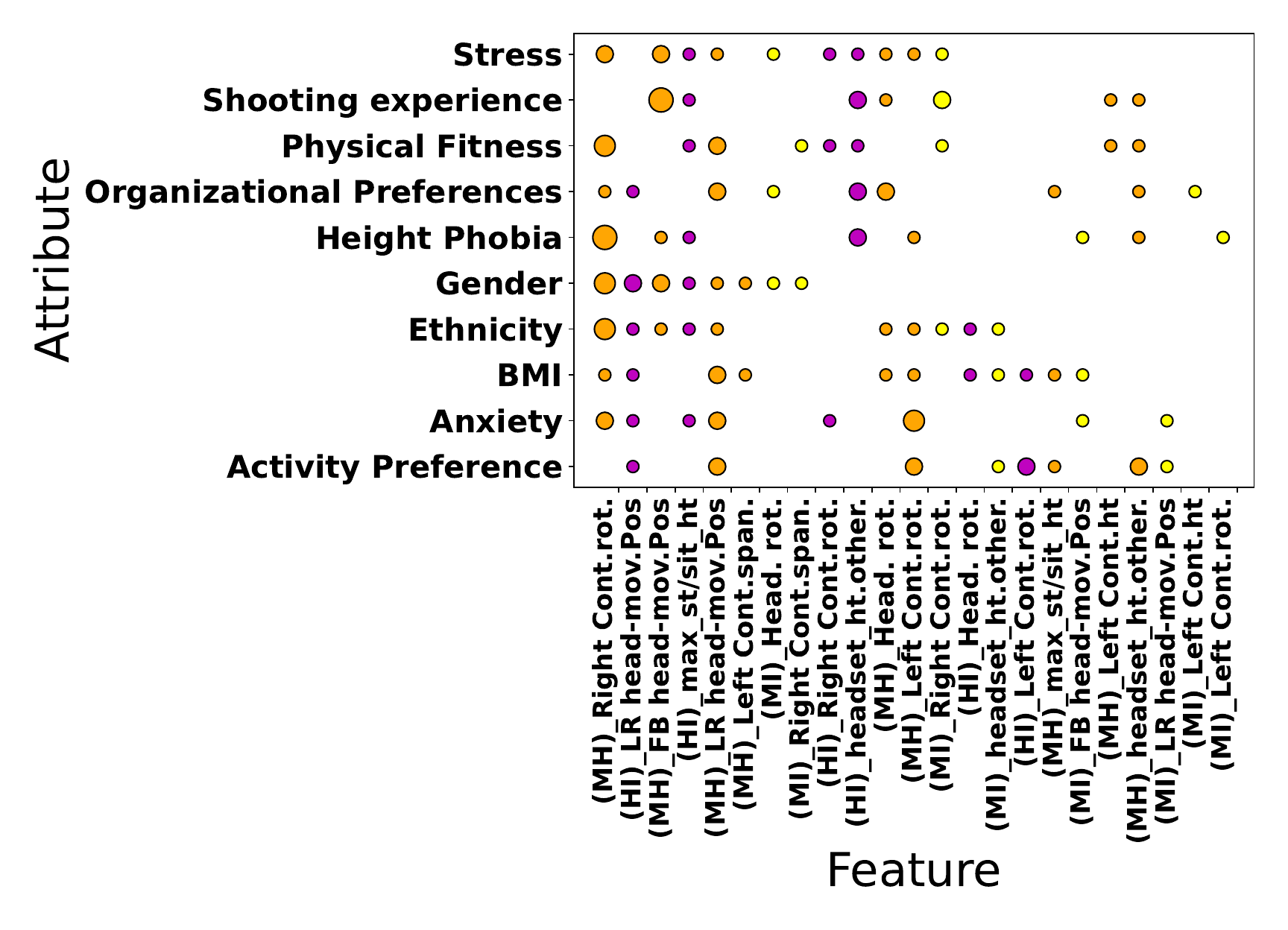}
        \caption{Knuckle Walking}
        \label{fig:min-time-devadv}
    \end{subfigure}
    \begin{subfigure}{0.49\textwidth}
        \centering
        \includegraphics[width=\linewidth]{fig/feature/BM/Archery_BM_feature.pdf}
        \caption{Archery}
        \label{fig:min-time-devadv_hand}
    \end{subfigure}
    \caption{\textbf{Feature Analysis for BM Group across Different App Groups.} Y-axis provides attribute names, X-axis represents corresponding top features for attribute inferences. Color code represents feature ranking: HI (high, pink), MH (medium-high, orange), MI (medium, yellow), while Circle size reflects feature frequency (\ie{} larger circles, higher occurrences).} \label{fig:feature_analysis_bm}
\end{figure*}

\begin{figure*}[ht!]
    \centering
    \begin{subfigure}{0.82\textwidth}
        \centering
        \includegraphics[width=.92\linewidth]{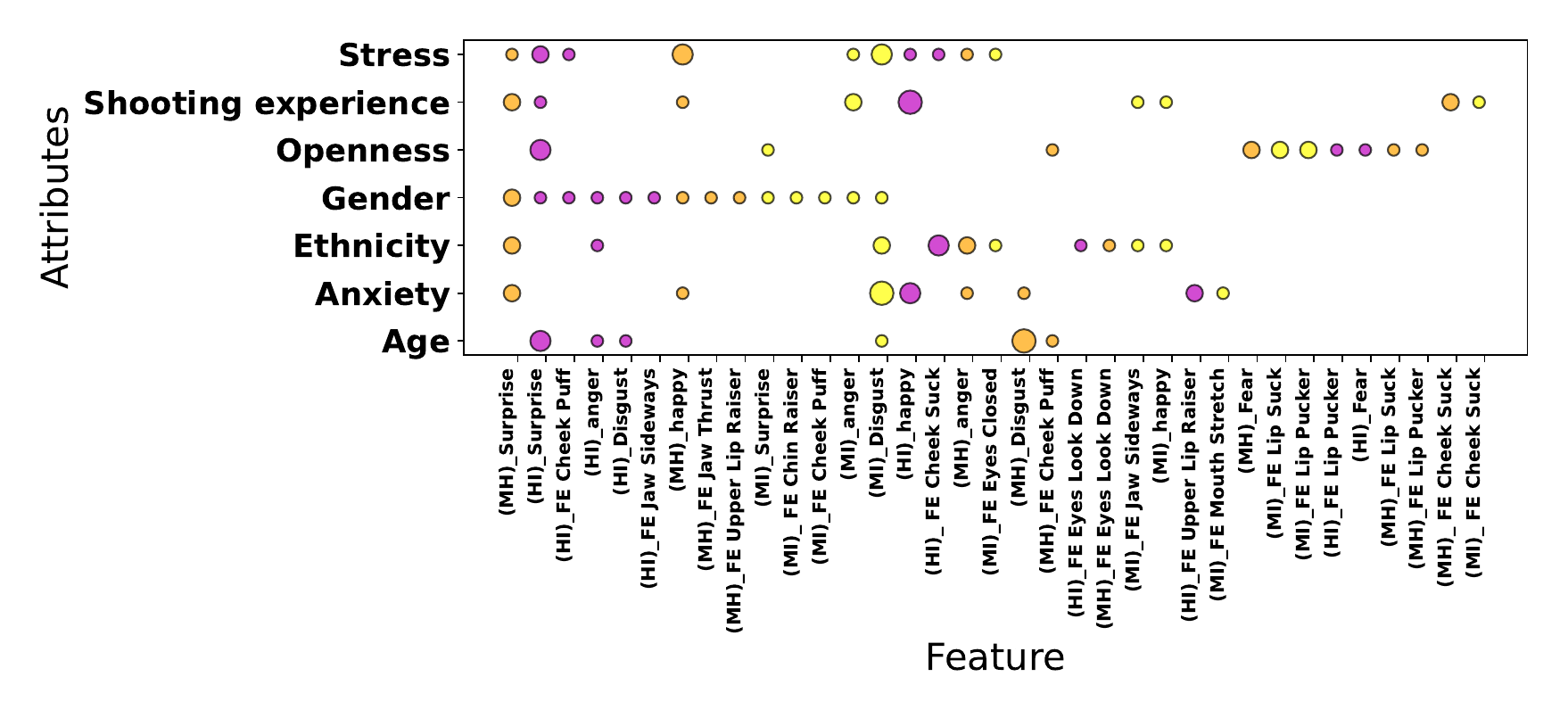}
        \caption{Social}
        \label{fig:AppAdv_AccBlockPerUser_motion}
    \end{subfigure}
    \begin{subfigure}{0.51\textwidth}
        \centering
        \includegraphics[width=\linewidth]{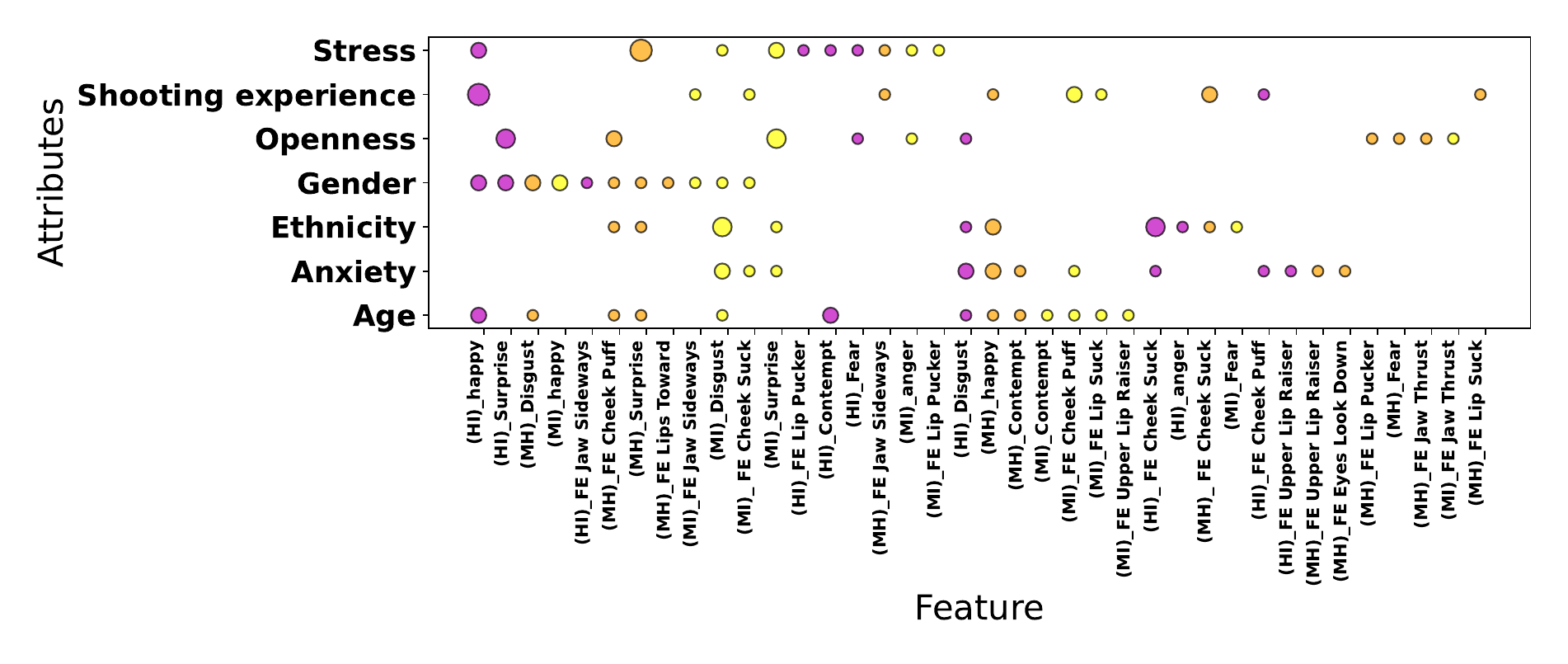}
        \caption{Archery}
        \label{fig:AppAdv_AccBlockPerUser_Eye}
    \end{subfigure}
    \begin{subfigure}{0.48\textwidth}
        \centering
        \includegraphics[width=.98\linewidth]{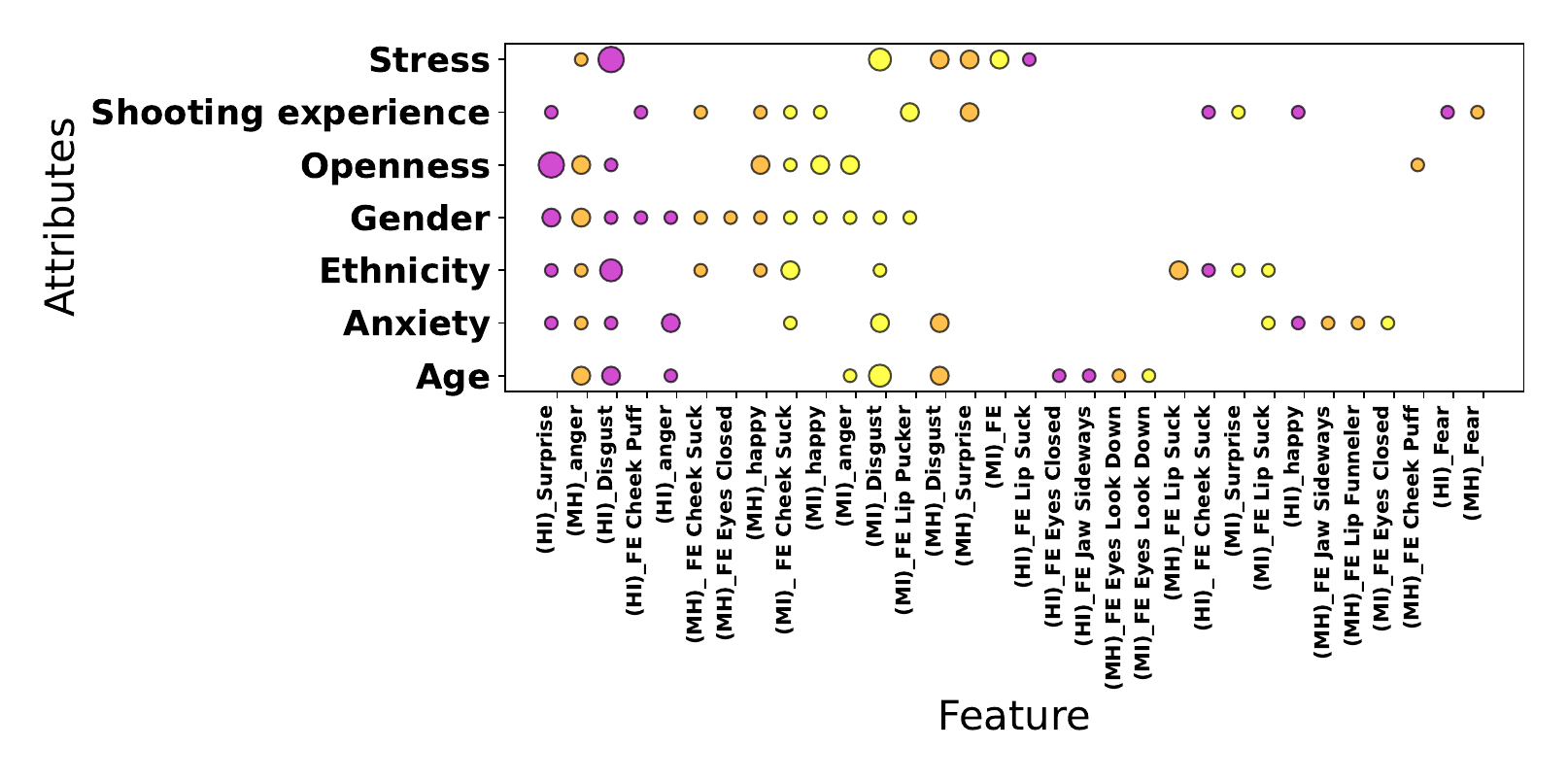}
        \caption{Interactive Navigation}
        \label{fig:AppAdv_AccBlockPerUser_Eye}
    \end{subfigure}
    \begin{subfigure}{0.49\textwidth}
        \centering
        \includegraphics[width=\linewidth]{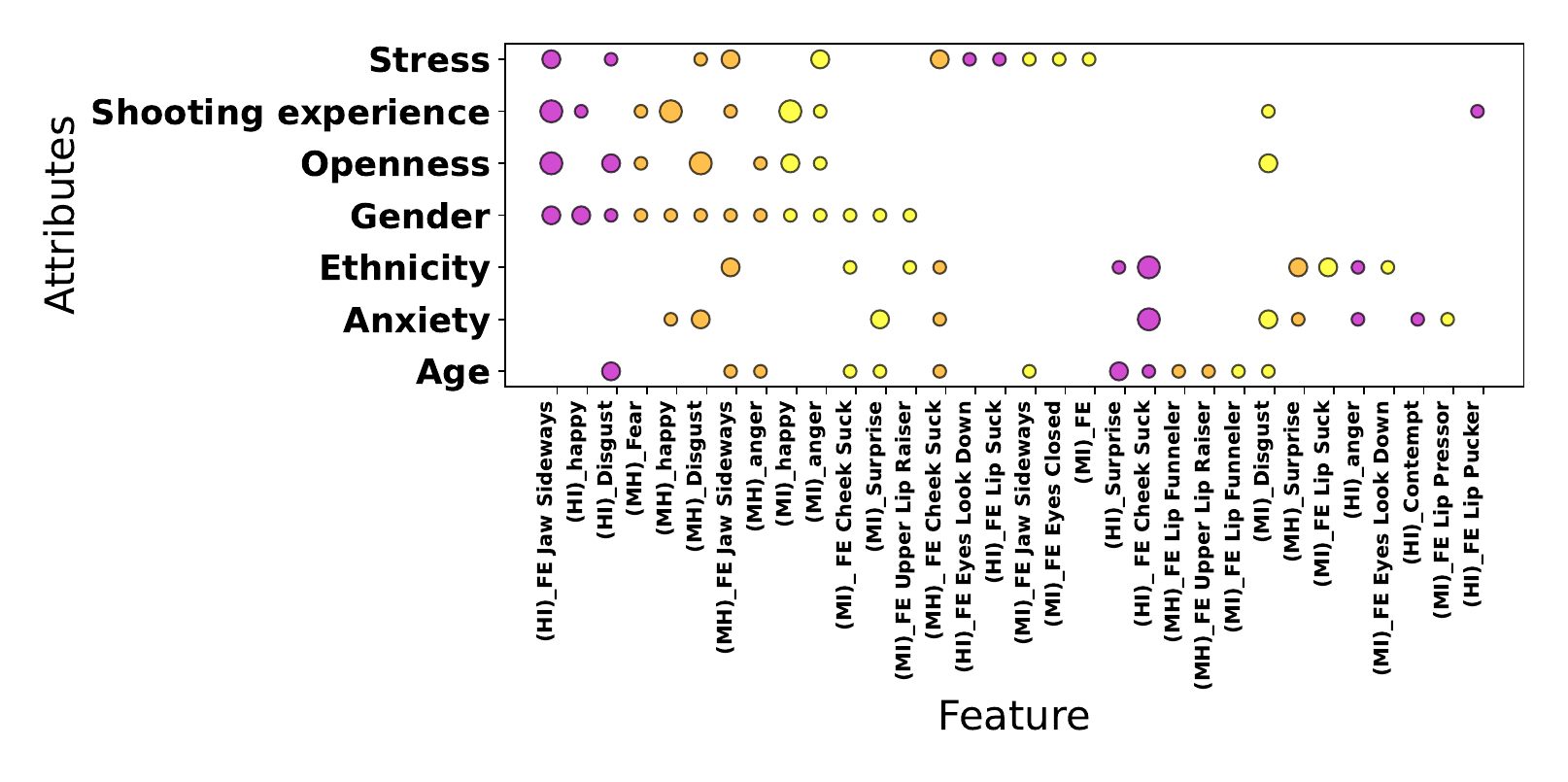}
        \caption{Shooting}
        \label{fig:AppAdv_AccBlockPerUser_hand}
    \end{subfigure}\hfill
    \begin{subfigure}{0.51\textwidth}
        \centering
        \includegraphics[width=1.02\linewidth]{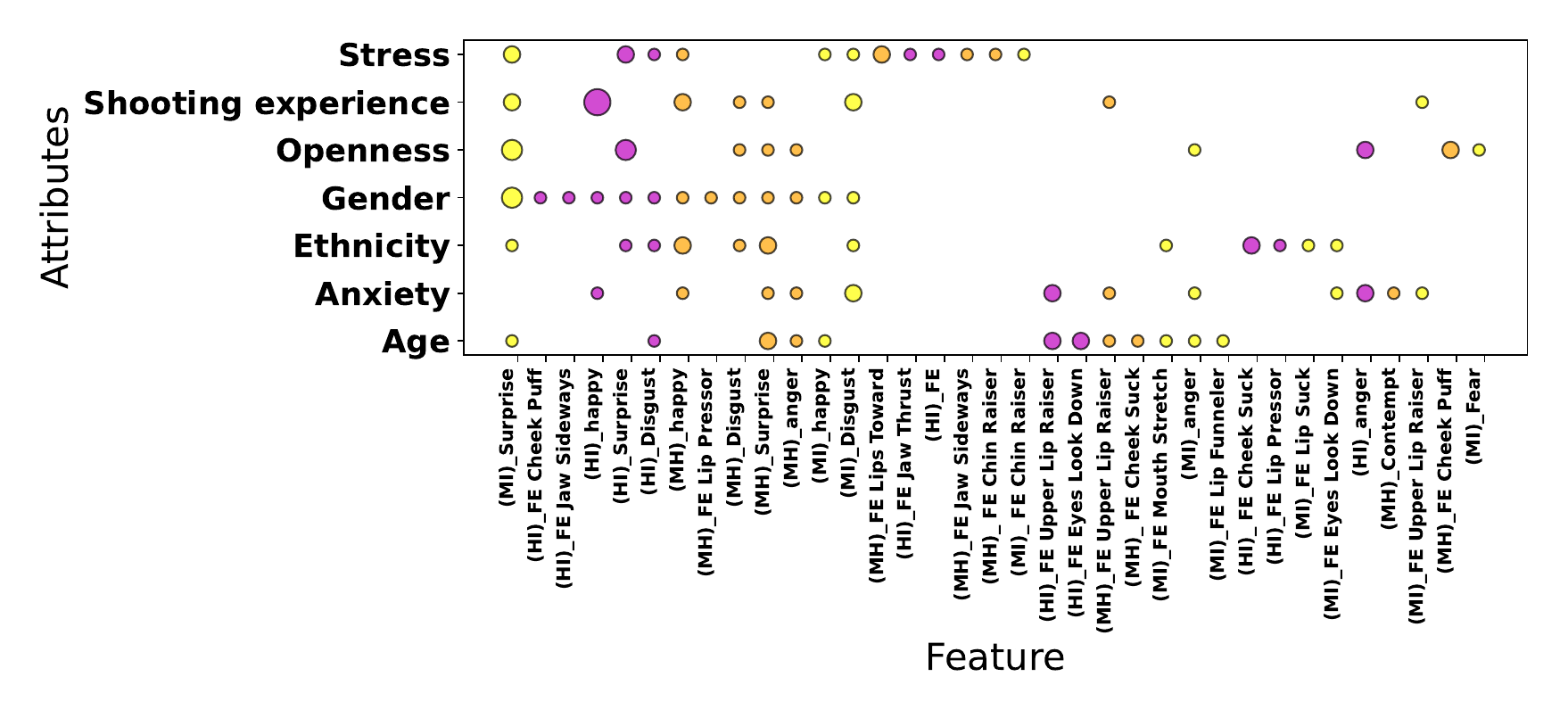}
        \caption{Rhythm}
        \label{fig:AppAdv_AccBlockPerUser_face}
    \end{subfigure}
    \label{fig:BM_feature}
        \begin{subfigure}{0.49\textwidth}
        \centering
        \includegraphics[width=1.05\linewidth]{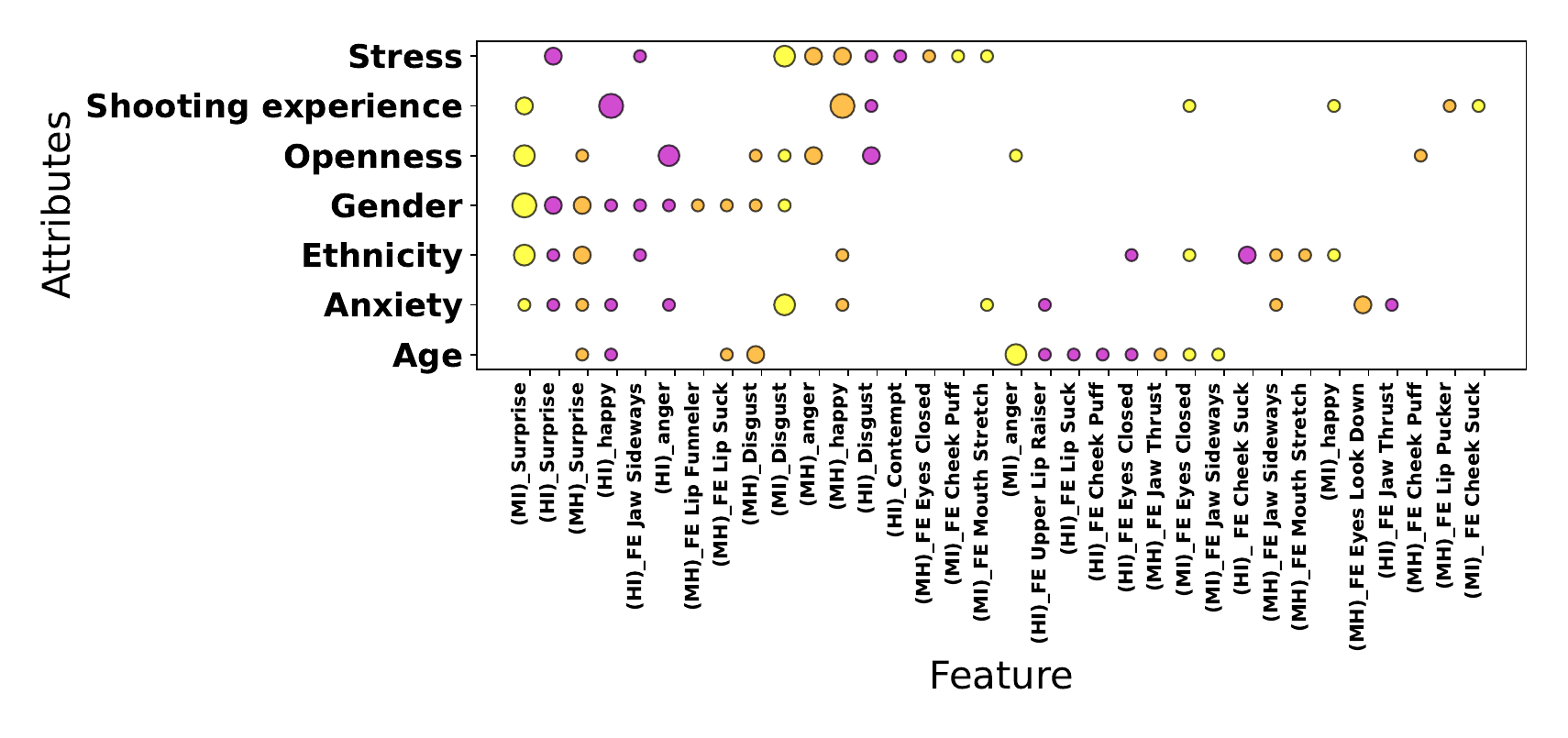}
        \caption{Knuckle-Walking}
        \label{fig:min-time-devadv}
    \end{subfigure}
    \hfill
    \begin{subfigure}{0.5\textwidth}
        \centering
        \includegraphics[width=.96\linewidth]{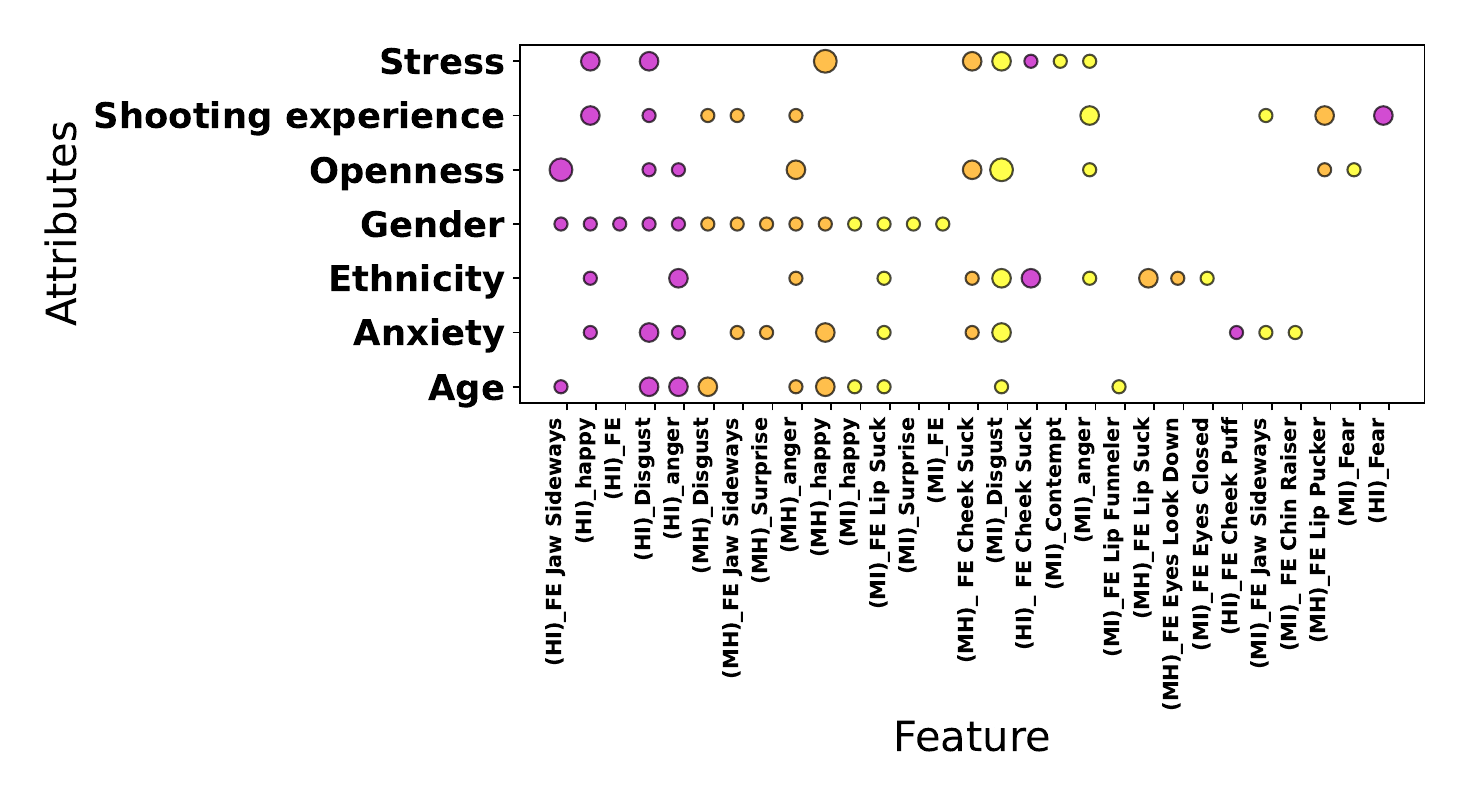}
        \caption{Flight Simulation}
        \label{fig:min-time-devadv_hand}
    \end{subfigure}
    \caption{\textbf{Feature Analysis for FE Sensor Group Across Different App Groups.} Y-axis provides attribute names, and X-axis represents corresponding top features for attribute inferences. Color code represents feature importance ranking: HI (high, pink), MH (medium-high, orange), and MI (medium, yellow), while circle size reflects feature frequency (\ie{} larger circles indicate higher occurrences).} \label{fig:feature_analysis_fe}
\end{figure*}

\begin{figure*}[t!]
    \centering
    \begin{subfigure}{0.67\textwidth}
        \centering
        \includegraphics[width=.95\linewidth]{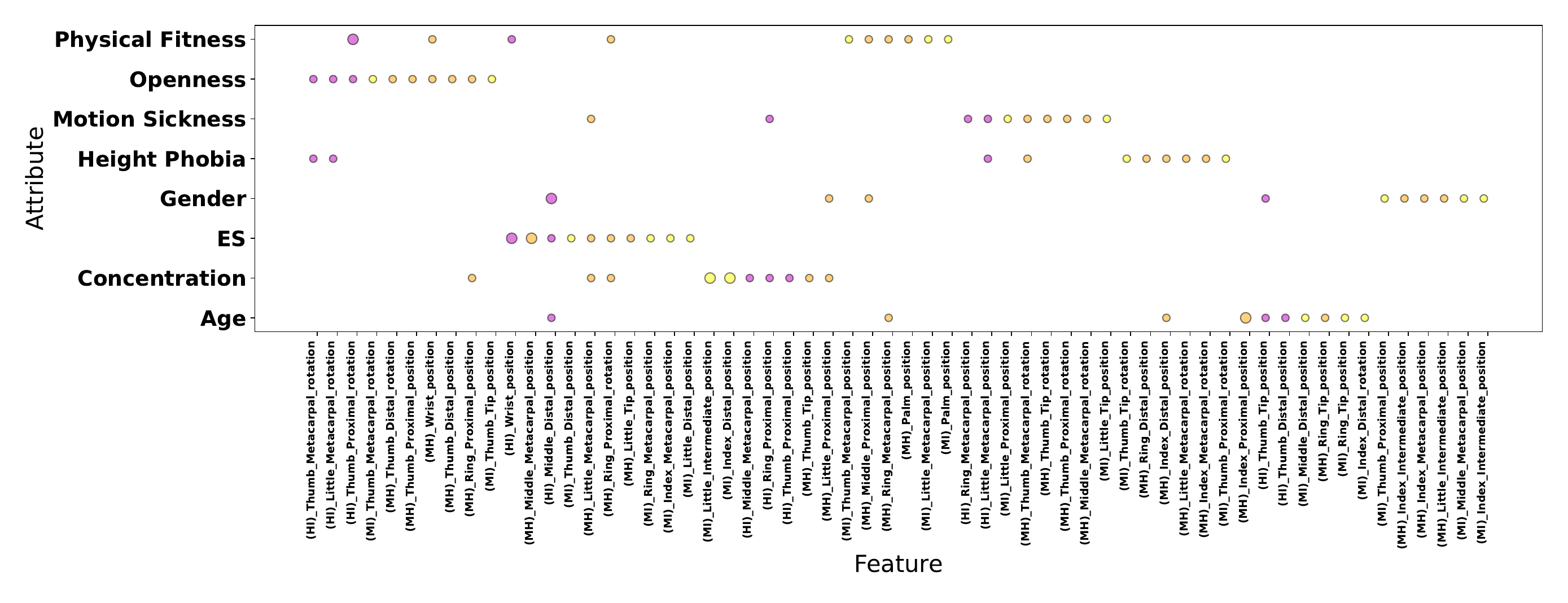}
        \caption{Social}
        \label{fig:AppAdv_AccBlockPerUser_motion}
    \end{subfigure}\hfill
    \begin{subfigure}{0.66\textwidth}
        \centering
        \includegraphics[width=.95\linewidth]{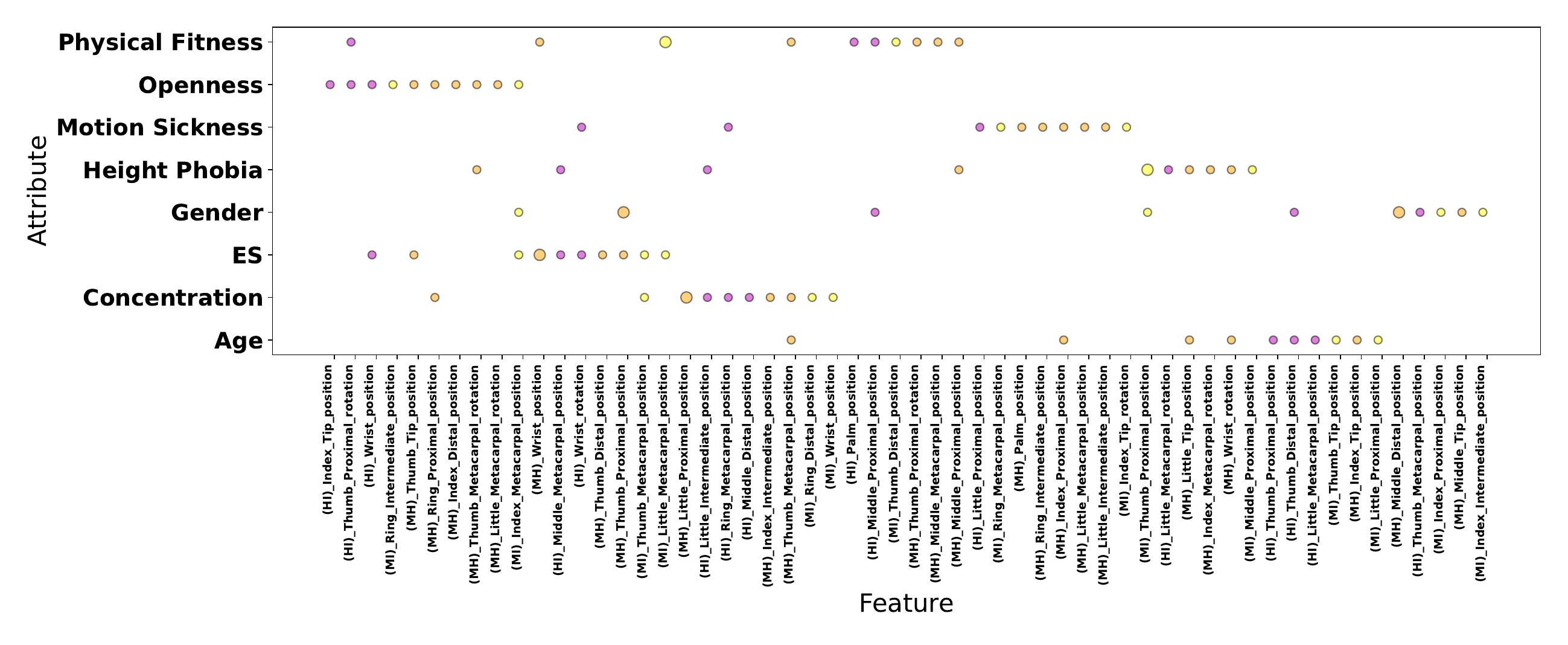}
        \caption{Flight Simulation}
        \label{fig:AppAdv_AccBlockPerUser_Eye}
    \end{subfigure}
    \begin{subfigure}{0.5\textwidth}
        \centering
        \includegraphics[width=\linewidth]{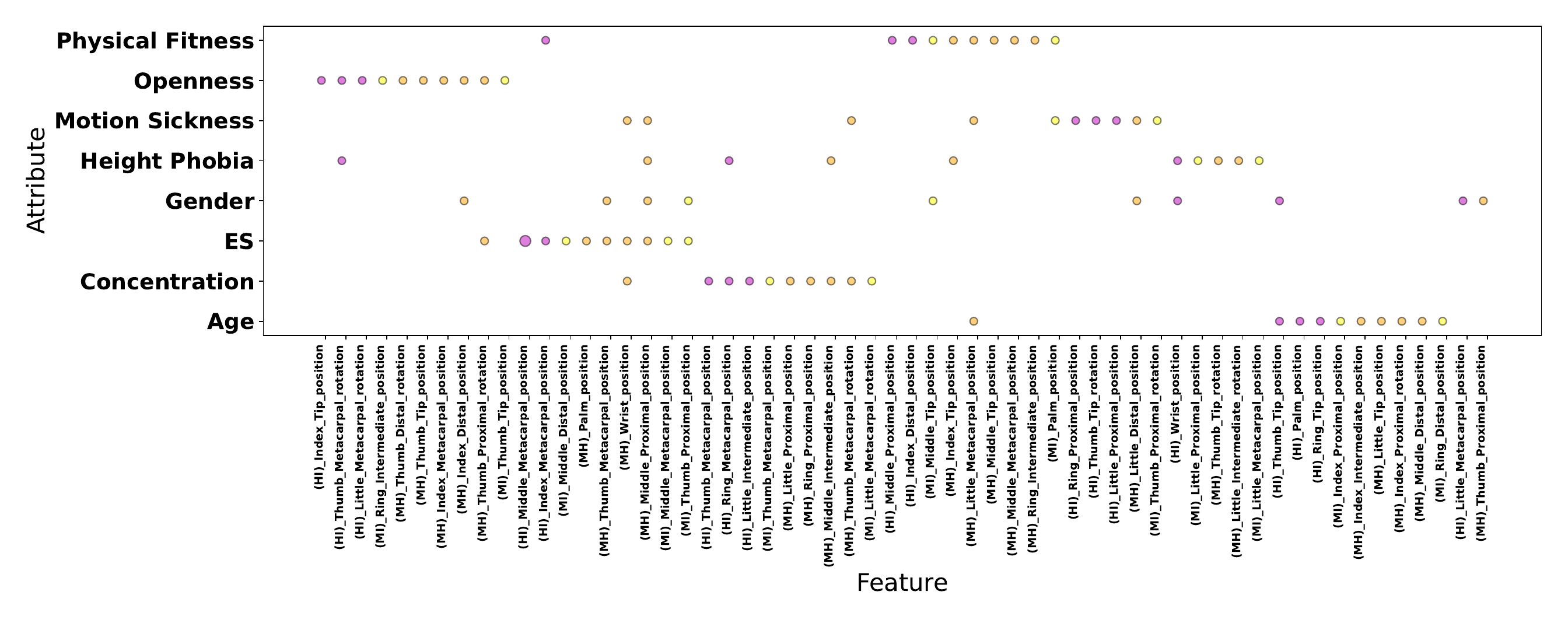}
        \caption{Shooting}
        \label{fig:AppAdv_AccBlockPerUser_hand}
    \end{subfigure}
    \begin{subfigure}{0.48\textwidth}
        \centering
        \includegraphics[width=.95\linewidth]{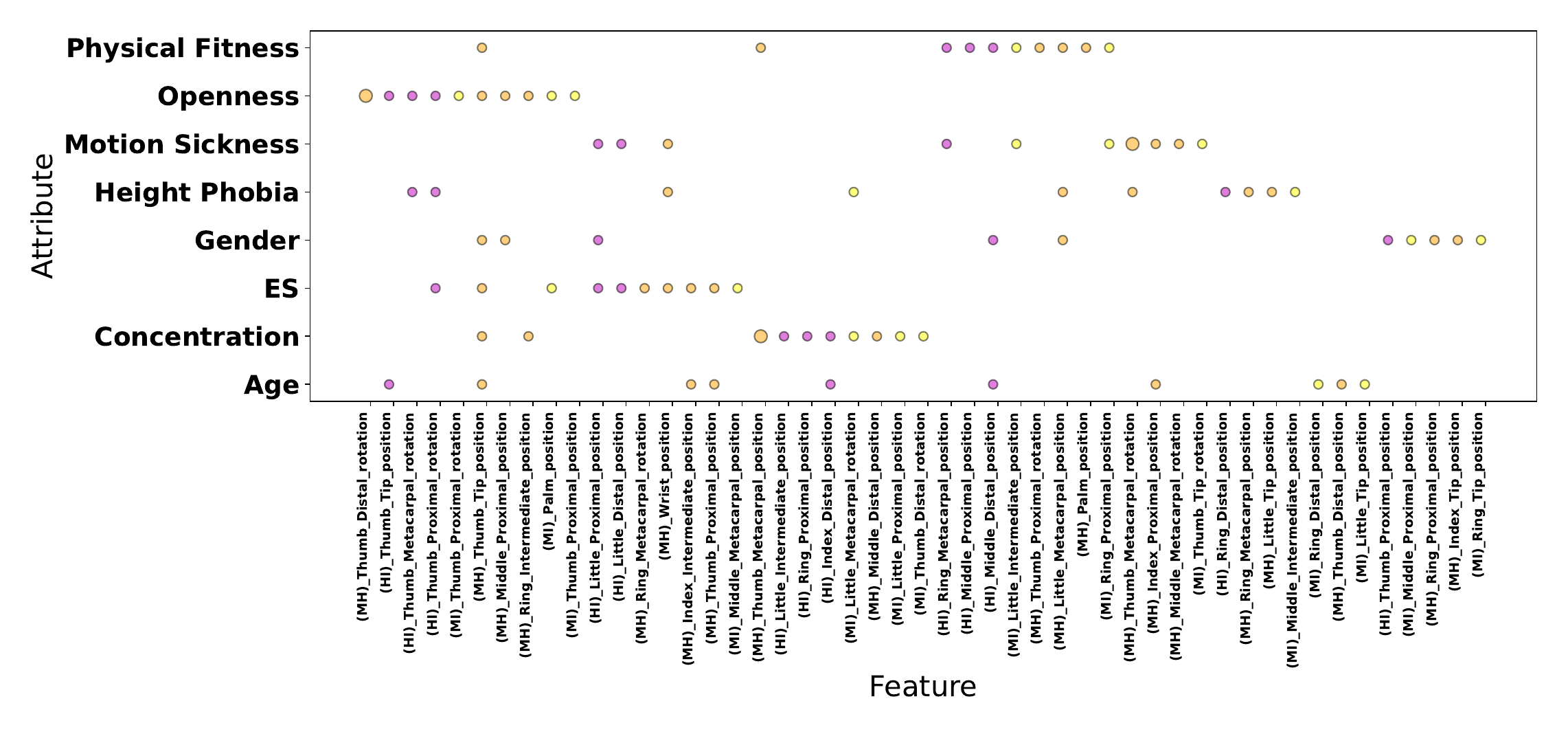}
        \caption{Interactive Navigation}
        \label{fig:AppAdv_AccBlockPerUser_hand}
    \end{subfigure}
    \begin{subfigure}{0.5\textwidth}
        \centering
        \includegraphics[width=\linewidth]{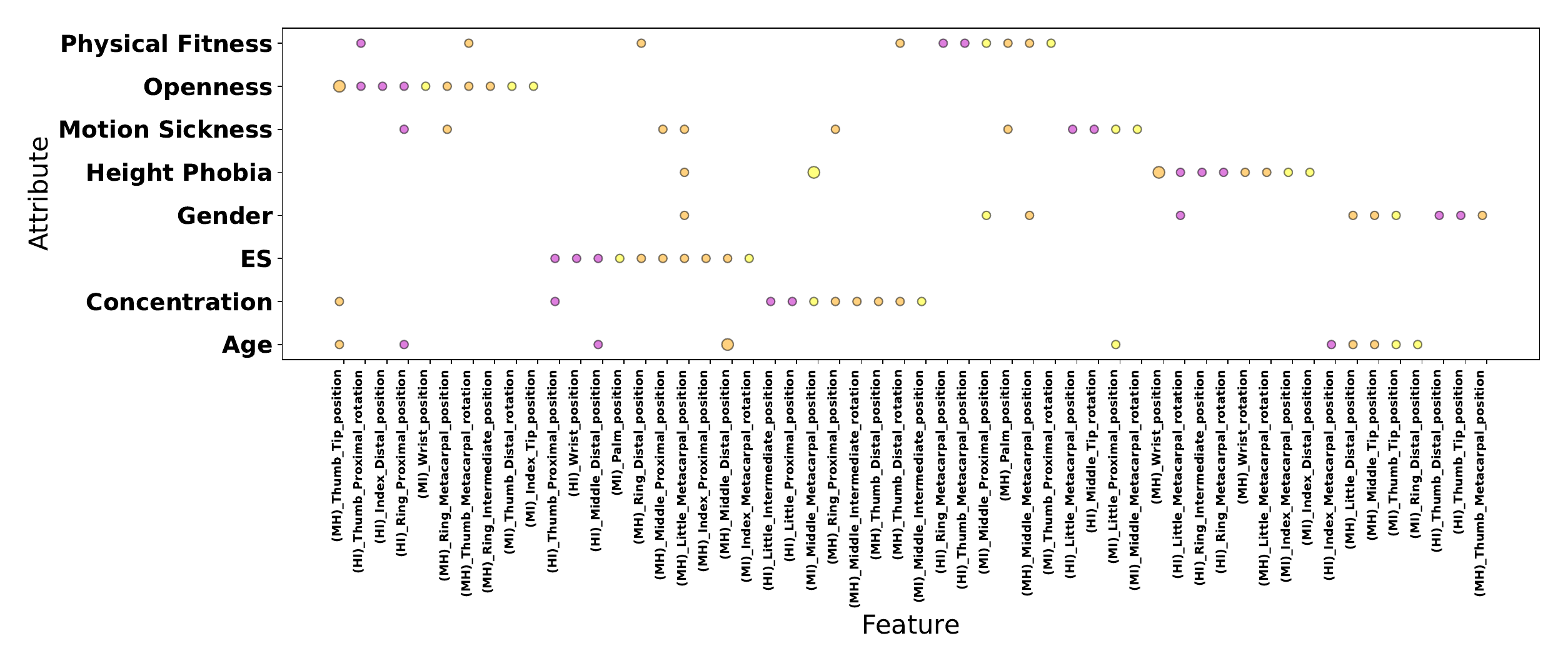}
        \caption{Rhythm}
        \label{fig:AppAdv_AccBlockPerUser_face}
    \end{subfigure}
    \label{fig:BM_feature}
        \begin{subfigure}{0.49\textwidth}
        \centering
        \includegraphics[width=.98\linewidth]{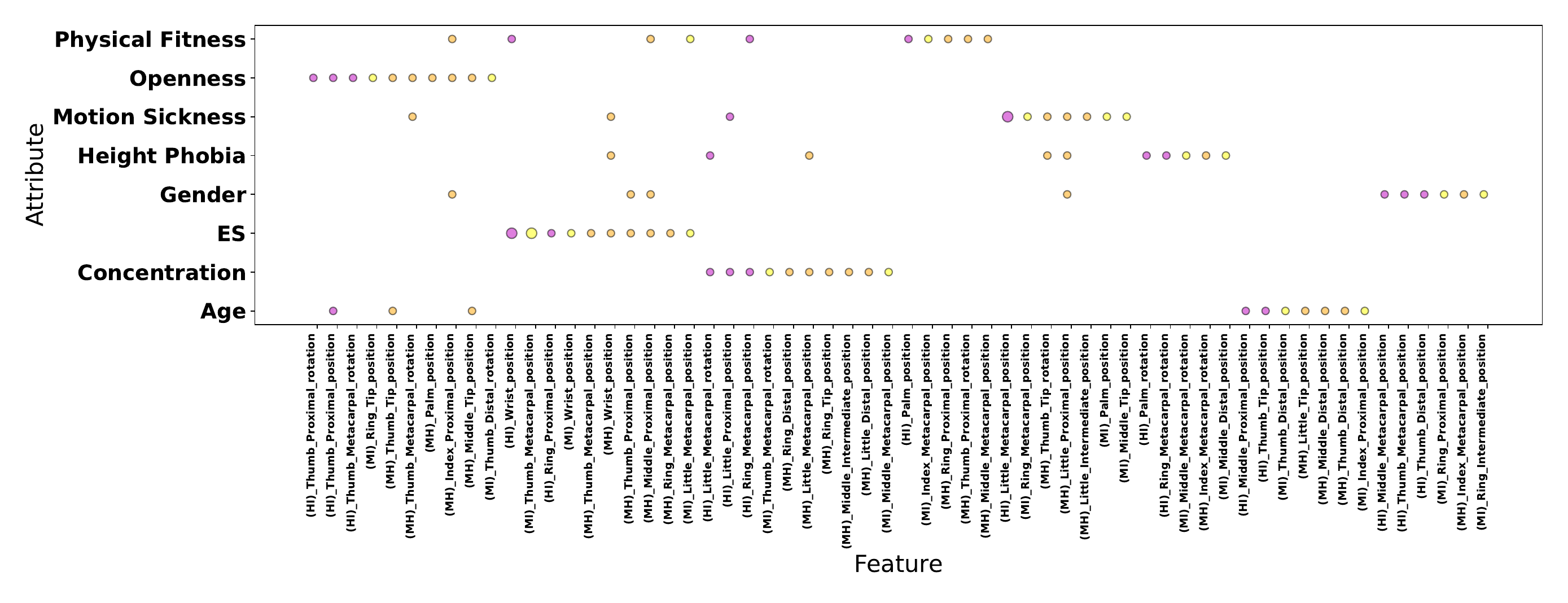}
        \caption{Knuckle-Walking}
        \label{fig:min-time-devadv}
    \end{subfigure}
    \begin{subfigure}{0.5\textwidth}
        \centering
        \includegraphics[width=\linewidth]{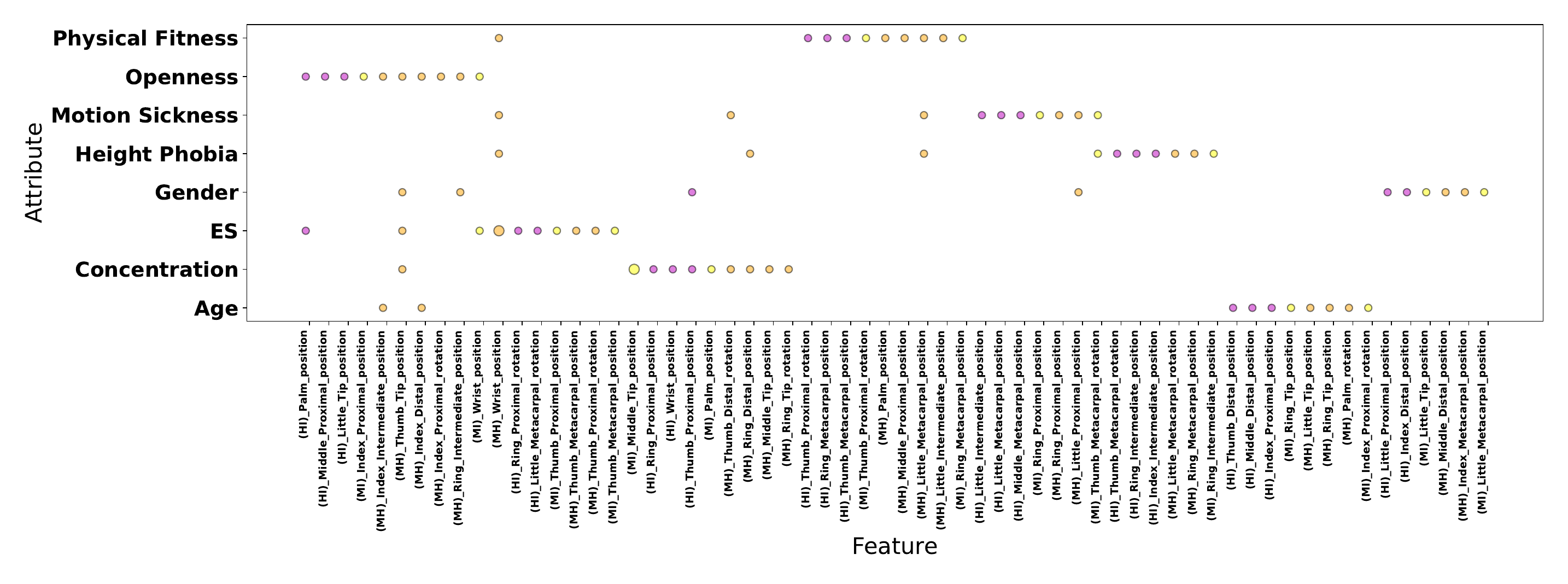}
        \caption{Archery}
        \label{fig:min-time-devadv_hand}
    \end{subfigure}
    \caption{\textbf{Feature Analysis for HJ Sensor Group Across Different App Groups.} Y-axis provides attribute names, and X-axis represents corresponding top features for attribute inferences. Color code represents feature importance ranking: HI (high, pink), MH (medium-high, orange), and MI (medium, yellow).} \label{fig:feature_analysis_hj}
\end{figure*}

This appendix outlines additional results for user attribute inferences and feature analysis to support Section~\ref{sec:evaluation}. The %
inference results for single-sensor adversaries (FE, EG, and HJ) correspond to Tables~\ref{tab:f1score-fe}, \ref{tab:f1score-eg}, and \ref{tab:f1score-hj}, and for multi-sensor adversaries (BM and FE, and BM, FE, and EG) correspond to Tables~\ref{tab:f1score-bm-fe} and \ref{tab:f1score-eg-bm-fe}.
Feature analysis results for attributes with high or moderately high risk across each app group per sensor groups in Figure~\ref{fig:feature_analysis_bm}, ~\ref{fig:feature_analysis_fe} and \ref{fig:feature_analysis_hj} for BM, FE and HJ respectively.


\appendices

\end{document}